\RequirePackage{amsthm}
\documentclass[default,iicol,sn-mathphys,Numbered]{sn-jnl}

\usepackage[utf8]{inputenc} 
\usepackage[T1]{fontenc}    
\usepackage[hyphenbreaks]{breakurl}
\usepackage{graphicx}%
\usepackage{multirow}%
\usepackage{amsmath,amssymb,amsfonts}%
\usepackage{appendix}%
\usepackage{xcolor}%
\usepackage{textcomp}%
\usepackage{manyfoot}%
\usepackage{booktabs}%
\usepackage{algorithm}%
\usepackage{algorithmicx}%
\usepackage{algpseudocode}%
\usepackage{listings}%
\usepackage{hyperref}
\usepackage{geometry} 
\usepackage{svg}
\usepackage{lmodern}
\usepackage{tabularx}
\usepackage{caption}
\usepackage{makecell}
\usepackage{float}
\usepackage[normalem]{ulem}
\useunder{\uline}{\ul}{}

\svgsetup{
    inkscapeexe = your_path/inkscape.exe, 
    inkscapelatex = false                 
}

\theoremstyle{thmstyleone}%
\theoremstyle{thmstyletwo}%

\theoremstyle{thmstylethree}%

\begin{document}

\title[Article Title]{Imaging foundation model for universal enhancement of non-ideal measurement CT}
\author[1]{\fnm{Rongjun} \sur{Ge}}\email{rongjun\_ge@seu.edu.cn}
\equalcont{These authors contributed equally to this work.}

\author[2]{\fnm{Yuxin} \sur{Liu}}\email{liuyuxin@seu.edu.cn}
\equalcont{These authors contributed equally to this work.}

\author[2]{\fnm{Zhan} \sur{Wu}}\email{101300254@seu.edu.cn}

\author[3]{\fnm{Shangwen} \sur{Yang}}\email{13584038729@126.com}

\author[4]{\fnm{Yuan} \sur{Gao}}\email{yuangao@qlu.edu.cn}

\author[5]{\fnm{Chenyu}\sur{You}}\email{chenyu.you@yale.edu}

\author[6]{\fnm{Ge} \sur{Wang}}\email{wangg6@rpi.edu}

\author[7,8]{\fnm{Shuo} \sur{Li}}\email{shuo.li11@case.edu}

\author*[8]{\fnm{Yuting} \sur{He}}\email{yuting.he4@case.edu}

\author*[2]{\fnm{Yang} \sur{Chen}}\email{chenyang.list@seu.edu.cn}

\affil[1]{School of Instrument Science and Engineering, Southeast University, China}

\affil[2]{School of Computer Science and Engineering, Southeast University, China}

\affil[3]{Department of Radiology, Nanjing Drum Tower Hospital, Affiliated Hospital of Medical School, Nanjing University}

\affil[4]{Shandong Fundamental Research Center for Computer Science, Qilu University of Technology (Shandong Academy of Sciences), China}

\affil[5]{Department of Electrical \& Computer Engineering, Yale University, USA}

\affil[6]{Department of Biomedical Engineering, Rensselaer Polytechnic Institute, USA}

\affil[7]{Department of Computer and Data Sciences, Case Western Reserve University, USA}

\affil[8]{Department of Biomedical Engineering, Case Western Reserve University, USA}

\abstract{
Non-ideal measurement computed tomography (NICT) employs suboptimal imaging protocols to expand CT applications. However, the resulting trade-offs degrade image quality, limiting clinical acceptability. Although deep learning methods have been used to enhance NICT images, their reliance on large training datasets and limited generalizability across diverse settings hinder practical use. We propose the multi-scale integrated \textbf{T}ransformer \textbf{AMP}lifier (\textbf{TAMP}), the first imaging foundation model for universal NICT enhancement. Pre-trained on 10.8 million physics-driven simulated NICT images, TAMP generalizes effectively across various NICT settings, defect degrees, and body regions. Moreover, a parameter-efficient fine-tuning strategy enables TAMP to adapt to specific clinical scenarios using only few slices. Extensive experiments, including radiologists and real-world validations, demonstrate that TAMP consistently improves image quality and clinical acceptability, underscoring its significant potential to advance CT imaging and broaden NICT applications in clinical practice.
}


\maketitle

\section{Introduction}\label{sec1}

\begin{figure*}[thbp] 
\centering
\includegraphics[width=\linewidth]{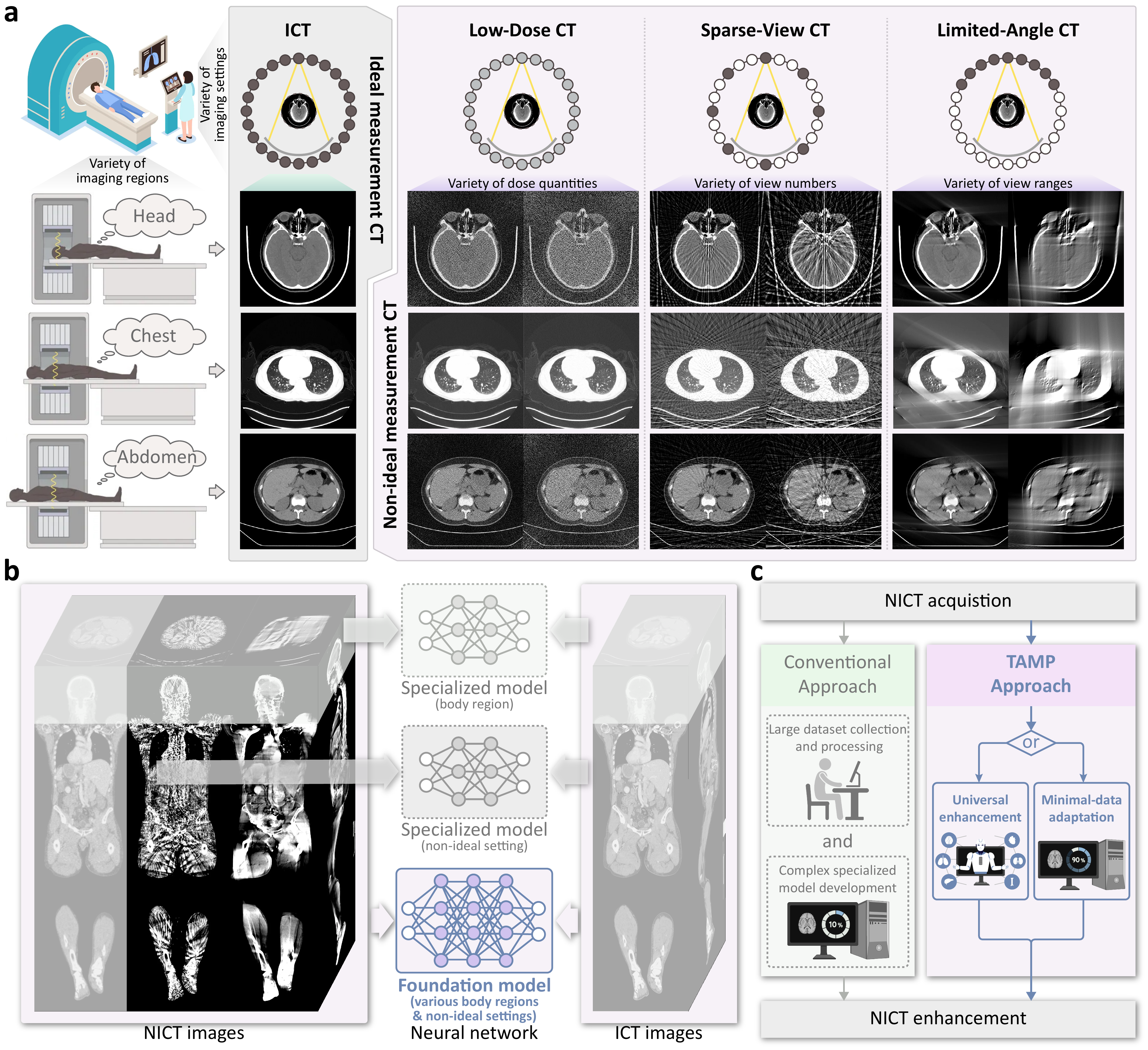}
\caption{Our TAMP is a universal non-ideal measurement computed tomography (NICT) enhancement foundation model that is able to enhance NICT images with various body regions and non-ideal settings and improve the efficiency of developing specialized NICT enhancement models. a) NICT expands the scope of CT applications with the advantages of radiation dose reduction, scanning acceleration, and adaptation of restricted scanning posture. However, the image quality of NICT is reduced, limiting its clinical effectiveness. b) Specialized NICT enhancement models that focus on specific body regions or non-ideal settings, are limited by the application scopes and construction costs. c) Our TAMP can directly enhance NICT images and adapt to specialized NICT enhancement tasks with low data and computational costs, improving imaging applications' development efficiency and effectiveness.}
\label{fig1_intro}
\end{figure*}

Non-ideal measurement computed tomography (NICT) employs suboptimal imaging protocols to expand the scope of CT applications. It employs imaging conditions that deviate from optimal standards \cite{candes2008introduction}, e.g., low-dose CT (LDCT) \cite{kalra2004strategies}, sparse-view CT (SVCT), and limited-angle CT (LACT) \cite{wang2023review}, offering benefits including reduced radiation exposure and compatibility with restricted patient postures. \textbf{However}, these suboptimal protocols compromise image quality, limiting clinical acceptance \cite{jiang2018super}. 
As shown in Fig.\hyperref[fig1_intro]{1a}, LDCT reduces radiation by lowering the tube current or voltage \cite{nakayama2005abdominal}; 
SVCT reduces radiation dose through sparse angle sampling \cite{bian2010evaluation}; 
and LACT acquires projections over a restricted angular range \cite{wu2003tomographic,chen2013limited}, enabling CT feasible in posture-restricted scenarios. 
These NICT settings have been widely applied in clinical practices (e.g., lung cancer screening, breast cancer diagnosis \cite{toyoda2008sensitivity,cui2015screening,boone2001dedicated}) and in medical device development (e.g., slow kVp switching dual energy CT \cite{szczykutowicz2010dual}, C-arm CT \cite{schafer2011mobile}). Compared with standard CT (named ideal measurement CT (ICT) in this paper), NICT suffers from incomplete information acquisition that results in loss of tissue details and increased noise and artifacts \cite{mackin2018effect}, challenging radiologists to accurately identify clinically relevant features and hindering its clinical utility.

Although numerous studies \cite{wang2020deep,chen2017low,jin2017deep,ma2023universal,yang2018low} have demonstrated that deep learning can enhance NICT image quality in specific scenarios, significant challenges remain. 1) Complex specialized model development. Each new NICT protocol requires extensive dataset collection, model architecture design, and training from scratch, extending the development cycle of NICT imaging devices (Fig.\hyperref[fig1_intro]{1c}). 2) Limited transfer learning capability. These studies typically target specific body regions (e.g., head, chest, abdomen) and NICT settings (LDCT \cite{chen2017low}, SVCT \cite{jin2017deep}, LACT \cite{wang2020deep}), as shown in Fig.\hyperref[fig1_intro]{1b}. When emerging NICT devices or updated protocols are introduced in clinical practice, these specialized models cannot effectively transfer their learned knowledge to new scenarios with limited available data \cite{liang2020generalizability,shan2019novel,zeng2022performance}. Overall, this specialized model development paradigm hinders the widespread and rapid clinical deployment of AI-integrated NICT technologies.

Foundation models (FMs) offer potential to address these challenges through two key capabilities (Fig.\hyperref[fig1_intro]{1c}) \cite{zhang2024generalist}: \textbf{a) Universal enhancement} enables rapid validation of emerging NICT protocols. FMs' generalization across diverse NICT settings supports training-free quality enhancement for a broad range of protocols, enabling rapid assessment of new device potential for clinical deployment. \textbf{b) Efficient adaptation} enables low-cost development of high-performance specialized models. FMs' transferable prior knowledge improves the emergence of specialized models with limited fine-tuning data, enabling low-cost deployment of models that surpass training-from-scratch approaches.

However, two main challenges have so far hindered their success in this domain: \textbf{a) Data quantity.} Ethical concerns \cite{cheplygina2019not} restrict the creation of large datasets for NICT FM training. The inherent radiation risks of CT scanning make it unethical to repeatedly scan individuals solely for data collection \cite{mayo2008radiation}. This limitation prevents the direct acquisition of large NICT datasets, thereby compromising the model's ability to generalize in universal scenarios \cite{li2020domain}. \textbf{b) Data variation.} Different physical processes in NICT settings lead to highly varied defect patterns. For example, LDCT images are characterized by fine-grained noise, whereas LACT images exhibit pronounced angular defects (Fig. \hyperref[fig1_intro]{1a}). This variability poses a challenge for universal NICT enhancement models, which struggle to accommodate a wide spectrum of defect patterns. Additionally, existing specialized NICT enhancement models focus on specific defect types within particular NICT settings, limiting their capacity for universal representation and learning during FM training.

In this paper, we propose the multi-scale integrated \textbf{T}ransformer \textbf{AMP}lifier (\textbf{TAMP}), an imaging FM for universal enhancement of NICT images in axial CT systems (e.g., diagnostic CT, cone-beam CT). TAMP leverages a physics-driven pre-training and parameter-efficient adaptation process for universal NICT enhancement ability and adaptation with low costs. The contributions of this work are summarized as follows:
\begin{itemize}
\item To the best of our knowledge, TAMP is the first imaging FM for universal NICT enhancement. It has a powerful generalization ability that is beneficial to the enhancement of diverse NICT images including the LDCT, SVCT, and LACT across the various body regions including the head, chest, abdomen, and lower-limbs. It will reduce the data and computational requirements in model development to enhance the subjective quality and clinical acceptability of NICT images, demonstrating strong potential for clinical application. 
Therefore, as shown in Fig.\hyperref[fig1_intro]{1c}, it has two advanced properties, i.e., universal enhancement and efficient adaptation.
\item \textbf{Universal enhancement:} We propose a physics-driven pre-training paradigm for large-scale training of the NICT enhancement FM (Fig.\hyperref[fig1_intro]{1b}). By simulating the defects that meet the physical principles of non-ideal measurement in the CT's projection domain \cite{kalra2004strategies,wang2023review}, NICT images are synthesized from ICT images for a large-scale NICT-ICT paired dataset. Then, a multi-scale integrated transformer network (MITNet) is designed to represent the multi-granularity defect features in the varied NICT data. Finally, a dual-domain enhancement learning (DDEL) is constructed to learn the \textit{universal NICT enhancement} both in image and projection domains. Based on the above methods, our TAMP will be able to be generalized to multiple NICT settings across different body regions. 

\item \textbf{Efficient adaptation:} We implement a parameter-efficient fine-tuning approach to optimize the performance of TAMP in specific scenarios, with low data and computational costs (Fig.\hyperref[fig1_intro]{1c}). This strategy employs low-rank adaptation (LoRA) \cite{hu2021lora}, enabling TAMP to adjust only a small number of parameters across the entire network, ensuring efficient and stable convergence without the need for excessive parameter training. With limited training samples (5 NICT-ICT image pairs) and iterations (20 epochs), TAMP rapidly adapts to specific NICT settings and body regions, demonstrating its practical viability for clinical data and time-constrained scenarios. 

\item We constructed and publicly released a large-scale simulated NICT dataset (SimNICT), providing researchers a valuable resource for exploring deep learning methods for NICT enhancement. The dataset comprises 10.8 million NICT-ICT image pairs simulated from 9,638 ICT volumes (3.6 million images), covering LDCT, SVCT, and LACT settings with varying defect degrees across head, chest, abdomen, and lower-limbs. SimNICT facilitates efficient data acquisition for NICT enhancement model development and establishes a benchmark for performance evaluation.
\end{itemize}

\section{Results}

\begin{figure*}[thbp] 
\centering
\includegraphics[width=0.9\linewidth]{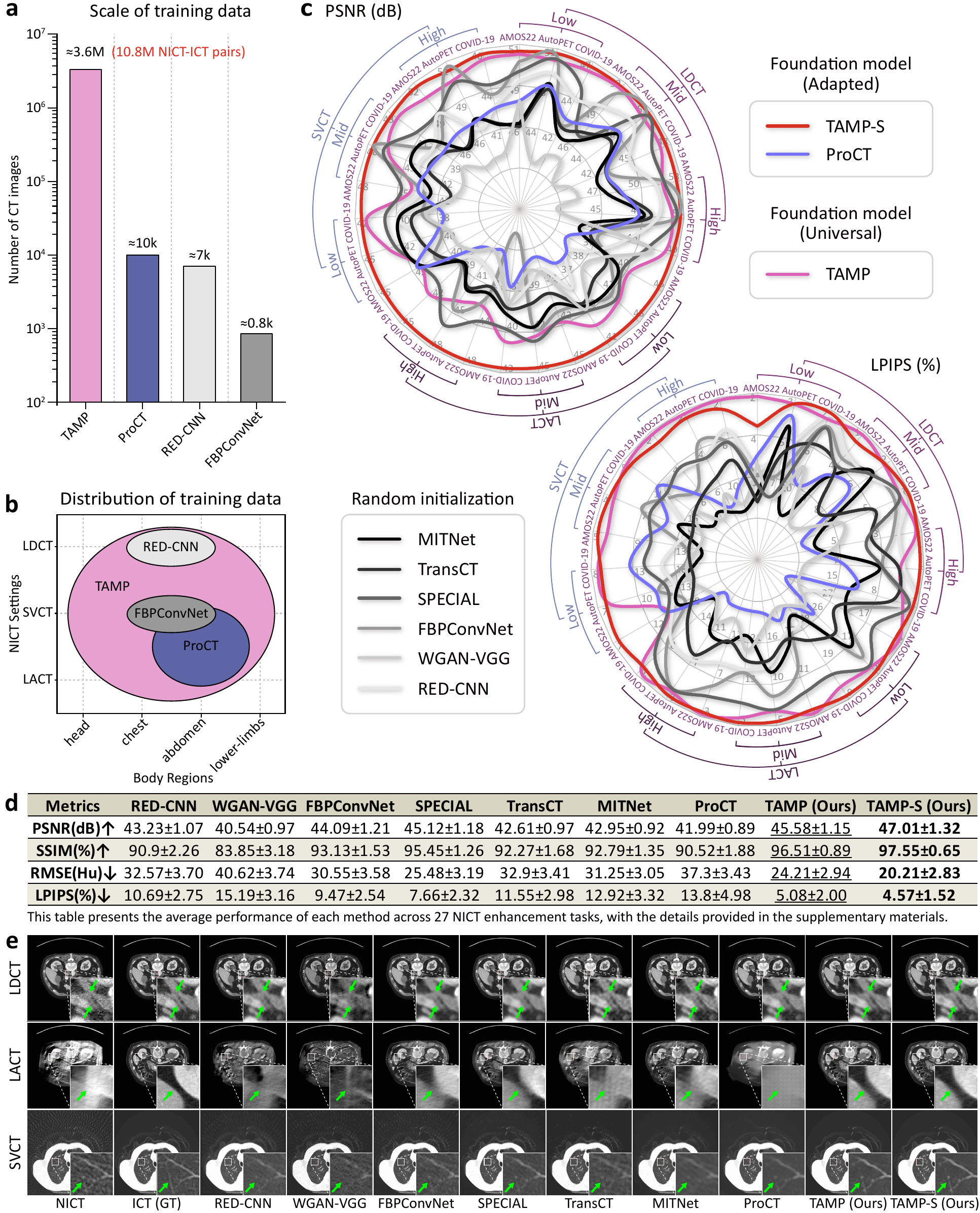}
\caption{Our TAMP enhances diverse NICT settings with varying defect degrees and body regions. a-b) Our TAMP leverages the largest training dataset (SimNICT), spanning diverse NICT settings and body regions. c-d) Our TAMP achieves universal NICT enhancement surpassing specialized models across three body regions, three NICT settings, and three defect degrees. e) Our TAMP significantly improves NICT image quality, preserving fine structural details and enhancing their clinical utility.}
\label{fig2_exp1} 
\end{figure*}

\subsection{SimNICT dataset with large NICT quantity and diversity}
As shown in Fig.\hyperref[fig2_exp1]{2a,2b}, our SimNICT is a large-scale NICT dataset containing 3.6 million images across diverse NICT settings and body regions. It is derived from ICT images sourced from ten publicly available CT datasets (Sec.\ref{data}), totaling 9,638 volumes. Three NICT settings (LDCT, SVCT, and LACT) and four body regions (head, chest, abdomen, and lower-limbs) are incorporated into SimNICT (Fig.\hyperref[fig2_exp1]{2b}), offering significantly greater diversity than existing datasets. We compared the data volume and variability against recent representative NICT enhancement methods, including FBPConvNet \cite{jin2017deep}, RED-CNN \cite{chen2017low}, and ProCT \cite{ma2023universal}. SimNICT provides over 360 times the data quantity of existing works, establishing it as the largest NICT enhancement dataset to date. It will pre-train the model for the enhancement of extensive NICT images, enabling universal enhancement capabilities. Although simulated, SimNICT replicates the physical principles of NICT imaging (Sec.\ref{SimNICT}), ensuring realistic defect generation. Real-world validation in Sec.\ref{result3} confirms the generalizability of SimNICT-trained models on clinical data.

\subsection{TAMP achieves universal NICT enhancement with powerful generalizability}\label{result1}
Our TAMP, trained on SimNICT, exhibits powerful generalizability and effectiveness, enabling direct enhancement of diverse NICT images without additional training. Following parameter-efficient fine-tuning with LoRA, TAMP can be further specialized to specific NICT settings and body regions (TAMP-S), achieving task-optimized performance (Fig.\hyperref[fig2_exp1]{2b}).

\textbf{\textit{Experimental Setting:}} The powerful generalizability and effectiveness of our TAMP are evaluated in 27 NICT enhancement tasks. Specifically, the tasks encompass NICT settings (LDCT, SVCT, LACT) with defect degrees of high, medium (hereafter abbreviated as mid), and low, covering body regions including the chest (from COVID-19 \cite{an2020ct}), abdomen (from AMOS22 \cite{ji2022amos}), and whole-body (from AutoPET \cite{gatidis2022whole}). Following data amount from related studies \cite{jin2017deep,chen2017low,ma2023universal}, we allocate 2,089, 8,669, and 19,613 CT images from three datasets into training (80\%) and testing (20\%) sets, independent of TAMP pre-training. We compare enhancement performance against one adapted FM, ProCT \cite{ma2023universal}, and six typical specialized NICT enhancement models, i.e., RED-CNN \cite{chen2017low} and WGAN-VGG \cite{yang2018low} (single-scale pure convolutional architecture), FBPConvNet \cite{jin2017deep} and SPECIAL \cite{hu2021special} (multi-scale pure convolutional architecture), TransCT \cite{zhang2021transct}, and the proposed MITNet (multi-scale transformer architecture). For each task, RED-CNN, WGAN-VGG, FBPConvNet, SPECIAL, TransCT and MITNet are trained from random scratch, TAMP enhances images without additional training, ProCT and TAMP-S are fine-tuned starting from their pre-trained parameters. We utilize the peak signal-to-noise ratio (PSNR) and root mean square error (RMSE) to evaluate the pixel-level accuracy of enhancement, and the structural similarity index measure (SSIM) and learned perceptual image patch similarity (LPIPS) \cite{zhang2018perceptual} to evaluate visual performance. The units for PSNR and RMSE are dB and Hu, and SSIM and LPIPS scores are scaled by 100 and reported as percentages to display finer details. Additional implementation details are provided in the \textit{Supplementary Materials}.

\textbf{\textit{Observations:}} As shown in Fig.\ref{fig2_exp1}, TAMP exhibits universal NICT enhancement capability with robust generalizability and effectiveness. There are two observations in Fig.\ref{fig2_exp1}: 

\textbf{1)} Our TAMP achieves superior performance across diverse NICT enhancement tasks, spanning multiple NICT settings, body regions, and defect degrees (Fig.\hyperref[fig2_exp1]{2c}). 
Without adaptation, our TAMP outperforms the compared methods in PSNR on 16 tasks (59.26\%) and LPIPS on 23 tasks (85.19\%). After adaptation, TAMP-S achieves PSNR improvements on all tasks and LPIPS improvements on 16 tasks (59.26\%), surpassing the compared methods in PSNR on 26 tasks (96.30\%) and LPIPS on all tasks. Specifically, the outstanding universal NICT enhancement capability of TAMP is demonstrated as follows:

\textbf{NICT settings:} 
Artifacts of varying scales and shapes manifest under different NICT settings. For example, while RED-CNN performs well on LDCT tasks, surpassing ProCT, TransCT, and WGAN-VGG in PSNR (Fig.~\ref{fig2_exp1}), it struggles with LACT tasks, where it only outperforms ProCT. This limitation arises because RED-CNN is primarily suited for small-scale artifacts, as its single high-resolution convolutional network has a limited effective receptive field (described in our \textit{Supplementary Materials}). In contrast, TAMP, with its multi-scale trasformer architecture and large-scale pre-training, handles diverse artifact scales and shapes under different NICT settings. TAMP significantly outperforms ProCT, TransCT, and WGAN-VGG on PSNR and LPIPS across all three NICT tasks. After adaptation, TAMP-S further improves performance, outperforming all compared methods on LPIPS across the three NICT tasks.

\textbf{Body regions:} 
CT images from different body regions exhibit distinct characteristics, which pose greater challenges for model generalization. Abdominal images show moderate values, cranial images display concentrated high values, and chest images present lower values in the pulmonary regions and moderate values in soft tissues. TAMP's transformer architecture expands network capacity, facilitating effective learning of diverse body region NICT representations during large-scale pre-training. In all nine AutoPET whole-body enhancement tasks, TAMP consistently outperforms RED-CNN, WGAN-VGG, FBPConvNet, TransCT and ProCT in LPIPS metrics. After adaptation, TAMP-S achieves better performance, surpassing all compared methods in LPIPS across these tasks.

\textbf{Defect degrees:} 
The defect degrees of NICT images significantly affects enhancement difficulty, yet TAMP consistently enhances image quality across all defect degrees. Compared to other methods, TAMP shows an average PSNR improvement of 5.24\%, 6.55\%, and 6.98\% across nine tasks for high, mid, and low defect degrees, respectively. These results demonstrate that TAMP not only performs well in challenging scenarios but also achieves more substantial improvements in less demanding tasks. After adaptation, TAMP-S further enhances these improvements to 7.85\%, 9.90\%, and 11.00\%, demonstrating its potential for widespread clinical needs.

\textbf{2)} Our TAMP effectively removes artifacts of various scales and shapes from NICT images (Fig.\hyperref[fig2_exp1]{2e}), significantly enhancing their image quality. The different physical processes in NICT settings lead to artifacts of varying scales and shapes in NICT images, placing higher demands on the model's comprehensive enhancement capabilities. 
The single-scale convolutional network structures of RED-CNN and WGAN-VGG focus more on image details but fail to represent large-scale artifact structures. The multi-scale convolutional channels of FBPConvNet and SPECIAL mitigate this issue, but their generalizability is still limited by the convolutional architectures. The introduction of transformers in TransCT and MITNet enhances the model's generalization capability, though this advantage is not observed with limited training data. TAMP, through MITNet's large-scale training, demonstrates strong adaptability to various artifacts as a NICT imaging FM. FM ProCT, due to its over-reliance on NICT sampling angle information, struggles to adapt to LACT and SVCT tasks with significant missing angles, while TAMP-S, with its general design, effectively adapts to different types of NICT enhancement tasks.

As shown in the first row of Fig.\hyperref[fig2_exp1]{2e}, small-scale speckle artifacts with discrete distributions in LDCT are removed to varying degrees by all methods, particularly RED-CNN, FBPConvNet, SPECIAL, TAMP, ProCT, and TAMP-S, which exhibit a smoother visual effect. However, the gaps between structures indicated by the upper arrow and the strip-like structures indicated by the lower arrow become difficult to recognize after enhancement by the compared methods, while our TAMP and TAMP-S effectively enhance these two areas, making their morphology closer to that of real tissues.
Moreover, large-scale continuous wedge-shaped artifacts in LACT severely disrupt the edges of the kidneys and liver (second row of Fig.\hyperref[fig2_exp1]{2e}), making them difficult to reconstruct based on adjacent image information by the compared methods. However, due to the multi-scale network architectures and NICT prior knowledge learned through large-scale pre-training, both TAMP and TAMP-S effectively reconstruct the disrupted image structures, making them easier to interpret.
Finally, as shown in the third row of Fig.\hyperref[fig2_exp1]{2e}, strip-shaped artifacts in SVCT disrupt the fine blood vessels of the lungs, while TAMP and TAMP-S reconstruct these structures with background denoising, making them easier to observe.
Consequently, TAMP’s ability to universally suppress various types of artifacts makes NICT images clearer in detail and more accurate in structure, providing greater value for clinical diagnosis.

\subsection{TAMP effectively reduces the cost for specialized NICT enhancement}\label{result2}

\begin{figure*}[thbp] 
\centering
\includegraphics[width=\linewidth]{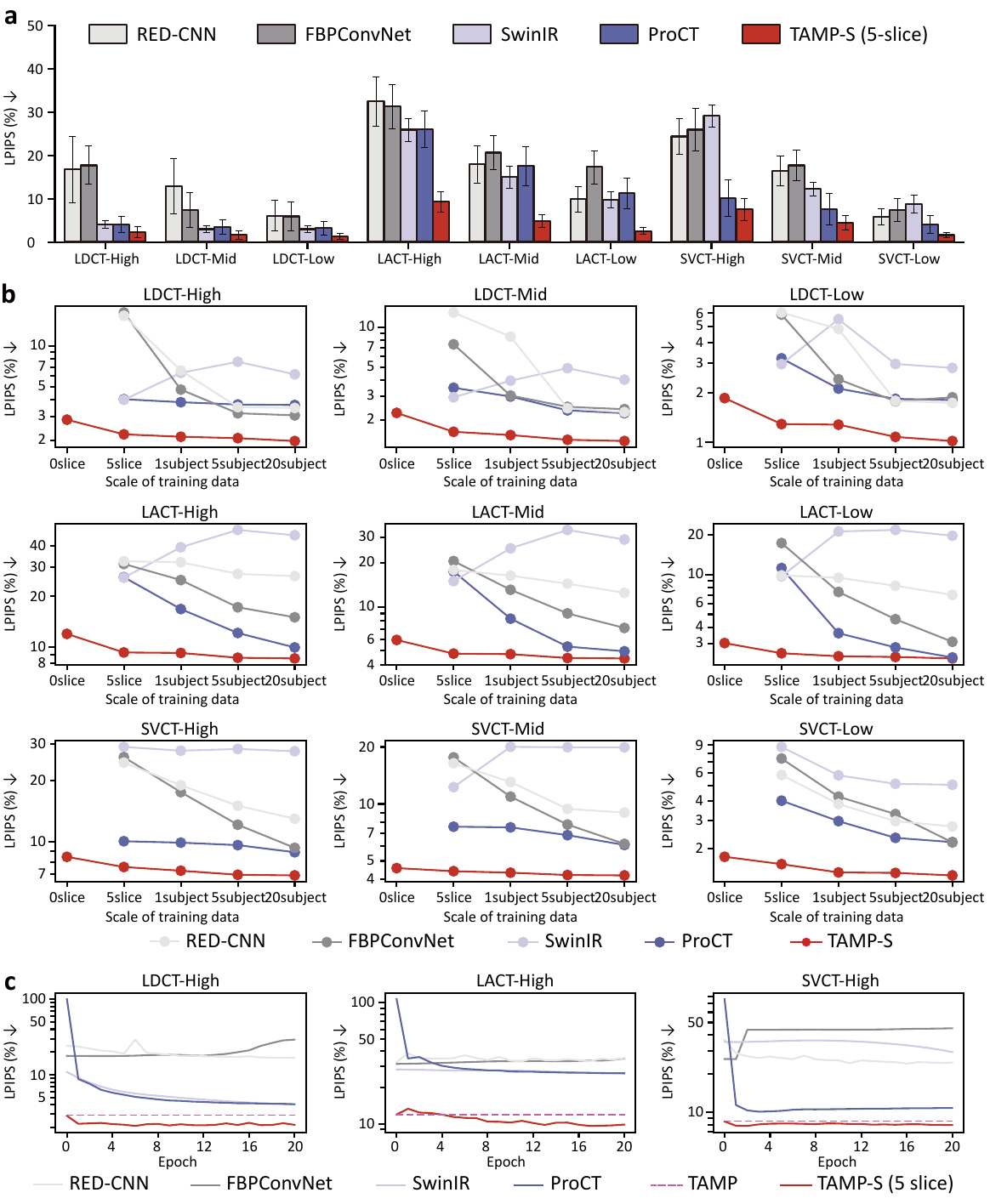}
\caption{TAMP achieves efficient adaptation to specific NICT settings with low data and computational costs. a) Only adapted with 5 slices, our TAMP has significantly outperformed the compared methods. b) Our TAMP can significantly reduce the training data requirement. It only needs 5 slices to achieve comparable or even better performance than the comparison methods with 20 volumes. c) Adapted with 5 slices, TAMP achieves convergence within a small number of epochs, thereby effectively reducing training iterations. }

\label{fig3_res2} 
\end{figure*}
Our TAMP, as an imaging FM, offers a prepared initialization that enables the efficient development of specialized models with only a few training data and computation. We conducted data validation experiments to explore the performance of specialized models developed using TAMP across varying data quantities. 

\textbf{\textit{Experimental Setting:}} We evaluated TAMP-S on nine NICT enhancement tasks by fine-tuning our TAMP with varying data quantities. For comparison, we used two specialized NICT enhancement models (RED-CNN, FBPConvNet), a pre-trained universal enhancement model for the SVCT and LACT tasks (ProCT), and a pre-trained natural image denoising FM (SwinIR \cite{liang2021swinir}). These tasks include three NICT settings (LDCT, SVCT, LACT) with three levels of defect degrees (high, mid, low). For each task, we fine-tune TAMP, ProCT and SwinIR, and train RED-CNN and FBPConvNet from scratch on five CT slices, one, five, and twenty subjects (abdomen regions from AMOS22 dataset) to evaluate the influence of data amount on enhancement performance.

\textbf{\textit{Observations:}} Our TAMP only requires very small data amount and training iterations in the adaptation of specialized NICT enhancement models, effectively reducing the cost in NICT imaging application development (Fig.\hyperref[fig3_res2]{3}). There are three observations in this experiment:

\textbf{1) }
Our TAMP-S achieves state-of-the-art performance across all tasks using five slices of training data (Fig.\hyperref[fig3_res2]{3a}). 
For the relatively simple tasks of LDCT-High, LDCT-Mid, and LDCT-Low, TAMP-S achieved LPIPS scores of 2.23, 1.62, and 1.29 with just five slices of training data, respectively, surpassing the best comparison methods by 44.39\%, 45.08\%, and 56.42\%, demonstrating its remarkably low data requirements for simpler tasks. 
For the most challenging task, LACT-High, TAMP-S still achieved a LPIPS of 9.24, improving the best score of the comparison method (25.87) by 64.28\%, demonstrating its robust performance in difficult scenarios. 
This excellent performance is due to its universal NICT enhancement capability as a FM that has undergone large-scale pre-training tailored for NICT enhancement. 
Although SwinIR was also pre-trained on large-scale data, the gap between its upstream task of grayscale image denoising and the downstream task of NICT image enhancement limits its performance. As a result, it is outperformed by FBPConvNet and RED-CNN, which were trained with random initialization, in task of SVCT-High and SVCT-Low. 

\textbf{2) } 
Our TAMP demonstrates efficient adaptation capabilities in diverse NICT enhancement tasks with lower data cost (Fig.\hyperref[fig3_res2]{3b}). 
TAMP-S, fine-tuned using only 5 slices of data, outperforms RED-CNN, FBPConvNet, and SwinIR trained on 20 volumes across all 9 NICT tasks, demonstrating its task-specific adaptation effectiveness. 
Moreover, as the amount of fine-tuning data increases, TAMP-S shows a stable improvement trend and consistently superior performance across varying quantities of training data in all NICT enhancement tasks. 
It is noteworthy that although the pre-trained SwinIR outperforms RED-CNN, FBPConvNet, and ProCT on the LDCT tasks (5 slices of training data), which are most similar to its upstream task, it is overtaken once the data volume exceeds 5 volumes. This is because SwinIR's network architecture and pre-trained representations are not well-suited for adapting to NICT enhancement tasks, limiting its adaptation capability in this domain. 

\textbf{3) }
The rapid and stable convergence performance of TAMP demonstrates its potential to reduce the consumption of computational time and resources (Fig.\hyperref[fig3_res2]{3c}). 
Using five data slices, TAMP demonstrates rapid performance enhancement within the initial 10 training epochs while maintaining subsequent stability without overfitting, even with extended training duration. 
In comparison, pretrained methods (e.g., SwinIR and ProCT) require more epochs for convergence, whereas randomly initialized methods (e.g., RED-CNN and FBPConvNet) show pronounced fluctuations or overfitting tendencies, thereby requiring additional time and computational resources to ensure optimal performance. 
TAMP's effective convergence stems from its LoRA-based fine-tuning strategy, enabling rapid task adaptation while preserving complete pretrained representations, thus ensuring stable performance without substantial fluctuations. 

\subsection{Real-world validation: TAMP enhances real-world NICT images}\label{result3}
\begin{figure*}[thbp] 
\centering
\includegraphics[width=\linewidth]{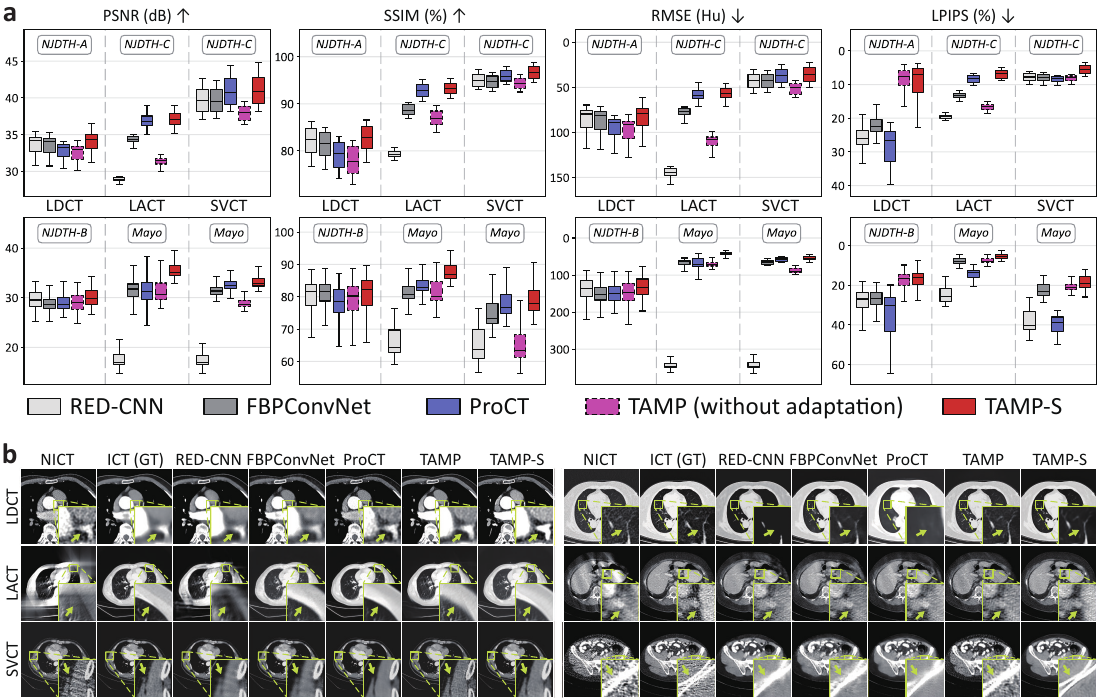}
\caption{Real-world validation demonstrates our great application potential in real-world NICT images. a) Our TAMP can directly enhance real-world NICT images and TAMP-S is adapted to specific NICT settings, achieving further improvement. b) Our TAMP has significant visual enhancement demonstrating its great clinical application potential in real-world NICT images.}
\label{fig4_exp3} 
\end{figure*}
Due to the difference between the real-world NICT imaging process and the simulated process in this work, there will be a domain gap between the simulation-based NICT enhancement training and real-world NICT enhancement. Therefore, this experiment further evaluates the enhancement capability of our TAMP on real-world NICT images and shows its applicability in real-world clinical practices.

\textbf{\textit{Experimental Setting:}} To evaluate TAMP's real-world enhancement capability across diverse clinical applications, we collected real-world NICT data from different anatomical regions and imaging protocols. The datasets include: \textbf{1) NJDTH-A (Routine Chest CT)}: LDCT data from Nanjing Drum Tower Hospital Jiangbei (NJDTH) for routine chest helical scanning with 6 cases (1,496 slices) acquired at low dose (80 kVp, 30 mAs) and high dose (120 kVp, 90 mAs), split into 5 training cases (1,234 slices) and 1 test case (262 slices), validating TAMP on standard thoracic imaging. \textbf{2) NJDTH-B (Coronary CTA)}: LDCT data from NJDTH for coronary CT angiography with 10 cases (2,802 slices) acquired at low dose (120 kVp, 50 mAs) and high dose (120 kVp, 200 mAs), split into 5 training cases (1,527 slices) and 5 test cases (1,275 slices), validating TAMP on cardiac imaging with complex vascular structures. \textbf{3) NJDTH-C (Cardiac SVCT/LACT)}: SVCT and LACT data from NJDTH for full-cycle cardiac imaging with 6 cases (750 slices) reconstructed from raw ICT projection data by undersampling at sparse views and limited angles, split into 5 training cases (625 slices) and 1 test case (125 slices), validating TAMP on geometric artifacts in dynamic cardiac scans. \textbf{4) Mayo (Abdominal SVCT/LACT)}: SVCT and LACT data with 10 cases reconstructed from publicly available projection data of Mayo Clinic Low Dose CT Grand Challenge by undersampling, split into 5 training cases (990 slices) and 5 test cases (737 slices), validating TAMP on abdominal imaging. These four datasets constitute 6 evaluation tasks across different anatomical regions and clinical applications: LDCT enhancement on routine chest (NJDTH-A) and coronary CTA (NJDTH-B), SVCT enhancement on cardiac (NJDTH-C) and abdominal (Mayo), and LACT enhancement on cardiac (NJDTH-C) and abdominal (Mayo). We compare RED-CNN, FBPConvNet, ProCT trained from scratch, against TAMP (zero-shot) and TAMP-S (fine-tuned) following Sec.\ref{result1} protocols. 

\textbf{\textit{Observations:}}
As shown in Fig.\hyperref[fig4_exp3]{4}, our TAMP and TAMP-S effectively enhance the quality of real-world NICT images across diverse clinical scenarios, based on two observations:

\textbf{1)} As an imaging FM, TAMP demonstrates robust simulation-to-reality generalization across multiple real-world datasets without additional training.
Quantitatively (Fig.\hyperref[fig4_exp3]{4a}), TAMP achieves competitive zero-shot performance across all real-world NICT tasks. In LDCT tasks, TAMP achieves median LPIPS scores of 7.65\% (NJDTH-A) and 16.54\% (NJDTH-B), substantially outperforming compared methods. For geometric artifact tasks, TAMP demonstrates effective zero-shot transfer on LACT (median PSNR: 31.53 dB on NJDTH-C, 30.61 dB on Mayo) and SVCT (median SSIM: 94.35\% on NJDTH-C, 63.37\% on Mayo), validating its learned universal representations. Qualitatively, as shown in Fig.\hyperref[fig4_exp3]{4b}, TAMP enhances LDCT and SVCT images by not only suppressing small-scale photon noise, but also by producing clearly distinguishable edges of structures such as muscles, blood vessels, and bones, resulting in smoother textures and enhanced structural clarity. Conversely, while compared methods achieve surface smoothness, they incur edge blurring and line discontinuity artifacts, leading to substantial structural detail degradation. 
Additionally, TAMP reconstructed the chest edges and soft tissues in LACT images, which are severely damaged areas where the original structures are difficult to infer from surrounding information. In contrast, RED-CNN failed to reconstruct them due to its single-scale convolutional network structure, which cannot adapt to such severe defects in LACT images.

\textbf{2)}
After parameter-efficient adaptation, TAMP-S achieves state-of-the-art performance across all real-world NICT tasks from diverse clinical scenarios.
Quantitatively (Fig.\hyperref[fig4_exp3]{4a}), TAMP-S achieves the best median performance in all 24 metric-task combinations (6 tasks $\times$ 4 metrics). Notably, for challenging geometric artifact reconstruction, TAMP-S demonstrates substantial improvements on LACT tasks with median PSNR of 37.01 dB (NJDTH-C, +0.26 dB vs ProCT) and 35.00 dB (Mayo, +3.26 dB vs FBPConvNet), and on SVCT tasks with median SSIM of 96.57\% (NJDTH-C, +0.73\% vs ProCT) and 77.97\% (Mayo, +1.39\% vs ProCT). For LDCT tasks, TAMP-S achieves median PSNR of 34.35 dB (NJDTH-A) and 29.91 dB (NJDTH-B), consistently outperforming baseline methods. These consistent improvements validate TAMP's effective parameter-efficient adaptation capability for real-world NICT enhancement.
Qualitatively, TAMP-S retains the visual advantages of TAMP while demonstrating more task-specific denoising and more accurate reconstruction of real-world NICT images (Fig.\hyperref[fig4_exp3]{4b}). 
It efficiently enhances the smoothness of homogenous regions in LDCT and SVCT images while maintaining the edge clarity advantage of TAMP, making structures easier to observe. Additionally, the severely damaged chest edges and soft tissues in LACT images are reconstructed more precisely, featuring clearer and more accurate shapes. 
By integrating these features, TAMP-S's visual performance closely resembles that of real-world CT images, which is due to its pre-training from large-scale CT data, allowing for a deeper understanding of CT features and reliable real-world noise reduction. 

\subsection{Radiologist validation: TAMP improves the clinical acceptability of NICT images}\label{result4}

\begin{figure*}[t] 
\centering
\includegraphics[width=\linewidth]{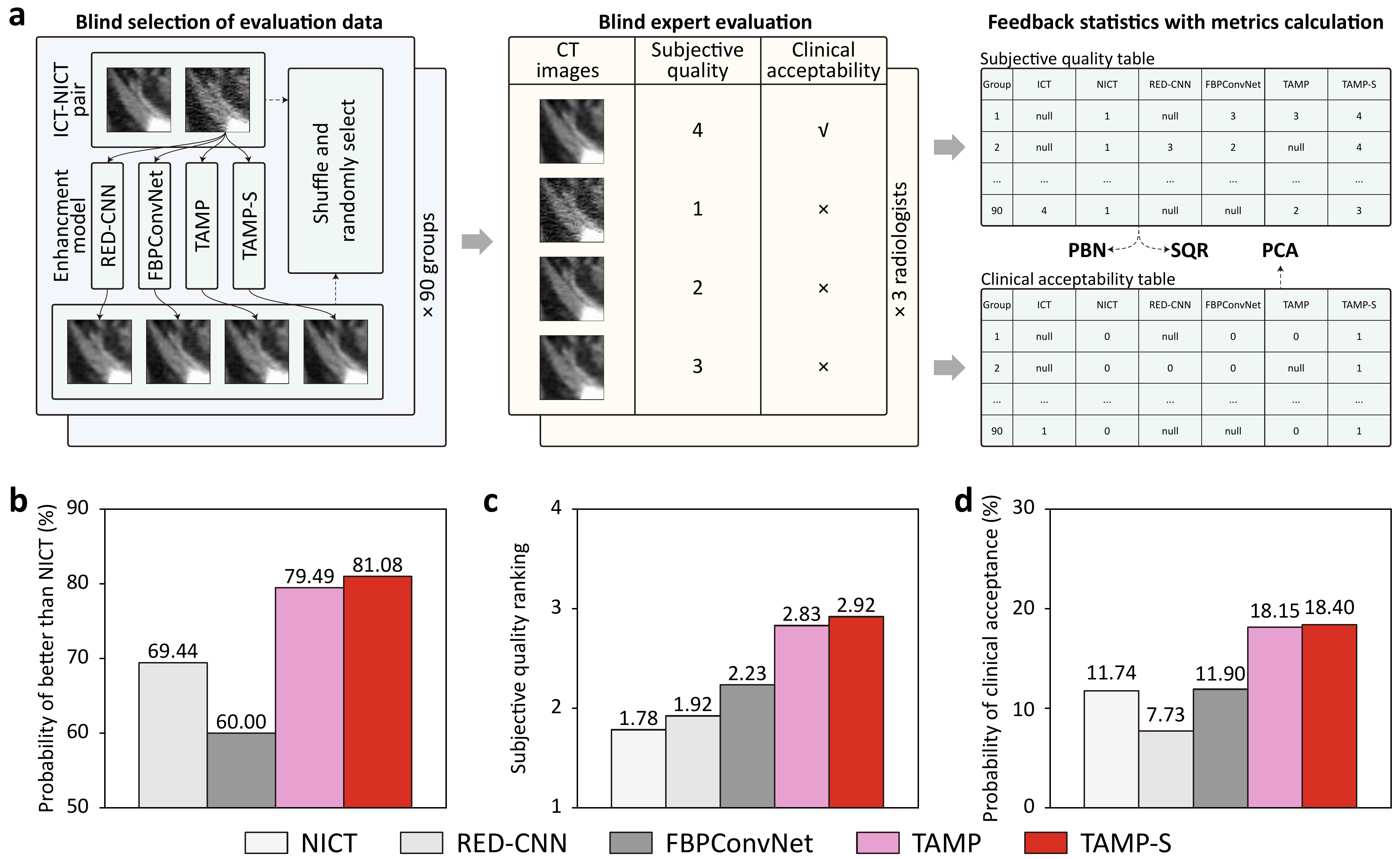}
\caption{The radiologist validation illustrates our superior clinical acceptance. a) The process of our radiologist validation. b-d) We designed three metrics to quantify this acceptance, i.e., the probability of being better than NICT, subjective quality ranking, and probability of clinical acceptance.}
\label{fig5_exp4} 
\end{figure*}
To evaluate TAMP's ability to enhance the clinical acceptability of NICT images, we conducted a radiologist validation study. Specifically, we invite three radiologists (one with over 10 years of experience and two with more than 5 years each) to blindly rank these images according to subjective quality and score them based on clinical acceptability. We then statistically analyzed the scoring data to evaluate the performance of each model. 

\textbf{\textit{Experimental Setting:}}
The experimental process of radiologist validation includes blind selection of evaluation data, blind expert evaluation, and feedback statistics with metrics calculation. 
One proficient (1P) and two competent radiologists (2C and 3C) from the Department of Radiology in the affiliated hospital of the medical school of Ningbo University were invited to score these images blindly. As shown in Fig.\hyperref[fig5_exp4]{5a}, four images were randomly selected from the NICT, ICT, FBPConvNet enhanced, RED-CNN enhanced, TAMP enhanced, and TAMP-S enhanced NICT to construct a validation group. Ninety groups were randomly selected for radiologists scoring. The score includes a ranking based on subjective quality (where the rank is derived by subtracting the score from 5, with higher values indicating better quality) and an assessment of clinical acceptability. To calculate the final score, the scoring data from the three radiologists are weighted according to their years of experience (weights of 0.5 for 1P, and 0.25 for 2C and 3C) and analyzed using three designed metrics: probability of better than NICT (PBN), subjective quality ranking (SQR), and probability of clinical acceptability (PCA). The PBN represents the enhancement degree for the enhanced NICT images via the methods. The SQR reflects the subjective ranking of the enhanced images' quality among the methods. The PCA represents the clinical acceptability of the various CT images. These metrics are defined as: $PBN(x) = \frac{1}{|S_{NICT_{x},NICT}|} \sum_{s \in S_{NICT_{x},NICT}} [R_{NICT_{x}}^{s} > R_{NICT}^s]$, $SQR(x) = \frac{1}{|S_{NICT_{x}}|} \sum_{s \in S} R_{NICT_{x}}^{s}$, and $PCA(x) = \frac{1}{|S_{NICT_{x}}|} \sum_{s \in S} [A_{NICT_{x}}^{s}]$, where $x$ denotes a model, $NICT_{x}$ represents the enhanced NICT image by model $x$, $S_{NICT_{x},NICT}$ denotes the set of groups where both $NICT_{x}$ and NICT are selected for scoring, $R_{NICT_{x}}^{s}$ and $R_{NICT}^s$ represent the subjective quality ratings, $A_{NICT_{x}}^{s}$ represents the clinical acceptability given by radiologists, and $[\cdot]$ denotes the Iverson bracket. More details of the radiologist validation process are described in the \textit{Supplementary Materials}.

\textbf{\textit{Observations:}}
Our results presented in Fig.\hyperref[fig5_exp4]{5} demonstrate that NICT images enhanced by TAMP have higher subjective quality and are more likely to be clinically accepted, based on three observations: 

\textbf{1)} 
Our TAMP significantly improves the subjective quality of NICT images, which has been recognized by radiologists. 
As shown in Fig.\hyperref[fig5_exp4]{5b}, the PBN for TAMP (79.49\%) significantly exceeds 50\% (the threshold indicating that the enhancement of NICT images has a positive effect), surpassing RED-CNN by 14.47\% and FBPConvNet by 32.48\%, indicating that the majority of NICT images enhanced by TAMP are recognized by radiologists for their improved subjective quality. After fine-tuning, the PBN for TAMP-S further improved to 81.08\%, reflecting its potential to achieve better subjective quality enhancements for NICT images in various specific scenarios. 
This robust subjective quality enhancement capability is due to TAMP's universal NICT representation achieved through large-scale pre-training, enabling consistent performance across various types of NICT images. 

\textbf{2)}
Furthermore, our TAMP effectively enhances the subjective quality of NICT images, which has also been recognized by radiologists. 
Enhanced by TAMP and TAMP-S, the subjective quality of NICT images is considered effectively improved by radiologists. 
As shown in Fig.\hyperref[fig5_exp4]{5c}, the SQR of TAMP (2.83) surpasses RED-CNN (1.92) by 47.40\% and FBPConvNet (2.23) by 26.91\%, reflecting a 58.99\% improvement in NICT images. After adaptation, the SQR of TAMP-S (2.92) increased by 3.18\%, achieving the best value among all methods and demonstrating its effectiveness in enhancing the quality of NICT images. 
Compared to the significance reflected by PBN, SQR focuses on the extent of image quality enhancement performance. 
For instance, FBPConvNet has a lower PBN than RED-CNN (60.00\% vs. 69.44\%) but a higher SQR (2.23 vs. 1.92), indicating that FBPConvNet is more effective for enhancing certain NICT images. 
However, our method surpasses the comparison methods in both PBN and SQR, reflecting its effective enhancement of subjective quality across various NICT images. 

\textbf{3)} Our TAMP improves NICT images, resulting in higher clinical acceptance among radiologists. 
As shown in Fig.\hyperref[fig5_exp4]{5d}, despite NICT's inherently limited clinical acceptability, TAMP and TAMP-S efficiently improve the PCA by 54.60\% and 56.73\% respectively, demonstrating significant improvements that are recognized by radiologists. 
An interesting observation is that although the SQR scores in Fig.\hyperref[fig5_exp4]{5c} demonstrate that the compared methods improve the subjective quality of NICT images, RED-CNN actually decreases the PCA of NICT images by a substantial 34.16\%, while FBPConvNet achieves only a slight improvement of 1.36\%. 
This is because, for clinical acceptance, radiologists are more concerned with factors beyond subjective quality, such as whether the regions of interest in the enhanced NICT images conform to the structural features of real-world CT images to provide reliable clinical diagnostic support. 
Thus, the improvement of NICT images in both SQR and PCA reflects the ability of TAMP and TAMP-S to enhance image subjective quality based on the features of real-world CT images, demonstrating significant potential for clinical application. 

\section{Discussion}\label{discussion}
In this paper, for the first time, we propose and validate TAMP, an imaging FM designed for the universal enhancement of NICT images. Our work pioneers the application of FM in the NICT enhancement domain, utilizing the pre-training and adaptation paradigm to advance research in universal NICT enhancement technologies and enable efficient model deployment for specific NICT enhancement scenarios. 

Our FM TAMP demonstrates superior generalization ability and versatility across a wide range of NICT enhancement tasks, significantly accelerating the deployment of NICT applications while reducing the costs in model construction. 
On the one hand, pre-trained on 10.8 million simulated NICT images, TAMP enhances diverse LDCT, LACT, and SVCT images with varying defects across body regions, achieving state-of-the-art performance in 18 tasks on PSNR and 23 tasks on LPIPS out of 27 NICT enhancement tasks (Sec.\ref{result1}). 
This robustness is a crucial advantage for medical imaging systems handling diverse datasets and unpredictable image quality. 
On the other hand, TAMP can be rapidly adapted to specific NICT enhancement scenarios through fine-tuning with the LoRA method, requiring only five slices of training data for excellent performance (Sec.\ref{result2}). 
This efficient adaptation is particularly valuable in data-scarce environments, where acquiring large datasets is impractical. Its generalization further highlights TAMP's advantage in real-world NICT scenarios (Sec.\ref{result3}). 

Technically, we designed MITNet and DDEL to encode diverse artifact patterns and learn universal representations for large-scale NICT pretraining. For network architecture, since NICT artifacts manifest at varied scales across protocols (noise in LDCT, streaks in SVCT, blurring in LACT), we designed scale-specific disentanglement and progressive integration mechanisms. MITNet employs parallel-to-serial transformers where parallel multi-scale embedding extracts scale-specific features, while serial progressive fusion enables coarse-to-fine refinement mimicking CT iterative reconstruction. For pretraining paradigm, foundation models require learning universal representations that capture both anatomical structures and degradation patterns. DDEL bridges image-domain structure learning and projection-domain physics-grounded degradation learning via dual-domain consistency, enabling the model to represent NICT degradation inversion. These designs establish the foundation for universal NICT enhancement beyond single-protocol models.

Clinically, TAMP addresses the practical need for versatile CT acquisition scenarios through its generalized enhancement capability. In contrast to specialized models limited to specific acquisition protocols, TAMP's inherent adaptability to low-dose, sparse-view, and limited-angle acquisitions enables more precise CT diagnostics across diverse clinical scenarios, meeting critical demands for patient-centric dose management and protocol flexibility. This widespread adaptability optimizes clinical workflows by maintaining diagnostic reliability in CT scanning-constrained scenarios, ensuring patient-centric clinical interventions and supporting more precise treatments. 

This study has two limitations. 
1) TAMP was pre-trained using simulated NICT images, which have a gap with real-world NICT images. Nonetheless, since both our NICT image simulation and DDEL strategy are physics-driven, adhering to the physical processes of NICT imaging, TAMP is generalizable to real-world NICT images with few data, as has been experimentally demonstrated in Sec.\ref{result3}. 
2) The transformer structure increases memory consumption during operation, although it provides TAMP with a universal NICT representation. Fortunately, techniques such as model pruning, knowledge distillation, and mixed precision training \cite{menghani2023efficient} have been employed to reduce memory usage, a challenge that will also be addressed in our future research. 
3) TAMP is designed for axial CT imaging systems where artifacts manifest in axial planes. Tomosynthesis (e.g., breast tomosynthesis), which performs acquisition and reconstruction in coronal or sagittal planes, is outside the current scope and will be explored in future work. 

\section{Methods}\label{method}

\begin{figure*}[thbp] 
\centering
\includegraphics[width=\linewidth]{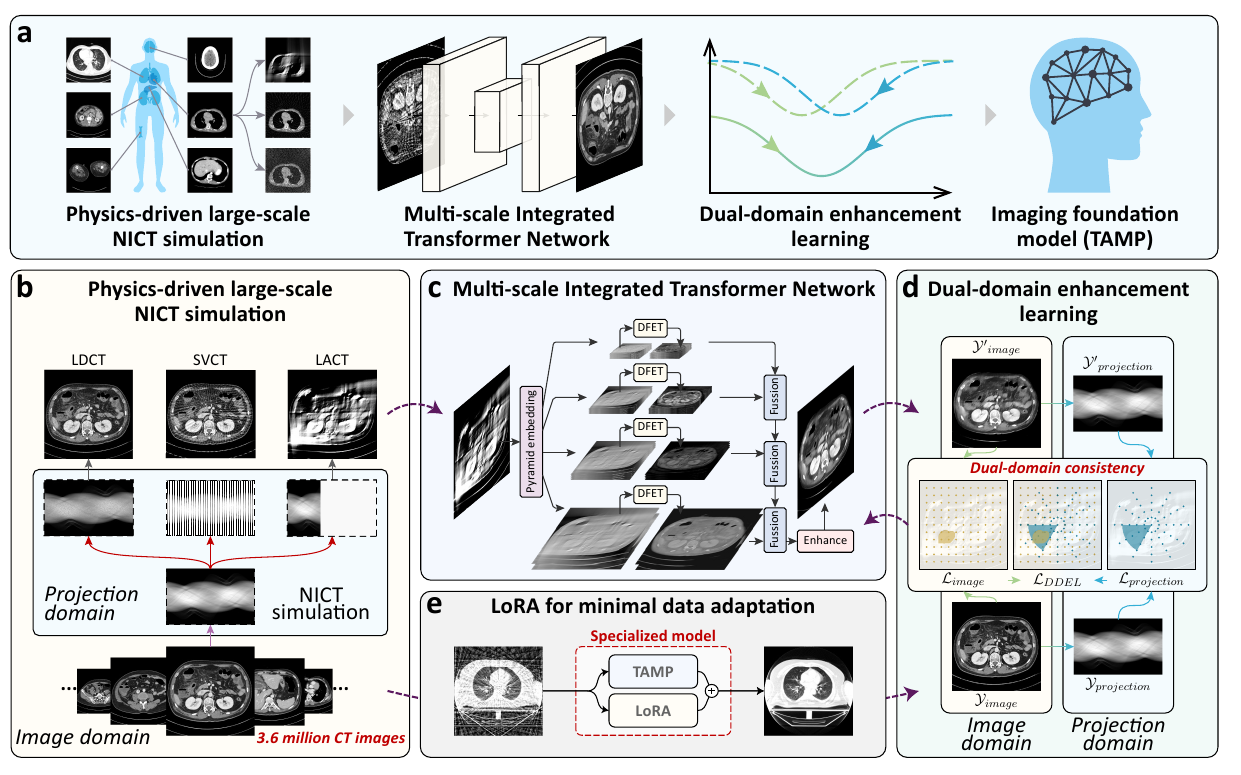}
\caption{The framework of our TAMP for universal enhancement of NICT images. a) It simulates large-scale NICT dataset (SimNICT), and trains our multi-scale integrated transformer network (MITNet) via our dual-domain enhancement learning (DDEL), thus achieving our TAMP model for universal NICT enhancement. b) By simulating the defects that meet the physical process of non-ideal measurement, NICT images are synthesized for our SimNICT. c) Our MITNet represents the multi-granularity defect features in the varied NICT data. d) Our DDEL trains the universal NICT enhancement both in image and projection domains. e) A parameter-efficient fine-tuning strategy, low-rank adaptation (LoRA), is employed for efficient adaptation.} 
\label{fig6_framework} 
\end{figure*}

As shown in Fig.\hyperref[fig6_framework]{6a}, our proposed imaging foundational model, TAMP, consists of a physics-driven large-scale NICT simulation, multi-scale integrated transformer network, and dual-domain enhancement learning, for universal enhancement of NICT. In this section, we illustrate the methods of our TAMP, and more specific details are described in our \textit{Supplementary Materials}.
 
\subsection{SimNICT: Physics-driven simulation for large-scale NICT dataset}\label{SimNICT}
SimNICT is the first dataset constructed for universal NICT enhancement model training. It starts from the ICT images from ten publicly CT datasets (Sec.\ref{data}) that encompass whole-body regions, including the head, chest, abdomen, and lower-limbs, and are simulated into LDCT, SVCT, and LACT under different defect degrees. By removing the volumes with low quality, our SimNICT dataset finally obtains 3,633,374 images from 9,638 ICT volumes. It simulates the NICT images with three NICT settings (LDCT, SVCT, and LACT) and different defect degrees, thus finally achieving over 10.8 million NICT-ICT image pairs.

As shown in Fig.\hyperref[fig6_framework]{6b}, we simulate the NICT images according to the physical processes of the non-ideal measurements. It projects the ICT images to the projection domain to simulate the maps of the CT raw signal. Then, according to the physical processes of LDCT, SVCT, and LACT, we simulate the defective copies of these CT raw signal maps. For the LDCT, according to the low anti-interference of low-dose radiation in the environment, we produce Gaussian noise on the maps to simulate the interfered measurement. For the SVCT, according to the sparse angle sampling, we reduce the views at equal intervals on the maps to simulate the sparse measurement. For the LACT, according to the restricted scanning angle range, we reduce the range of views on the maps to simulate the angular-defected measurement. Finally, the defective CT raw signal maps are back-projected to the image domain, thus achieving the NICT images with real-world defects. We utilize the ASTRA Toolbox\footnote{ASTRA: \url{https://astra-toolbox.com/index.html}} to achieve this physics-driven simulation process.

\subsection{MITNet: Multi-scale integrated transformer network for the representation of varied defect patterns}\label{MITNet}
Our MITNet advances scale-independent multi-scale feature representation through a distinctive parallel-to-serial transformer \cite{vaswani2017attention} architecture, enabling unified encoding of varied NICT artifacts across diverse protocols. Since NICT degradations from different protocols manifest at varying scales and shapes, sequential multi-scale encoding introduces interference from preceding scale features. As shown in Fig.\hyperref[fig6_framework]{6c}, our parallel-to-serial architecture addresses this through two aspects: 1) Parallel multi-scale embedding: Different patch scales and linear layers independently extract embedding features at multiple pyramid levels in parallel, isolating scale-specific information without cross-scale interference. These features will be further represented in multiple deep feature extraction transformer (DFET) blocks \cite{liang2021swinir} which is a stack of Swin transformer Layers to learn to repair the defect information in each level. 2) Serial progressive fusion: Features are progressively fused from lowest to highest resolution, mimicking the coarse-to-fine refinement process of CT iterative reconstruction. In each fusion stage, the feature maps from different pyramid levels are concatenated and put into a convolution-ReLU layer for multi-scale feature integration. Finally, these features are input into several convolution layers to predict the final enhanced NICT images. This design enables artifact removal and detail reconstruction adapted to different NICT categories, establishing a unified foundation model capable of handling varied degradation patterns.

\subsection{Dual-domain enhancement learning for imaging pre-training}\label{DDEL}
We design a dual-domain enhancement learning (DDEL) that bridges physical simulation and representation learning through dual-domain consistency constraints \cite{wang2023review} (Fig.\hyperref[fig6_framework]{6d}). Our DDEL achieves this through two aspects: 1) Dual-domain constraints: In the image domain, multi-granularity losses including MSE loss $\mathcal{L}^{I}_{MSE}$, SSIM loss $\mathcal{L}^{I}_{SSIM}$ \cite{zhao2016loss}, and VGG loss $\mathcal{L}^{I}_{VGG}$ \cite{johnson2016perceptual} learn real anatomical features. In the projection domain, the enhanced and target images are mapped via the Operator Discretization Library\footnote{ODL: \url{https://github.com/odlgroup/odl}} for sinogram maps, and the MSE loss $\mathcal{L}^{P}_{MSE}$ learns diverse NICT degradation patterns. 2) Bridging through bidirectional mappings: By constraining $Y'=f' \circ f(Y) \approx Y$, we bridge the physical simulation $f$ (SimNICT) and representation learning $f'$ (MITNet), ensuring learned representations are grounded in both anatomical reality and physical degradation processes. Totally, these four losses are weighted and summed for the training loss, i.e., $\mathcal{L}_{train}=w_{1}\mathcal{L}^{I}_{MSE}+w_{2}\mathcal{L}^{I}_{SSIM}+w_{3}\mathcal{L}^{I}_{VGG}+w_{4}\mathcal{L}^{P}_{MSE}$. 

\subsection{Parameter-efficient fine-tuning for efficient adaptation}
We employ LoRA, a parameter-efficient fine-tuning method, to adapt our TAMP for specialized NICT enhancement tasks, requiring very little training data and computation. (Fig.\hyperref[fig6_framework]{6e}). It only tunes a small number of parameters in the whole network, thus greatly reducing the risk of over-fitting caused by training too many parameters. Therefore, it enables only very little data to stimulate the professional performance of our TAMP in specific scenarios achieving efficient adaptation. Specifically, following the implementation of LoRA \cite{hu2021lora}, we add the bypasses of low-rank matrix on the linear layers and convolutional layers in the whole network to adapt their representation to target scenarios. During adaptation, the parameters in the LoRA bypasses are tuned and the original parameters in the TAMP are fixed. We utilize the same training loss $\mathcal{L}_{train}$ as the pre-training stage for adaptation. 

\subsection{Details of pre-training and adaptation}
Our TAMP is implemented by PyTorch\footnote{PyTorch: \url{https://pytorch.org/}} for its pre-training and adaptation. To reduce the time consumption of inputting and outputting (IO) large-scale data on disk during training, a queued training process is designed. It loads $N$ (we set $N=5$) NICT volumes into memory as a queue and shuffles the slices for learning. To avoid over-fitting for a fixed defect pattern in the training process, we ensure that all NICT settings exist in the queue. Once all the slices have been iterated, the oldest volume is removed from the queue, and a new NICT volume is loaded, thus effectively reducing the IO cost. We take the Adan \cite{xie2024adan} as our optimizer, which is an outstanding optimizer targetedly designed for FM training. It accelerates the convergence speed and reduces the loss fluctuation in the learning process of transformer networks. For pre-training, we set the learning rate $\delta$, b1, and b2 as 5e-4, 0.5, and 0.999, and the learning rate becomes 0.95 times, i.e., $\delta=0.95\delta$ after the training of every 100 queues for finer fitting. For adaptation, we set the same learning rate $\delta$, b1, and b2 as the pre-training, and the learning rate becomes 0.9 times, i.e., $\delta=0.9\delta$ after the training of every 1 queue. We set the batch size as 5 and the input size as $512\times512$ to train the NICT images in a high resolution. In our experiment, considering the scale difference of loss values, we set the weights $w_{1}, w_{2}, w_{3}, w_{4}$ of losses in the training loss as 1, 5e-3, 1e-4, and 5e-4.

\section{Data Availability}\label{data}
The datasets and code used in this work are publicly available to ensure reproducibility. 1) SimNICT pretraining dataset: This work enrolled ten publicly available datasets, including COVID-19-NY-SBU\footnote{COVID-19-NY-SBU: \url{https://wiki.cancerimagingarchive.net/pages/viewpage.action?pageId=89096912}} \cite{saltz2021stony}, STOIC\footnote{STOIC: \url{https://stoic2021.grand-challenge.org/}} \cite{revel2021study}, MELA\footnote{MELA: \url{https://mela.grand-challenge.org/}} \cite{shuang_song_2022_6575197}, LUNA\footnote{LUNA: \url{https://luna16.grand-challenge.org/}} \cite{armato2011lung}, LNDb\footnote{LNDb: \url{https://lndb.grand-challenge.org/}} \cite{pedrosa2021lndb}, HECKTOR22\footnote{HECKTOR22: \url{https://hecktor.grand-challenge.org}} \cite{andrearczyk2023automatic}, CT\_COLONOGRAPHY\footnote{CT\_COLONOGRAPHY: \url{https://wiki.cancerimagingarchive.net/pages/viewpage.action?pageId=3539213}} \cite{johnson2008accuracy}, AutoPET\footnote{AutoPET: \url{https://autopet.grand-challenge.org/}} \cite{gatidis2022whole}, AMOS22\footnote{AMOS22: \url{https://amos22.grand-challenge.org/}} \cite{ji2022amos}, and CT Images in COVID-19\footnote{CT Images in COVID-19: \url{https://www.cancerimagingarchive.net/collection/ct-images-in-covid-19/}} \cite{harmon2020artificial}. 8 out of 10 datasets are released on HuggingFace\footnote{SimNICT: \url{https://huggingface.co/datasets/YutingHe-list/SimNICT}} with data hosted on Internet Archive\footnote{Internet Archive: \url{https://archive.org/details/@tamp_research_group}}; AutoPET and HECKTOR22 are excluded due to original licensing restrictions. 
2) Synthetic evaluation datasets: Complete evaluation data with training/testing splits for all 27 benchmark tasks are available on HuggingFace\footnote{Synthetic evaluation datasets: \url{https://huggingface.co/datasets/YutingHe-list/SimNICT/tree/main}}. 
3) Real-world evaluation data: Clinical data from Nanjing Drum Tower Hospital cannot be released due to patient privacy (Ethics approval: AF/SC-08/03.0, 2025-0068-02); however, trained model weights and testing code are provided on GitHub. Additionally, the Mayo Clinic LDCT dataset is publicly available\footnote{AAPM LDCT: \url{https://www.aapm.org/GrandChallenge/LowDoseCT/}}.

\section{Code Availability}
Our TAMP will be released at \url{https://github.com/YutingHe-list/TAMP}.


\newpage
\setcounter{section}{0}\renewcommand\thesection{\Alph{section}}

\section{More details of SimNICT dataset} 
SimNICT is the dataset used for TAMP pre-training, derived from CT images across ten publicly available CT datasets to simulate a diverse range of NICT images. These source datasets cover various body regions, including the head, chest, abdomen, and lower limbs. By filtering out low-quality data, SimNICT ultimately includes 9,513 volumes with over 3.6 million slices for TAMP pre-training, while an additional 125 volumes with over 30 thousand slices from the AMOS22, AutoPET, and CT Images in COVID-19 datasets are used for downstream validation experiments. Detailed information is provided in Table \hyperref[SimNICT_source_info]{1}.

\begin{table*}[thbp]
\centering
\caption{
Source Datasets for SimNICT Pre-training and Validation Experiments.
}\label{SimNICT_source_info}
\resizebox{\linewidth}{!}{
\begin{tabular}{cccccc}
\toprule
Dataset name & Body regions & Original volumes & Pre-train/Downstream volumes & link & license \\
\midrule
COVID-19-NY-SBU & Chest & 1,384 & 459/0 & \href{https://wiki.cancerimagingarchive.net/pages/viewpage.action?pageId=93258257}{\checkmark} & \href{https://creativecommons.org/licenses/by/4.0/}{CC BY 4.0} \\
STOIC & Chest & 2,000 & 2,000/0 & \href{https://stoic2021.grand-challenge.org/}{\checkmark} & \href{https://creativecommons.org/licenses/by-nc/4.0/}{CC BY-NC 4.0} \\
MELA & Chest & 1,100 & 1,100/0 & \href{https://mela.grand-challenge.org/}{\checkmark} & \href{https://creativecommons.org/licenses/by/4.0/}{CC BY 4.0} \\
LUNA & Chest & 888 & 888/0 & \href{https://luna16.grand-challenge.org/}{\checkmark} & \href{https://creativecommons.org/licenses/by/4.0/}{CC BY 4.0} \\
LNDb & Chest & 294 & 294/0 & \href{https://lndb.grand-challenge.org/}{\checkmark} & \href{https://creativecommons.org/licenses/by-nc-nd/4.0/}{CC BY-NC-ND 4.0} \\
HECKTOR22 & Head, neck & 882 & 882/0 & \href{https://hecktor.grand-challenge.org/Overview/}{\checkmark} & \href{https://creativecommons.org/licenses/by/4.0/}{CC BY-NC-ND 4.0} \\
CT\_COLONOGRAPHY & Abdomen & 1,730 & 1,730/0 & \href{https://pubs.rsna.org/doi/10.1148/radiol.2512080200}{\checkmark} & \href{https://creativecommons.org/licenses/by/4.0/}{CC BY 4.0} \\
CT Images in COVID-19 & Chest & 771 & 736/35 & \href{https://www.covid19-ct-dataset.com/}{\checkmark} & \href{https://creativecommons.org/licenses/by/4.0/}{CC BY 4.0} \\
AutoPET & Head, chest, abdomen, lower-limbs & 1,014 & 979/35 & \href{https://autopet.grand-challenge.org/}{\checkmark} & \href{https://wiki.cancerimagingarchive.net/download/attachments/4556915/TCIA\%20Restricted\%20License\%2020220519.pdf?api=v2}{TCIA Restricted} \\
AMOS22 & Abdomen & 500 & 445/55 & \href{https://amos22.grand-challenge.org/}{\checkmark} & \href{https://creativecommons.org/licenses/by/4.0/}{CC BY 4.0} \\
\bottomrule
\end{tabular}
}
\end{table*}

\subsection{Simulation of NICT for pre-training}\label{Simulation of NICT for pre-training}
The SimNICT dataset used for TAMP pre-training simulates NICT images by introducing defects during the CT image projection and reconstruction processes (corresponding to Section Result1 of the paper). This process models a fan-beam CT with a full 720-degree angular view, implemented using code from The Operator Discretization Library (ODL) \footnote{ODL: \url{https://github.com/odlgroup/odl}}, as represented by:
\[
P_{\text{ICT}} = P_{\text{proj}}(I_{\text{ICT}}, \theta_{\text{Full}}), \quad \theta_{\text{Full}} = \{ 1, 2, \dots, 720 \}, 
\]
where \(I_{\text{ICT}}\) represents the ideal measurement CT image from SimNICT, and \(P_{\text{ICT}}\) denotes the simulated projection data. The process of introducing defects and reconstruction for the NICT images of three settings is presented below.

1) \textbf{LDCT}. The simulation method for LDCT images follows the approach used in the 2016 NIH-AAPM-Mayo Clinic Low Dose CT Grand Challenge \cite{NIH2016LowDoseCT, yu2012development}.
In this noise simulation process, quantum noise \( x \) is modeled based on photon counting statistics, following a Poisson distribution. For high photon counts, this is approximated by a Gaussian distribution. Noise \( x \) is introduced as a normally distributed random variable applied to adjust the initial measurement \(P_{\text{ICT}}\) for simulating lower dose conditions, expressed as:
\[
P_{\text{LD}} = P_{\text{ICT}} + \sqrt{\frac{1 - a}{a} \cdot \frac{\exp(P_{\text{ICT}})}{N_{0}}} \cdot x,
\]
where \( a = \frac{N_{\text{LD}}}{100} \), and \( a \in (0, 1) \) is the scaling factor that adjusts the dose reduction through the parameter \( N_{\text{LD}} \), and \( N_{0} = 1 \times 10^6 \) is the initial incident photon count. The data is then reconstructed into the LDCT image as:
\[
I_{\text{LD}} = \mathrm{P}^{-1}(P_{\text{LD}}, \theta_{\text{Full}}).
\]

2) \textbf{SVCT}. The simulation of SVCT is achieved by reconstructing the \( P_{\text{ICT}} \) projection data, which is evenly sampled based on a specified number of angles through the parameter \( N_{\text{SV}} \), as follows:
\[
I_{\text{SV}} = \mathrm{P}^{-1}(P_{\text{ICT}}, \theta_{\text{SV}}), 
\]
\[
\theta_{\text{SV}} = \left\{ \theta_i \mid \theta_i = \left\lfloor \frac{720 \cdot i}{N_{\text{SV}}} \right\rfloor, \, i = 1, 2, \dots, N_{\text{SV}} \right\}.
\]

3) \textbf{LACT}. The simulation of LACT is achieved by reconstructing the \( P_{\text{ICT}} \) projection data, which is continuously sampled based on a range of angles through the parameter \( N_{\text{LA}} \), as follows:
\[
I_{\text{LA}} = \mathrm{P}^{-1}(P_{\text{ICT}}, \theta_{\text{LA}}),
\]
\[
\theta_{\text{LA}} = \{ 1, 2, \dots, N_{\text{LA}} \}.
\]

During the construction of SimNICT, each source CT slice is simulated into one LDCT, one SVCT, and one LACT image. In this simulation process, the parameters \( N_{\text{LD}} \), \( N_{\text{SV}} \), and \( N_{\text{LA}} \) are randomly sampled within a range to enhance the diversity of the SimNICT data, thereby improving the model’s adaptability to different defect levels, with the range settings presented in Table \hyperref[NICT_simulation_condition]{2}.

\begin{table}[thbp] 
\caption{The settings of \( N_{\text{LD}} \), \( N_{\text{SV}} \), and \( N_{\text{LA}} \) for NICT simulation in the SimNICT dataset and validation experiment with three types of defect degrees.}

\label{NICT_simulation_condition}
\footnotesize
\centering
\begin{tabular}{ccccc}
\toprule
{Condition} & Defect degree & \( N_{\text{LD}} \) & \( N_{\text{SV}} \) & \( N_{\text{LA}} \) \\
\midrule
{SimNICT} & random & [5, 75] & [15, 360] & [75, 270] \\
\multirow{3}{*}{\begin{tabular}[c]{@{}c@{}}Validation \\ experiment\end{tabular}} & low & 60 & 300 & 150 \\
& mid & 40 & 120 & 120 \\
& high & 20 & 60 & 90 \\
\bottomrule
\end{tabular}
\end{table}

\subsection{Simulation of NICT for validation experiments}
In the validation experiments, to specifically evaluate the capability of various methods on different types of NICT enhancement tasks, we configured three defect degrees (Low, Mid, and High) for each NICT setting (LDCT, SVCT, and LACT) in the simulation by adjusting \( N_{\text{LD}} \), \( N_{\text{SV}} \), and \( N_{\text{LA}} \), as detailed in Table \hyperref[NICT_simulation_condition]{2}.

\section{More details of real-world NICT}
We collect and reconstruct real-world NICT from diverse clinical sources including Nanjing Drum Tower Hospital Jiangbei and the Mayo Clinic Low Dose CT Grand Challenge to evaluate the adaptation of TAMP to real-world NICT. The details of the relevant information and reconstruction parameters are as follows. 

\subsection{Detail information of real-world data}
The real-world validation dataset comprises four sources from diverse clinical scenarios and imaging protocols: NJDTH-A, NJDTH-B, and NJDTH-C from Nanjing Drum Tower Hospital Jiangbei (NJDTH), and Mayo from the Mayo Clinic Low Dose CT Grand Challenge.

\textbf{NJDTH-A:} LDCT paired data from six patients (1 male and 5 females, aged 49-79 years) with 1,496 slices. Patients underwent dual scans at low dose (80 kVp, 30 mAs) and high dose (120 kVp, 90 mAs). The scan protocol was set to Helical mode with a reconstruction kernel of B\_VSHARP\_C, which represents the standard clinical helical trajectory used in routine diagnostic CT imaging.

\textbf{NJDTH-B:} LDCT paired data from ten patients with 2,802 slices. Patients underwent dual scans at low dose (120 kVp, 50 mAs) and high dose (120 kVp, 200 mAs), providing validation for current-reduction-based LDCT protocols.

\textbf{NJDTH-C:} Raw projection data from six patients (5 males and 1 female, aged 43-79 years) with 750 slices acquired at 120 kVp and 39 mA. The scan protocol was set to AXIAL mode with a reconstruction kernel of C\_SOFT\_BA, enabling the collection of raw projection data necessary for subsequent sparse-view and limited-angle reconstruction experiments.

\textbf{Mayo:} Publicly available projection data from the Mayo Clinic Low Dose CT Grand Challenge with ten patients. SVCT and LACT variants were reconstructed by undersampling the projection data at sparse views and limited angles, providing independent validation across different institutions.

\subsection{Reconstruction of real-world data}
\textbf{NJDTH-A and NJDTH-B:} For LDCT paired data, to address respiratory-induced misalignment between LDCT and ICT, we performed image registration using Elastix \cite{klein2009elastix,shamonin2014fast}. The registration employed lung-specific parameters from ElastixModelZoo\footnote{Parameters: \url{https://github.com/SuperElastix/ElastixModelZoo/blob/master/models/Par0004}}, using ICT as fixed and LDCT as moving images, followed by cropping of non-overlapping regions.

\textbf{NJDTH-C:} For SVCT and LACT, using the TIGRE toolbox \cite{biguri2016tigre} in MATLAB, we reconstructed SVCT and LACT from raw projection data following the downsampling protocol in Section~\ref{Simulation of NICT for pre-training} with \( N_{\text{SV}} = 240 \) and \( N_{\text{LA}} = 120 \).

\textbf{Mayo:} For SVCT and LACT, we first converted the helical CT projection data to fan-beam projections using helix2fan conversion, then performed iterative reconstruction using the ASTRA Toolbox with the CGLS\_CUDA algorithm for 50 iterations, with \( N_{\text{SV}} = 180 \) and \( N_{\text{LA}} = 140 \).

\section{More details of Experiment}
\subsection{Radiologist validation application}
\begin{figure*}[ht] 
\centering
\includegraphics[width=\linewidth]{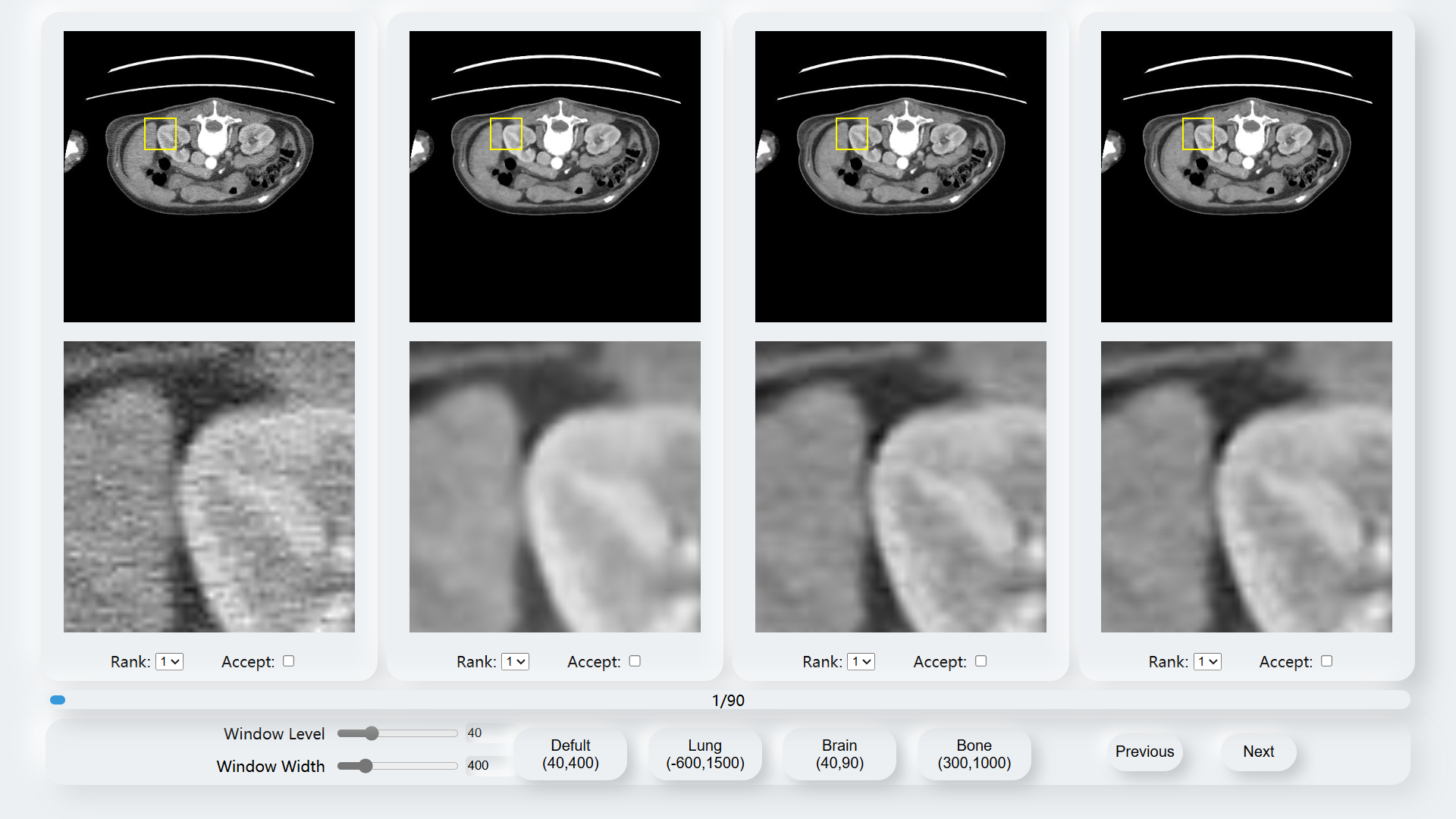}
\caption{
The scoring website for the radiologist validation.
}
\label{exp4_screenshot} 
\end{figure*}

\begin{figure*}[ht] 
\centering
\includegraphics[width=\linewidth]{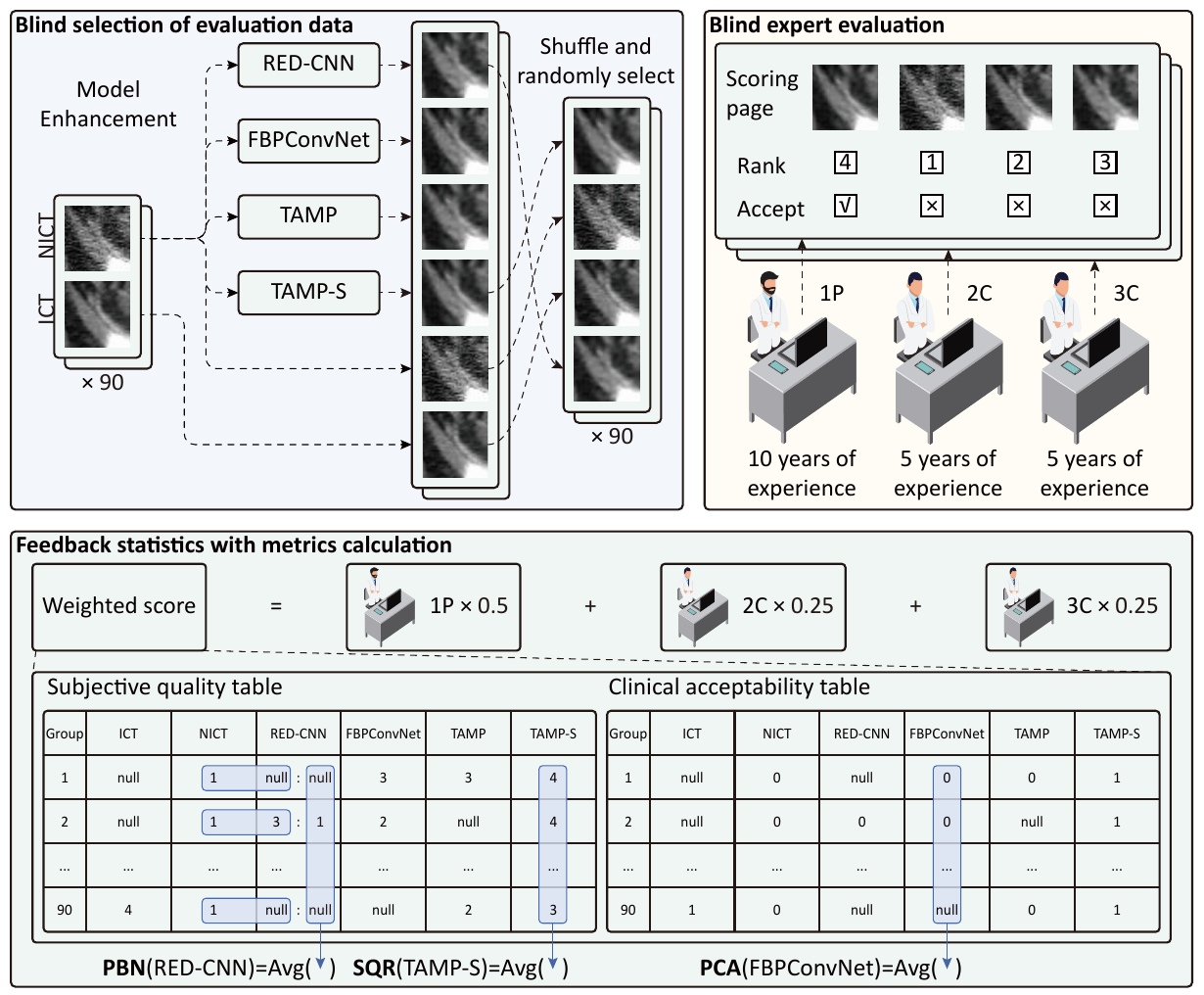}
\caption{
The process of our radiologist validation.
}
\label{sup_expert} 
\end{figure*}

In the radiologist validation experiment, the enhancement results of each method on NICT images are presented to radiologists for scoring. To facilitate the experts' observation and scoring of the CT images, we developed an application with an interactive website to enable a more convenient and accurate evaluation, as shown in Fig. \ref{exp4_screenshot}.
We use the Flask Toolbox\footnote{Flask: \url{https://github.com/pallets/flask}} to load and present CT images in an interactive webpage. The webpage displays a group of four CT images at a time, allowing radiologists to rank the images based on quality and assess their clinical acceptability. Radiologists can also easily adjust window width and level, and zoom in on specific regions for detailed observation. Moreover, the detail of the process of our radiologist validation are shown in Fig. \ref{sup_expert}.

\subsection{Evaluation metrics}
\subsubsection{NICT image quality}
\label{NICT image quality}
To evaluate the enhanced NICT image quality, we utilize four image quality assessment metrics: root mean square error (RMSE) \cite{sheikh2006statistical}, peak signal-to-noise ratio (PSNR) \cite{sheikh2006statistical}, structural similarity index measure (SSIM) \cite{wang2004image}, and learned perceptual image patch similarity (LPIPS) \cite{zhang2018perceptual}. These metrics are defined as follows:

1) \textbf{RMSE} calculates the average difference between the pixel intensities of the ground truth image \( x \) and the predicted image \( \hat{x} \), representing the pixel-wise error in the image, defined by:
\[
RMSE(x, \hat{x}) = \sqrt{ \frac{1}{N}  \sum_{i=1}^{N} (x(i) - \hat{x}(i))^2 },
\]
where \( N \) is the total number of pixels in the image.

2) \textbf{PSNR} calculates the ratio of the maximum possible pixel value to the noise (RMSE), representing the peak signal-to-noise ratio and indicating the overall quality of the image, defined by:
\[
PSNR(x, \hat{x}) = 20 \log_{10} \left( \frac{M-1}{RMSE(x, \hat{x})} \right),
\]
where \( M = 4096 \) for the image with units of Hu.

3) \textbf{SSIM} calculates the similarity between the luminance, contrast, and structure of the two images, representing the perceptual quality and structural similarity between the ground truth and predicted images, defined by:
\[
SSIM(x, \hat{x}) = \frac{(2 \mu_{x} \mu_{\hat{x}} + c_{1}) (2 \sigma_{x\hat{x}} + c_{2})}{(\mu_{x}^2 + \mu_{\hat{x}}^2 + c_{1}) (\sigma_{x}^2 + \sigma_{\hat{x}}^2 + c_{2})}, 
\]
where \( \mu_{x} \) and \( \mu_{\hat{x}} \) are the mean values of \( x \) and \( \hat{x} \) respectively, \( \sigma_{x}^2 \) and \( \sigma_{\hat{x}}^2 \) are the variances of \( x \) and \( \hat{x} \), and \( \sigma_{x \hat{x} } \) is the covariance between \( x \) and \( \hat{x} \). The constants \( c_1 = (k_1 L)^2 \) and \( c_2 = (k_2 L)^2 \), with \( k_1 = 0.01 \) and \( k_2 = 0.03 \) by default. 

4) \textbf{LPIPS} calculates the perceptual distance between image patches using deep neural network features, representing the perceptual similarity based on learned feature spaces, defined by:
\[
LPIPS(x, \hat{x}) = \sum_{l=1}^{L} w_l \cdot \|\phi_l(x) - \phi_l(\hat{x})\|_2,
\]
where \( L \) is the total number of layers in the network, \( w_l \) is the weight assigned to the \( l \)-th layer, while \( \phi_l(x) \) and \( \phi_l(\hat{x}) \) are the feature maps from the \( l \)-th layer for images \( x \) and \( \hat{x} \), respectively, and \( \| \cdot \|_2 \) denotes the L2 norm.

\subsubsection{Model’s Clinical Application Value}
In the radiologist validation experiment, to provide an intuitive assessment of each model’s clinical application value, we designed three metrics to statistically analyze the scoring data from radiologists: probability of better than NICT (PBN), subjective quality ranking (SQR), and probability of clinical acceptability (PCA). These metrics are defined as follows:

1) \textbf{PBN} calculates the probability that the model’s enhanced image is rated higher than the original NICT image, describing the significance of the model in improving the quality of NICT, defined by:
\begin{footnotesize}
\[
PBN(x) = \frac{1}{|S_{NICT_{x},NICT}|} \sum_{s \in S_{NICT_{x},NICT}} [R_{NICT_{x}}^{s} > R_{NICT}^s],
\]
\end{footnotesize}
where \( x \) denotes a model, \( NICT_{x} \) represents the NICT image enhanced by model \( x \), \( S_{NICT_{x},NICT} \) denotes the set of groups where both \( NICT_{x} \) and NICT are selected for scoring, \( | \cdot | \) indicates the cardinality of the groups, and \( R_{NICT_{x}}^{s} \) and \( R_{NICT}^s \) represent the subjective quality ratings of the \( NICT_{x} \)  and NICT, respectively, in group \( s \).

2) \textbf{SQR} calculates the average subjective quality rating of the model’s enhanced images, describing the efficiency of the model in improving the quality of NICT, defined by:
\[
SQR(x) = \frac{1}{|S_{NICT_{x}}|} \sum_{s \in S} R_{NICT_{x}}^{s},
\]
where \( S_{NICT_{x}} \) denotes the set of groups where \( NICT_{x} \) is selected for scoring.

3) \textbf{PCA} calculates the probability of clinical acceptability of the model’s enhanced images, describing the capability of the model in making NICT images more acceptable for clinical use, defined by:
\[
PCA(x) = \frac{1}{|S_{NICT_{x}}|} \sum_{s \in S} [A_{NICT_{x}}^{s}],
\]
where \( A_{NICT_{x}}^{s} \) represents the clinical acceptability of the model’s enhanced CT image in group \( s \), given by radiologists. \([ \cdot ]\) denotes the Iverson bracket, which is 1 if the statement is true and 0 if false.

\section{More technical details of TAMP}
\subsection{Loss function in DDEL}
Dual-domain enhancement learning (DDEL) is designed with composite loss functions to enable TAMP to learn both image and projection domain features of NICT images, expressed as:
\[
\mathcal{L}_{\text{train}} = w_{1}\mathcal{L}^{I}_{\text{MSE}} + w_{2}\mathcal{L}^{I}_{\text{SSIM}} + w_{3}\mathcal{L}^{I}_{\text{VGG}} + w_{4}\mathcal{L}^{P}_{\text{MSE}},
\]
where \( w_{1}, w_{2}, w_{3}, w_{4} \) are set to 1.0, \( 5.0 \times 10^{-3} \), \( 1.0 \times 10^{-4} \), and \( 5.0 \times 10^{-4} \), respectively. The four loss terms are defined as follows:

1) \textbf{\(\mathcal{L}^{I}_{\text{MSE}}\)} is commonly used in training self-supervised image denoising models \cite{zhang2017learning} and aims to minimize pixel-wise differences between the predicted and ground truth images, serving as a basic measure of enhancement quality. It is defined as:
\[
\mathcal{L}^{I}_{\text{MSE}} = \frac{1}{N_I} \sum_{i=1}^{N_I} \left( I_{\text{Pred}}(i) - I_{\text{ICT}}(i) \right)^2, 
\]
where \( I_{\text{Pred}} \) is the predicted image, \( I_{\text{ICT}} \) is the ground truth image, and \( N_I \) is the total number of pixels in the image.

2) \textbf{\(\mathcal{L}^{I}_{\text{SSIM}}\)} calculates the structural similarity index, which assesses image similarity based on luminance, contrast, and structure \cite{zhao2015loss}. This loss function improves perceptual image quality by aligning structural features. It is defined as:
\[
\mathcal{L}^{I}_{\text{SSIM}} = 1 - SSIM(I_{\text{Pred}}, I_{\text{ICT}}),
\]
where \( SSIM(I_{\text{Pred}}, I_{\text{ICT}}) \), as defined in Section \ref{NICT image quality}, measures the structural similarity between the predicted and ground truth images.

3) \textbf{\(\mathcal{L}^{I}_{\text{VGG}}\)} computes the difference between the feature maps extracted from a pre-trained VGG19 network, encouraging the model to preserve high-level perceptual features \cite{yang2018low}. This loss helps to improve the perceptual quality of the enhanced images. It is defined as:
\[
\mathcal{L}^{I}_{\text{VGG}} = \frac{1}{N_\phi} \sum_{i=1}^{N_\phi} \| \phi(I_{\text{Pred}})(i) - \phi(I_{\text{ICT}})(i) \|_1,
\]
where \( \phi \) represents the feature maps extracted from the 35th layer of a pre-trained VGG19 network, and \( N_\phi \) is the number of features in the layer.

4) \textbf{\(\mathcal{L}^{P}_{\text{MSE}}\)} calculates the pixel-wise mean squared error in the projection domain, ensuring that the model’s predicted projections are consistent with the original projections. This loss ensures proper alignment in the projection domain. It is defined as:
\[
\mathcal{L}^{P}_{\text{MSE}} = \frac{1}{N_P} \sum_{i=1}^{N_P} \left( P(I_{\text{Pred}})(i) - P(I_{\text{ICT}})(i) \right)^2,
\]
where \( P \) denotes the projection operation, and \( N_P \) is the number of projections. The projection operation uses the same simulation environment and toolkit (ODL) as in Section \ref{Simulation of NICT for pre-training}, and is executed on CUDA, allowing the computed loss to be backpropagated through the model for weight updates.

\subsection{More details of TAMP pre-training}
\begin{figure}[H]
\centering
\includegraphics[width=0.9\linewidth]{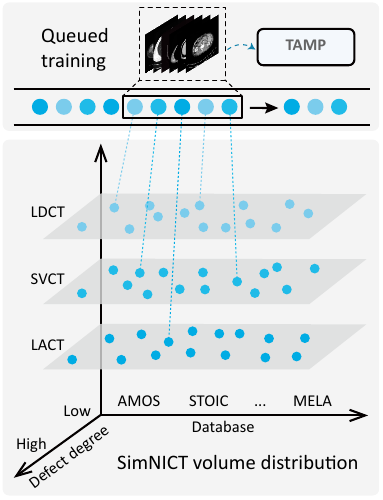}
\caption{The queued training process designed for TAMP pre-training.}
\label{queued training} 
\end{figure}

TAMP is pre-trained on SimNICT datasets to achieve universal NICT enhancement capability. Firstly, to reduce the time consumption of inputting and outputting (IO) large-scale data on disk during training, a queued training process is designed. As shown in Fig. \ref{queued training}, it loads $N$ (we set $N=5$) NICT volumes into memory as a queue and shuffles the slices for learning. To avoid over-fitting for a fixed defect pattern in the training process, we ensure that all NICT settings exist in the queue. Once all the slices have been iterated, the oldest volume is removed from the queue, and a new NICT volume is loaded, thus effectively reducing the IO cost. Secondly, to improve model convergence, we use warm-up at the beginning of the TAMP pre-training. The hyperparameter settings are provided in Table \hyperref[pre-train hyperparameters]{3}.

\begin{table*}[thbp]
\centering
\caption{
Hyperparameter settings used in the TAMP pre-training.
}
\label{pre-train hyperparameters}
\begin{tabular}{ll}
\toprule
\textbf{Hyperparameter} & \textbf{Value} \\
\midrule
Optimizer                                   & Adan \cite{xie2024adan} \\
Warm-up epochs                              & 10 \\
Full epochs                                 & 28,539 \\
Initial learning rate at warm-up            & 0.00005 \\
Learning rate after warm-up                 & 0.0005 \\
Learning betas                              & [0.98, 0.92, 0.99] \\
Epochs per learning rate decay              & 100 \\
Learning rate decay rate                    & 0.95 \\
Batch size on each GPU                      & 5 \\
Training environment                        & 2× NVIDIA RTX 3090 GPUs \\
\bottomrule
\end{tabular}
\end{table*}

\subsection{More details of TAMP adaptation}
We employ LoRA \cite{hu2021lora}, a parameter-efficient fine-tuning method, to adapt our TAMP to specialized NICT enhancement tasks with few training data and computation. LoRA tunes only a small number of parameters, reducing the risk of overfitting and enabling professional performance with limited data. Specifically, we add low-rank matrix bypasses to the linear and convolutional layers in the network, tuning only the parameters in the LoRA bypasses while keeping the original TAMP parameters fixed. The same optimizer, learning rate, learning betas, batch size, and training environment are used as in pre-training after the warm-up phase, with the learning rate halved every 10 epochs.

\subsection{More details of MITNet architecture}
\begin{figure*}[htbp]
\centering
\includegraphics[width=0.8\linewidth]{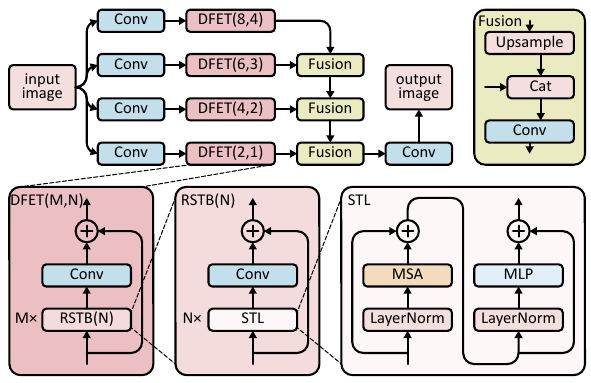}
\caption{Network architecture of MITNet.}
\label{MITNet architecture} 
\end{figure*}

Our MITNet constructs a transformer \cite{vaswani2017attention} architecture that is compatible with multi-scale features to represent the varied defect patterns in different NICT images. As shown in Fig. \ref{MITNet architecture}, MITNet first uses convolutions with varying kernel sizes to extract features from NICT images at four scales. It then employs multiple Deep Feature Extraction Transformer (DFET) blocks \cite{liang2021swinir}, with Swin Transformer layers and convolutional operations alternated across multiple levels, to refine defects at each scale. This is followed by a step-by-step fusion process, restoring features to the original NICT image scale, and culminating in an enhancement phase that produces the final output image.

\section{More experiment results}\label{result detail}
\subsection{Additional qualitative results}
\begin{figure*}[p]
\centering
\includegraphics[width=\linewidth]{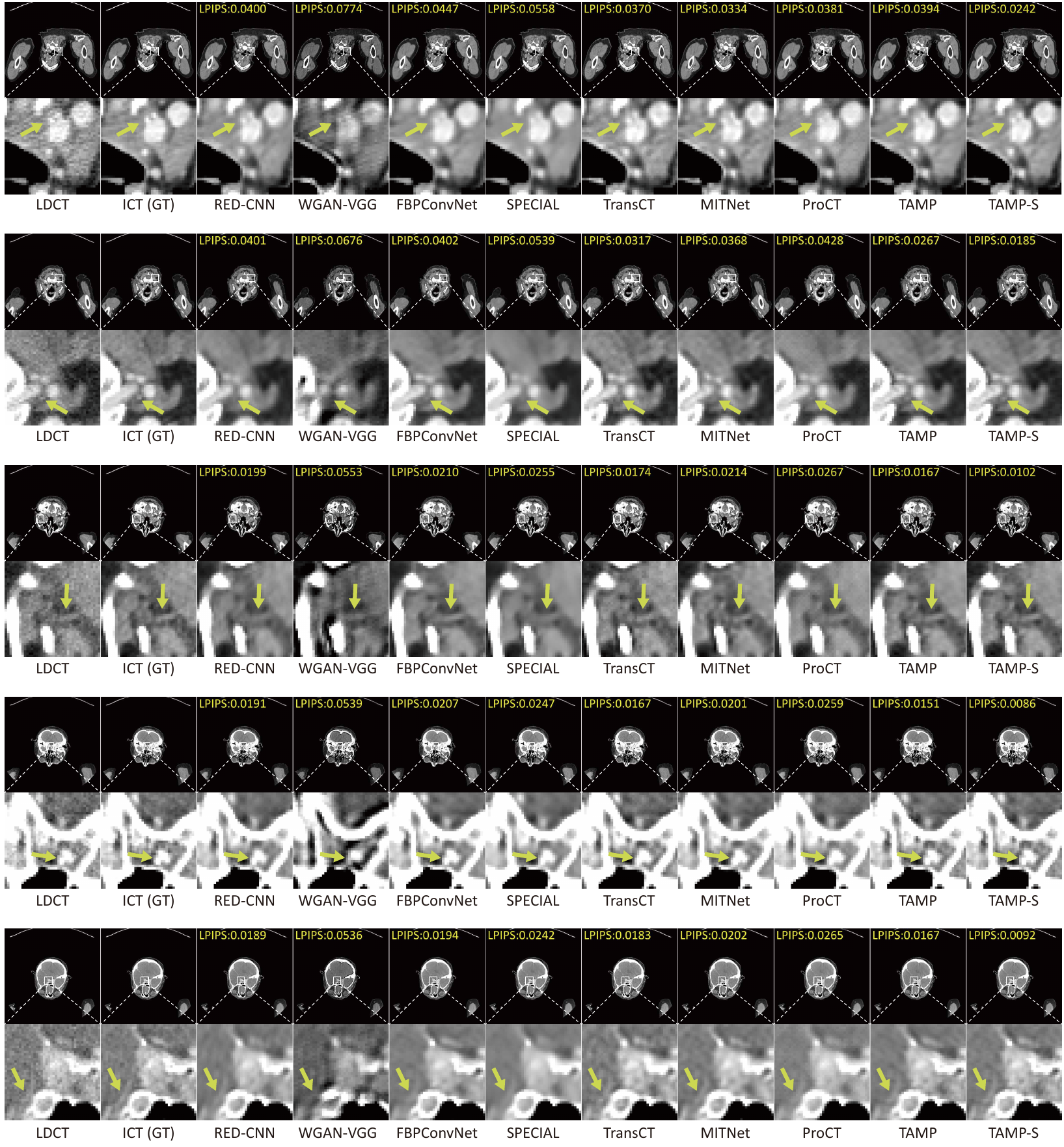}
\caption{Our TAMP has powerful universal enhancement capabilities, using the head LDCT as an example.}
\label{TAMP enhance head LDCT} 
\end{figure*}

\begin{figure*}[p]
\centering
\includegraphics[width=\linewidth]{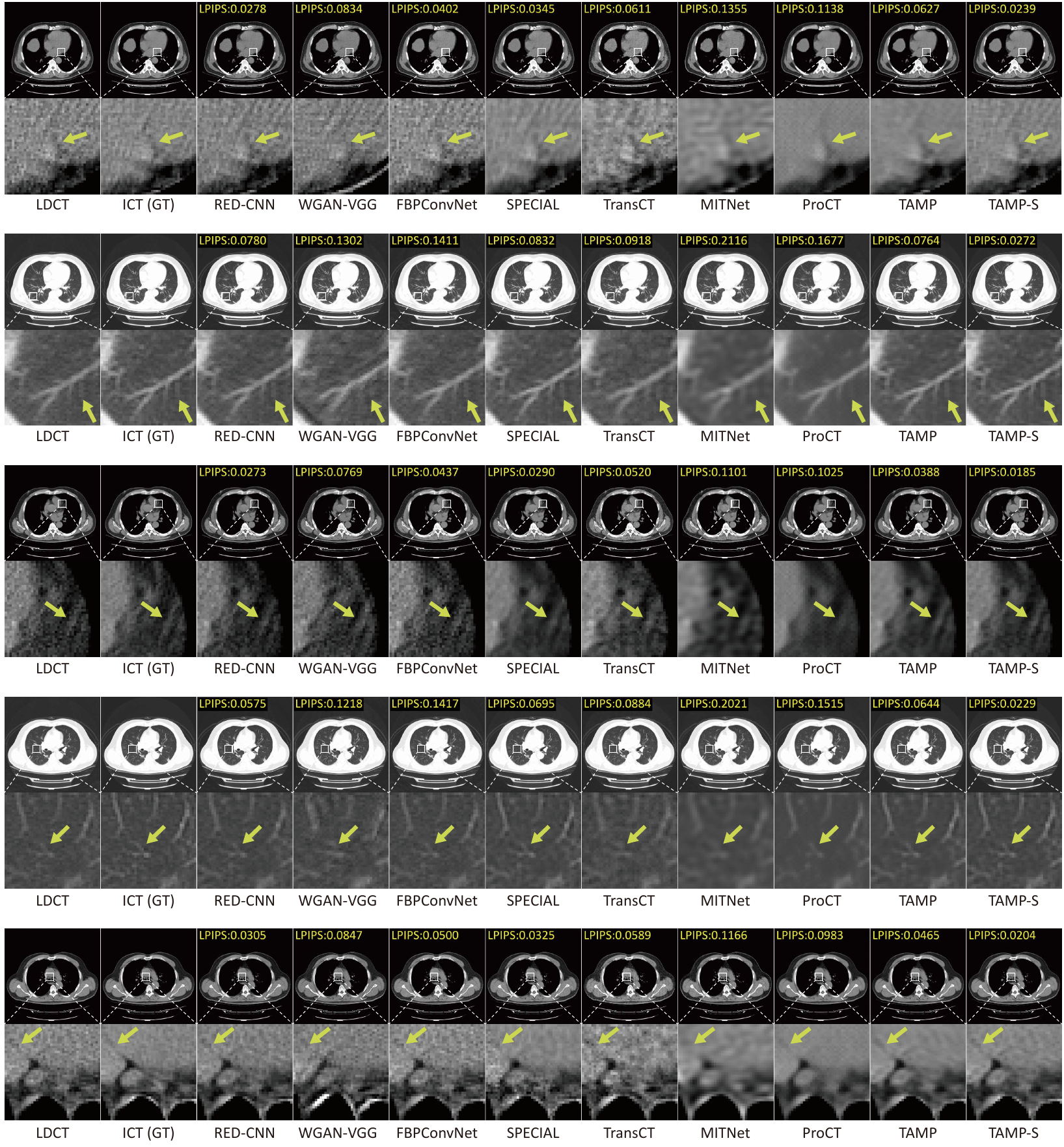}
\caption{Our TAMP has powerful universal enhancement capabilities, using the chest LDCT as an example.}
\label{TAMP enhance chest LDCT} 
\end{figure*}

\begin{figure*}[p]
\centering
\includegraphics[width=\linewidth]{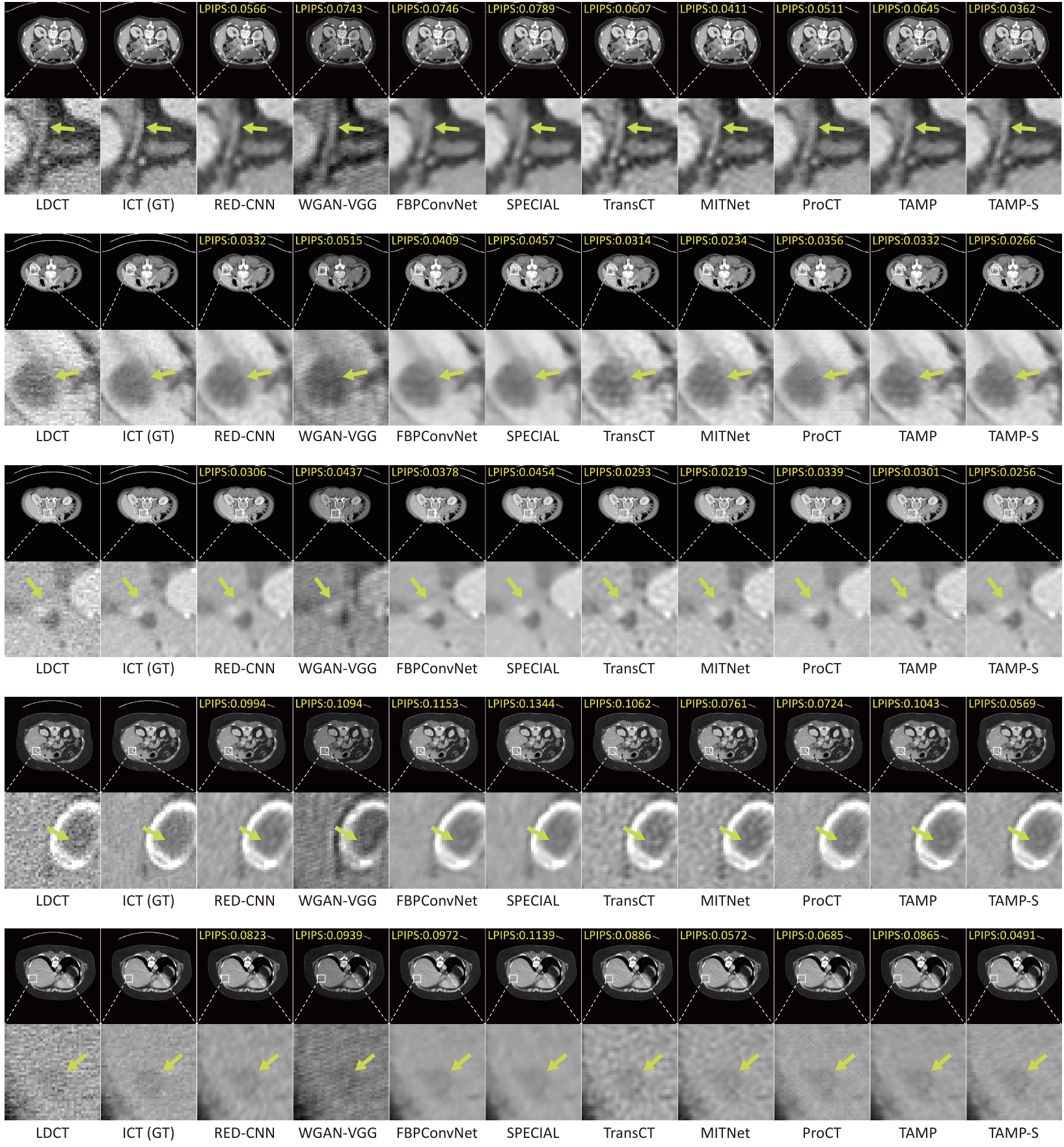}
\caption{Our TAMP has powerful universal enhancement capabilities, using the abdomen LDCT as an example.}
\label{TAMP enhance abdomen LDCT} 
\end{figure*}

\begin{figure*}[p]
\centering
\includegraphics[width=\linewidth]{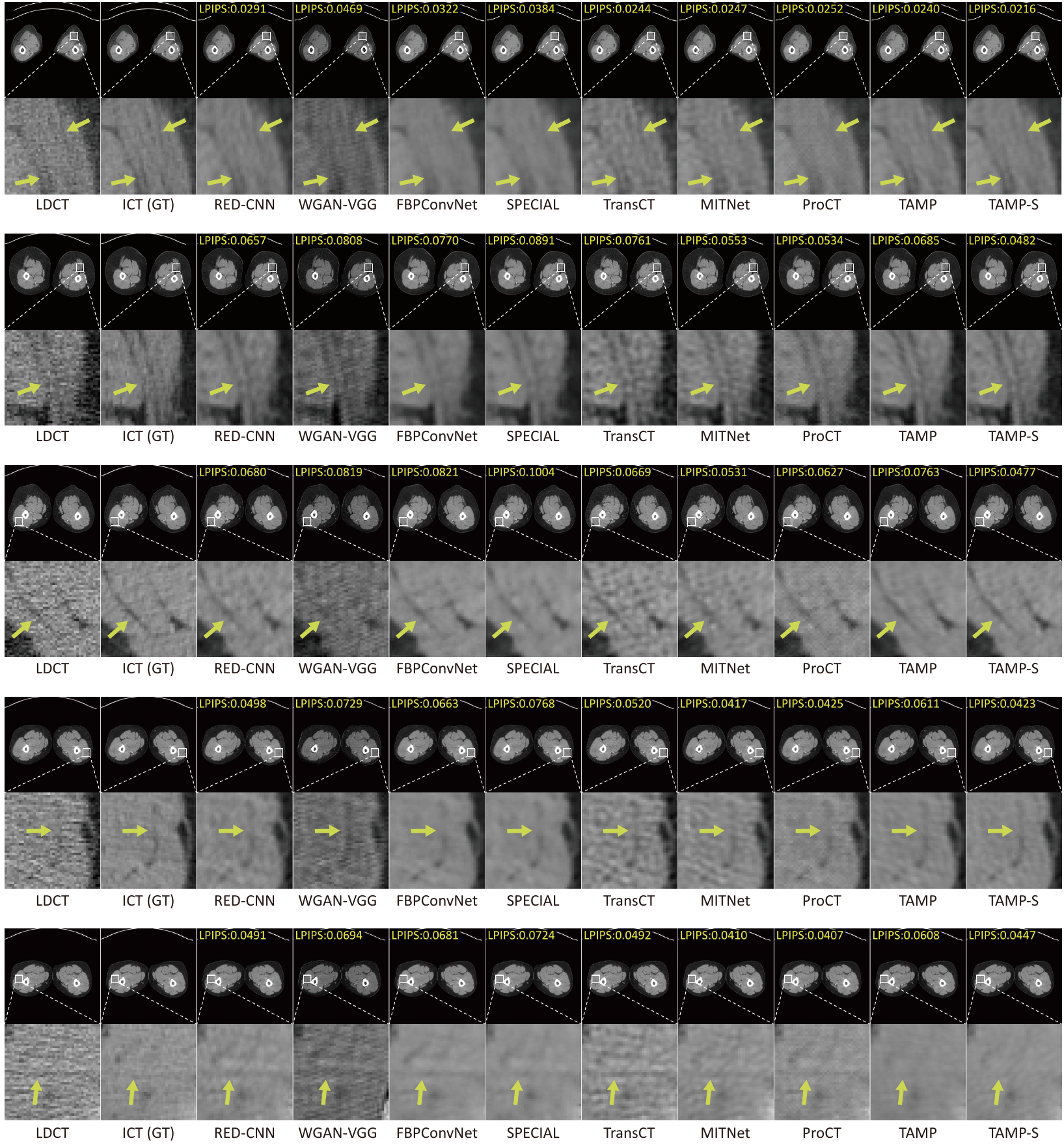}
\caption{Our TAMP has powerful universal enhancement capabilities, using the lower-limbs LDCT as an example.}
\label{TAMP enhance lower-limbs LDCT} 
\end{figure*}

\begin{figure*}[p]
\centering
\includegraphics[width=\linewidth]{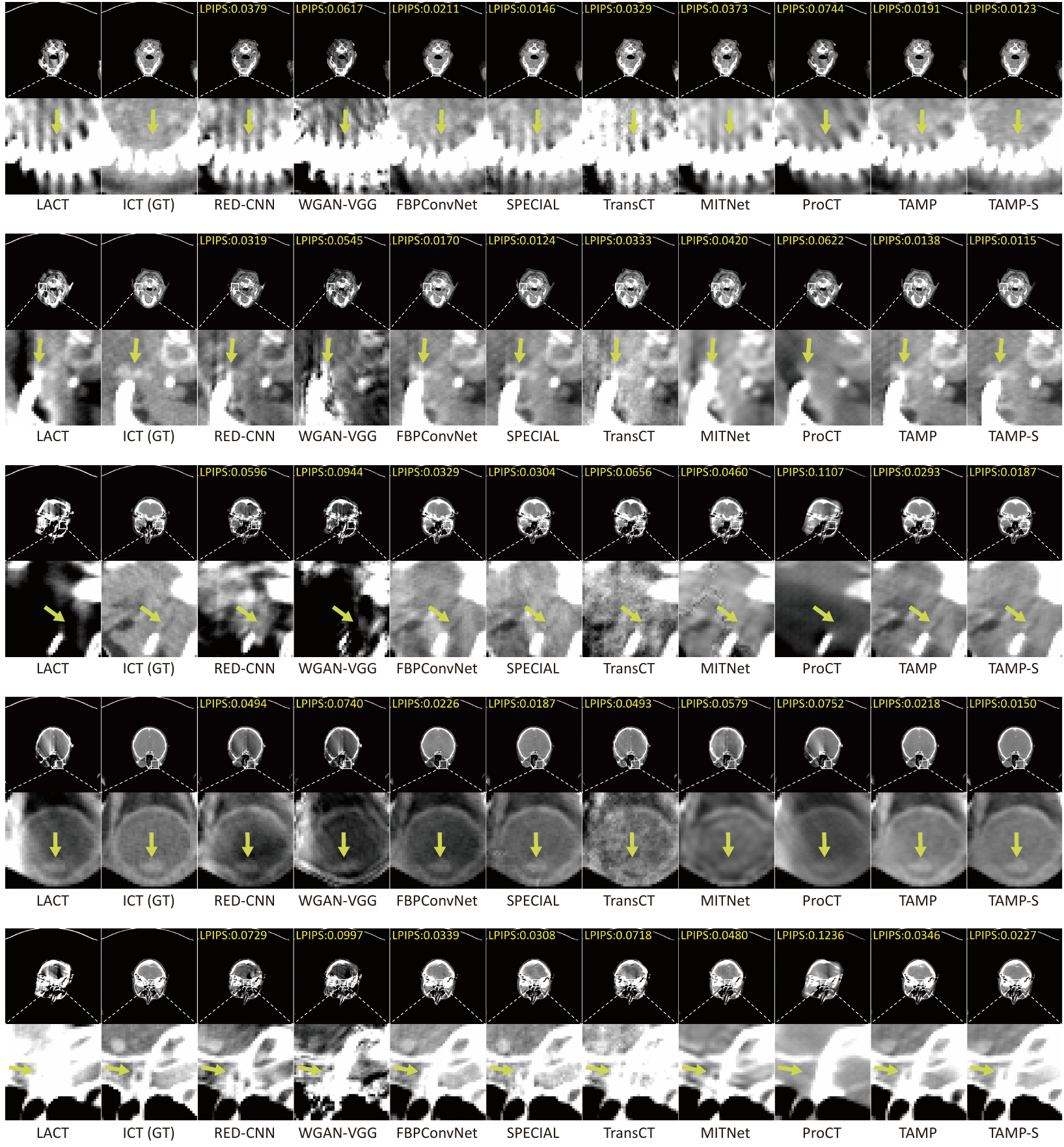}
\caption{Our TAMP has powerful universal enhancement capabilities, using the head LACT as an example.}
\label{TAMP enhance head LACT} 
\end{figure*}

\begin{figure*}[p]
\centering
\includegraphics[width=\linewidth]{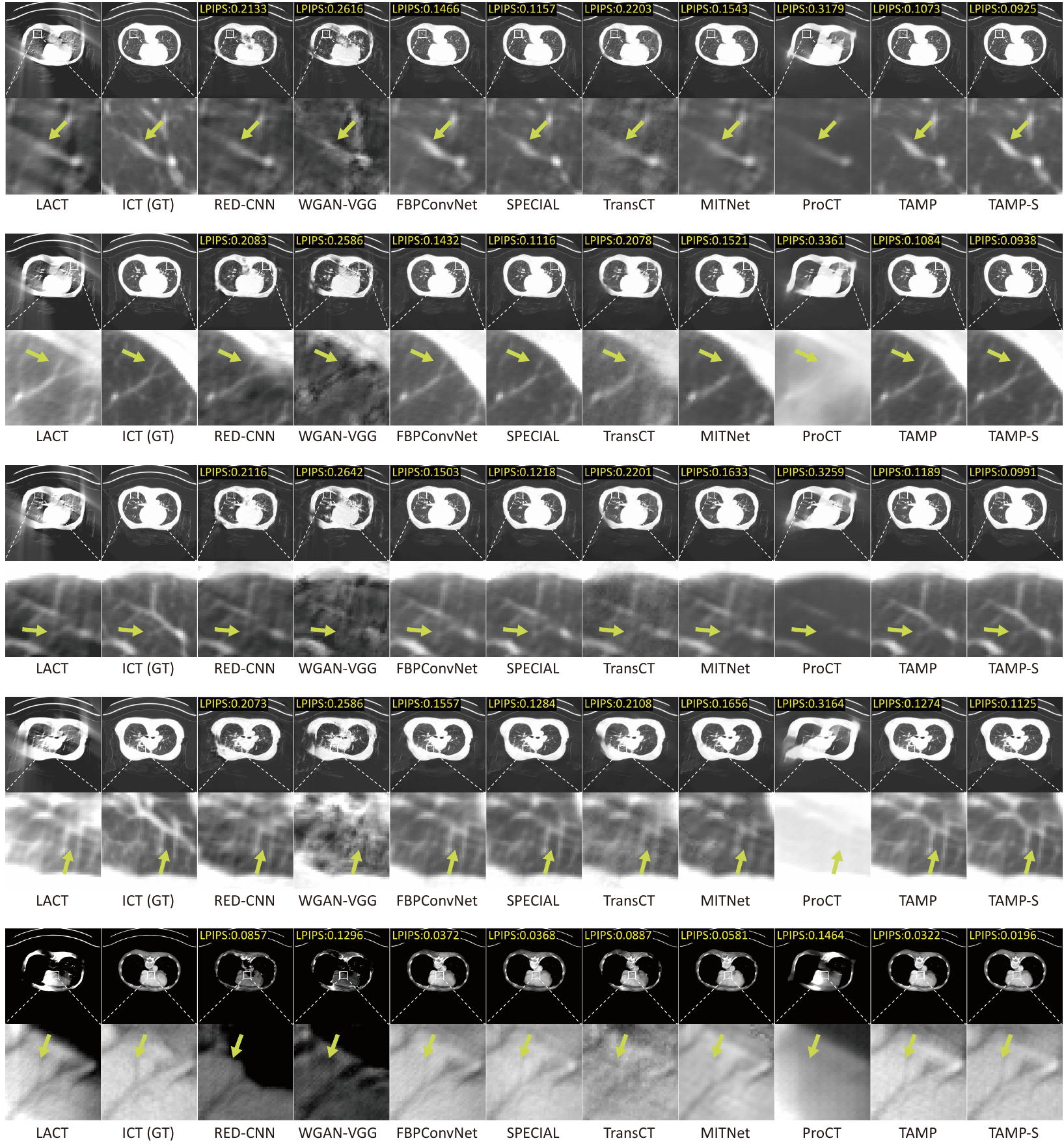}
\caption{Our TAMP has powerful universal enhancement capabilities, using the chest LACT as an example.}
\label{TAMP enhance chest LACT} 
\end{figure*}

\begin{figure*}[p]
\centering
\includegraphics[width=\linewidth]{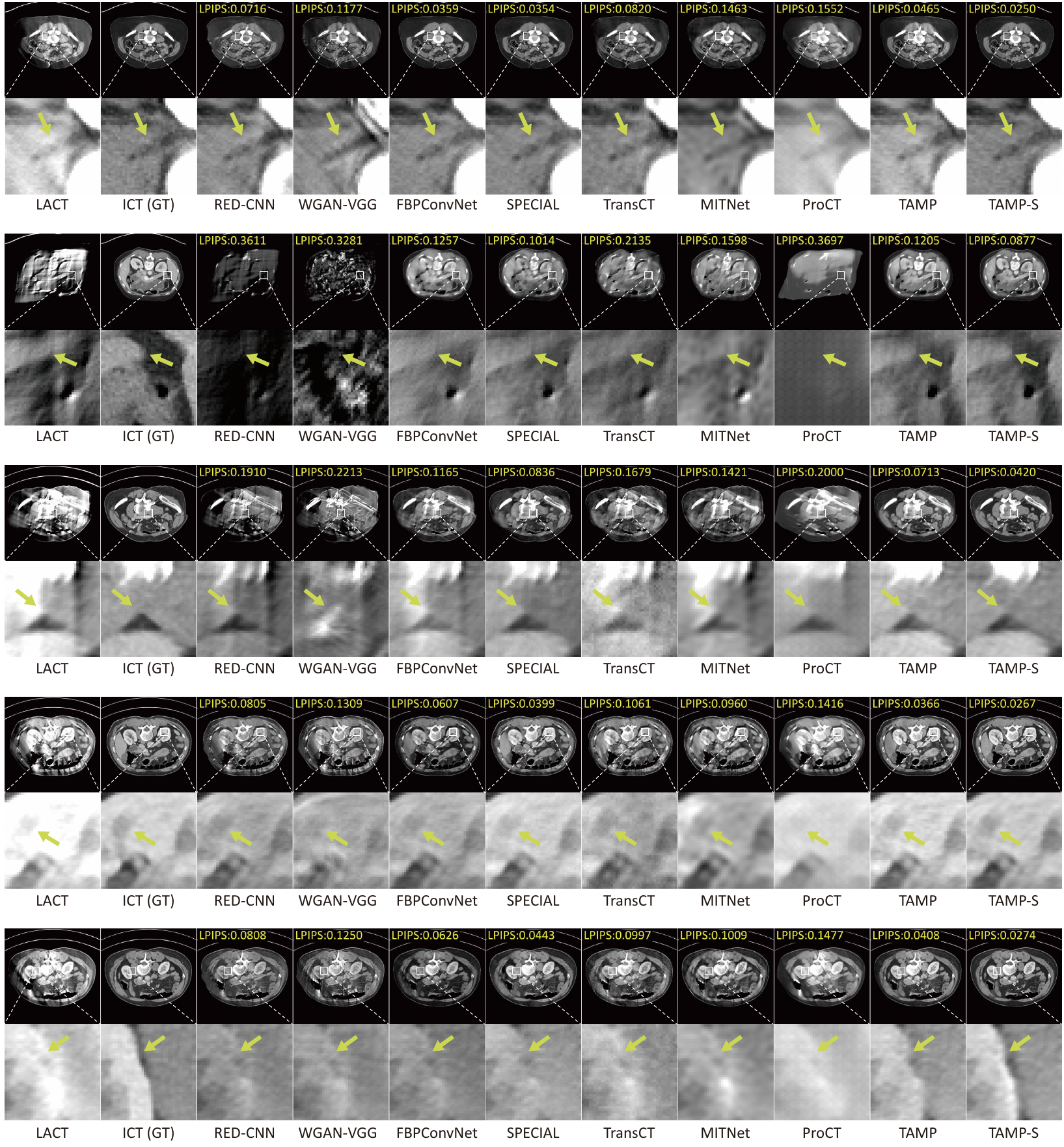}
\caption{Our TAMP has powerful universal enhancement capabilities, using the abdomen LACT as an example.}
\label{TAMP enhance abdomen LACT} 
\end{figure*}

\begin{figure*}[p]
\centering
\includegraphics[width=\linewidth]{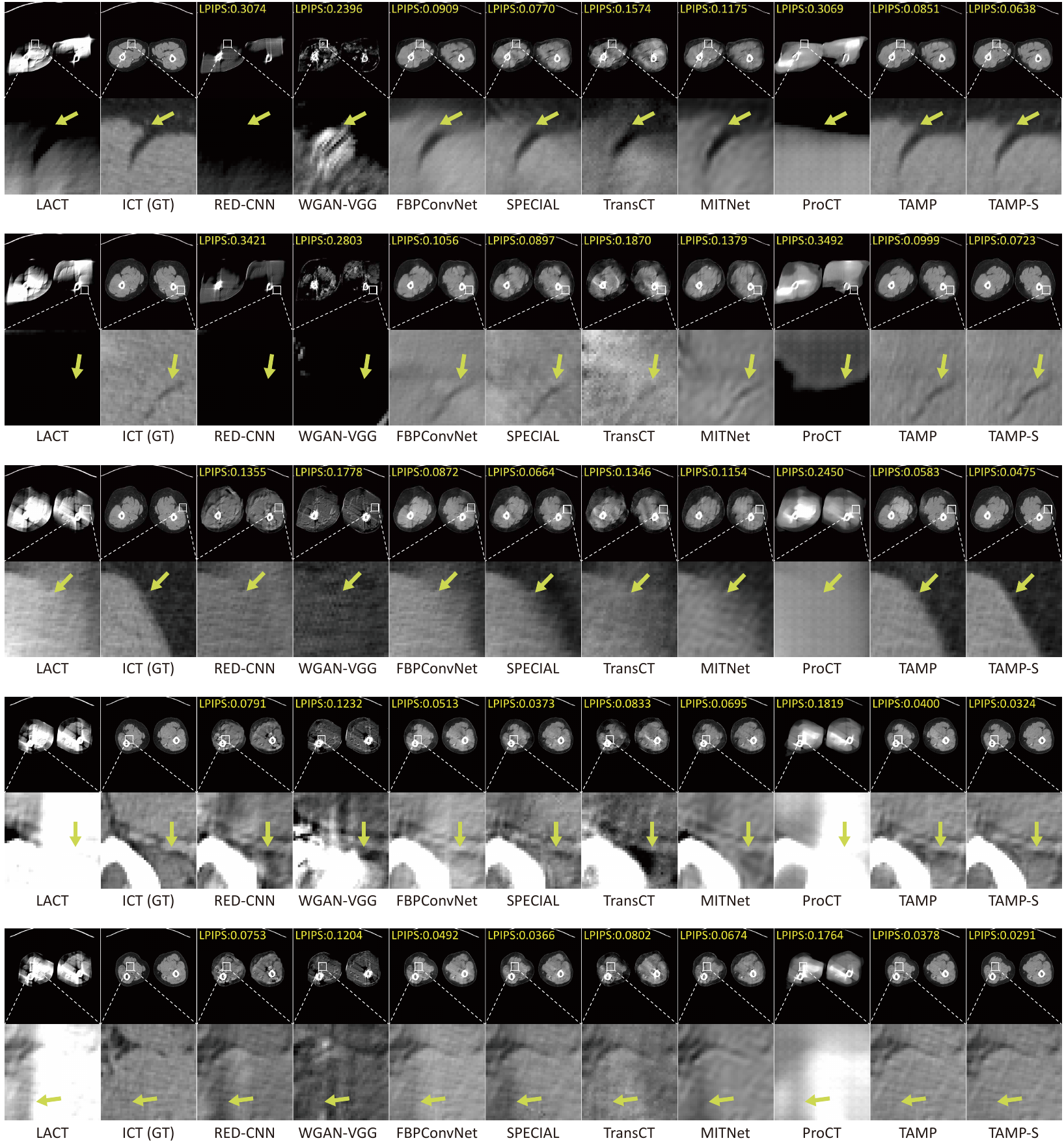}
\caption{Our TAMP has powerful universal enhancement capabilities, using the lower-limbs LACT as an example.}
\label{TAMP enhance lower-limbs LACT} 
\end{figure*}

\begin{figure*}[p]
\centering
\includegraphics[width=\linewidth]{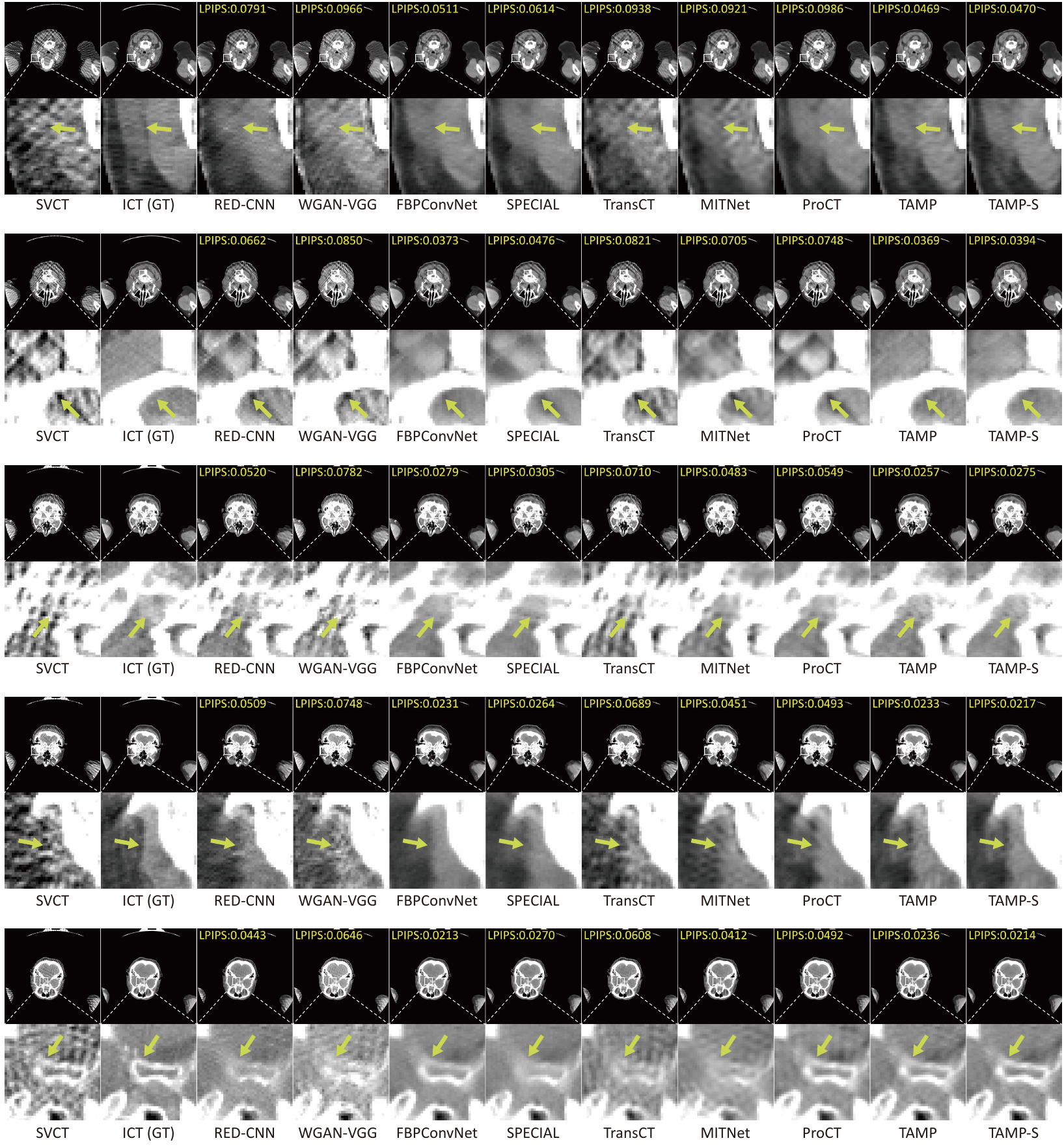}
\caption{Our TAMP has powerful universal enhancement capabilities, using the head SVCT as an example.}
\label{TAMP enhance head SVCT} 
\end{figure*}

\begin{figure*}[p]
\centering
\includegraphics[width=\linewidth]{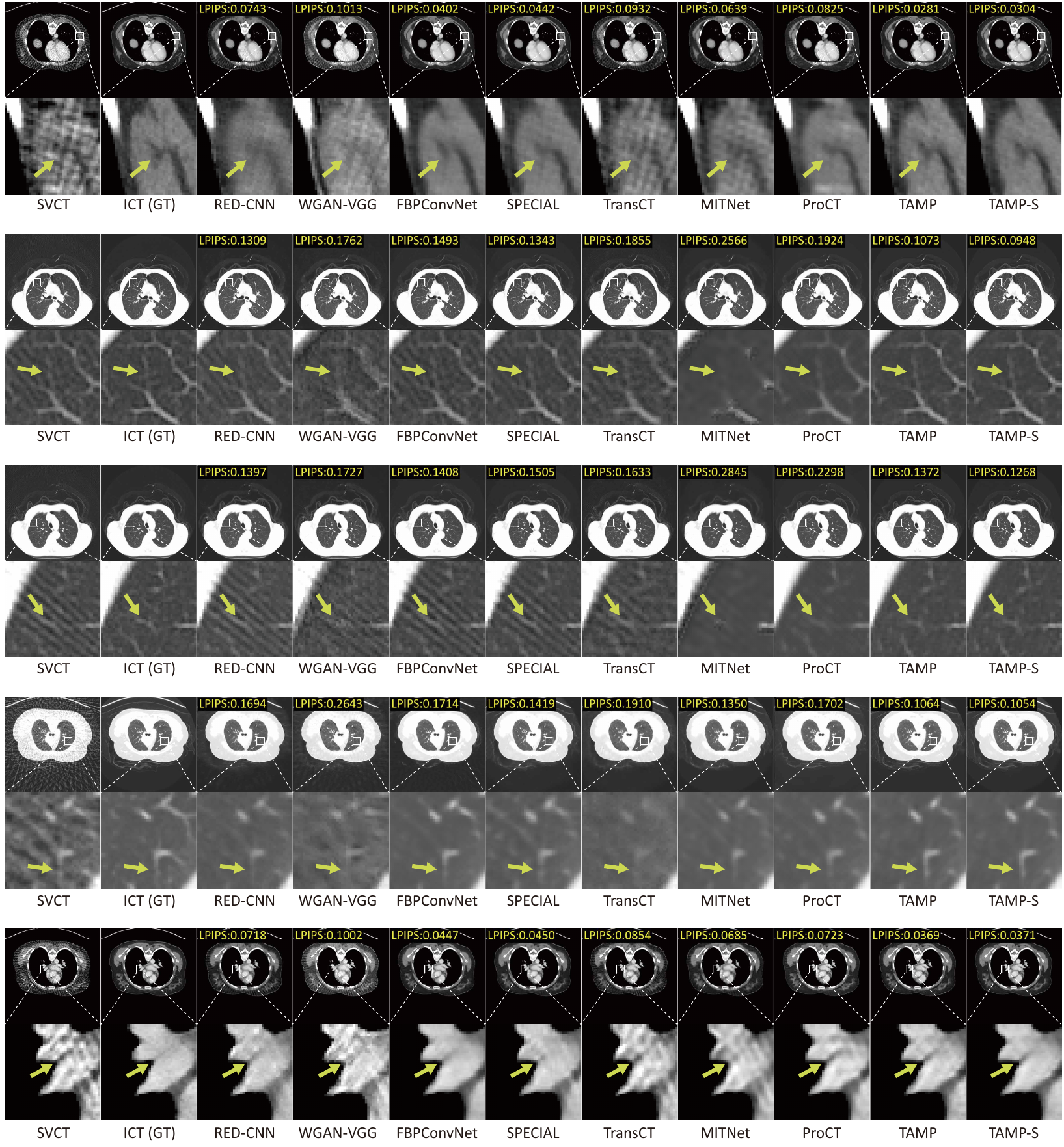}
\caption{Our TAMP has powerful universal enhancement capabilities, using the chest SVCT as an example.}
\label{TAMP enhance chest SVCT} 
\end{figure*}

\begin{figure*}[p]
\centering
\includegraphics[width=\linewidth]{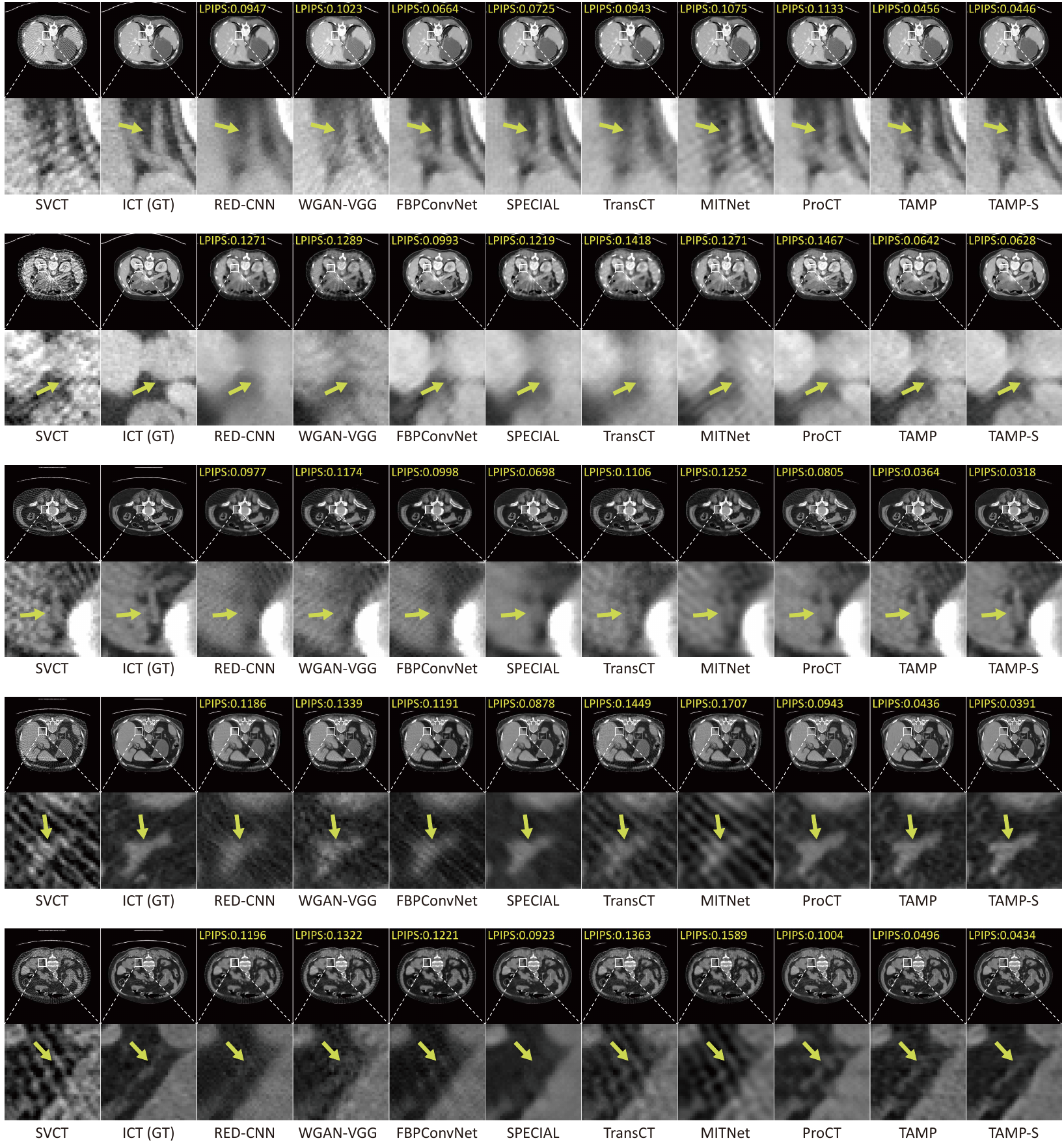}
\caption{Our TAMP has powerful universal enhancement capabilities, using the abdomen SVCT as an example.}
\label{TAMP enhance abdomen SVCT} 
\end{figure*}

\begin{figure*}[p]
\centering
\includegraphics[width=\linewidth]{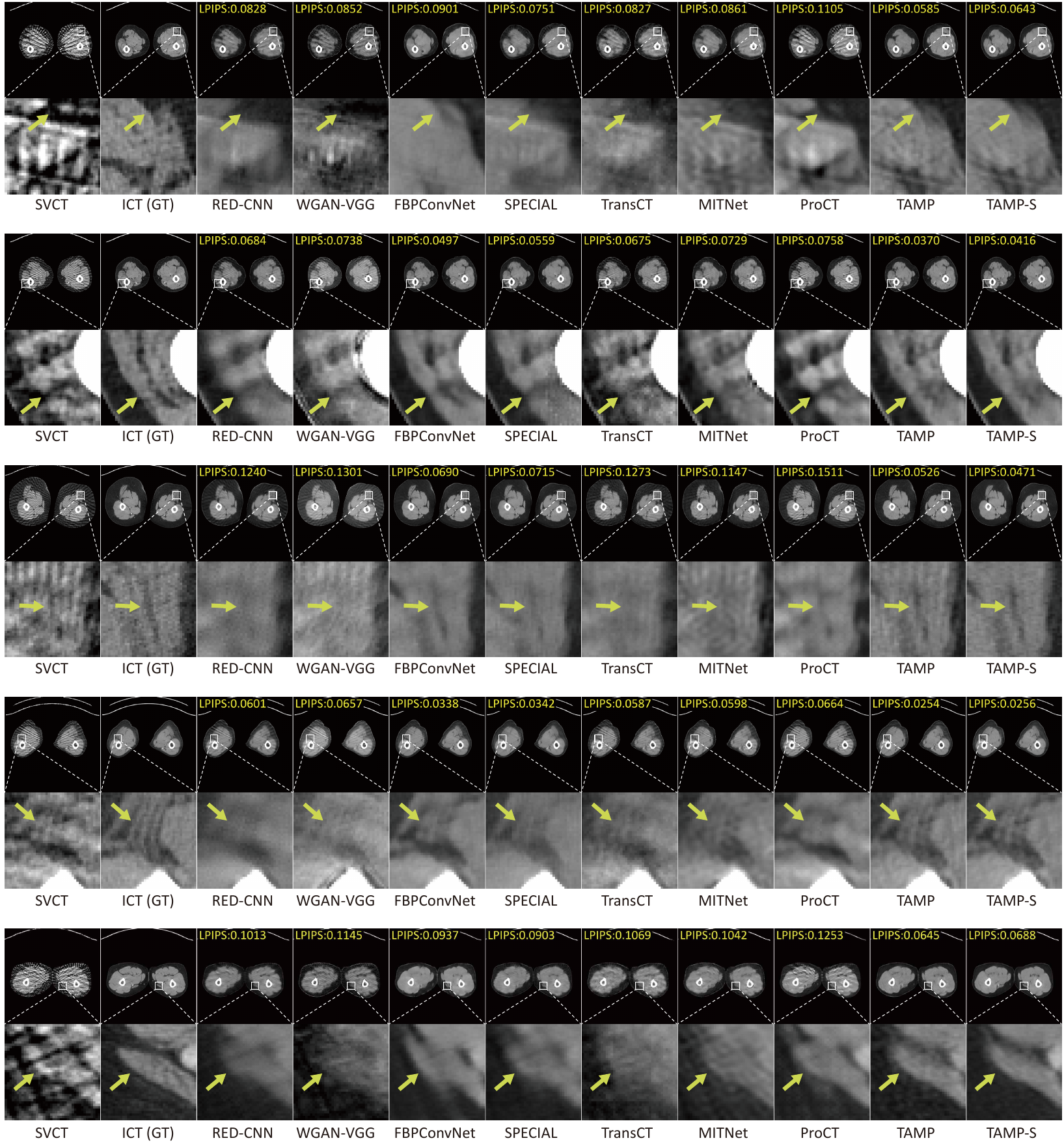}
\caption{Our TAMP has powerful universal enhancement capabilities, using the lower-limbs SVCT as an example.}
\label{TAMP enhance lower-limbs SVCT} 
\end{figure*}

\begin{figure*}[p]
\centering
\includegraphics[width=\linewidth]{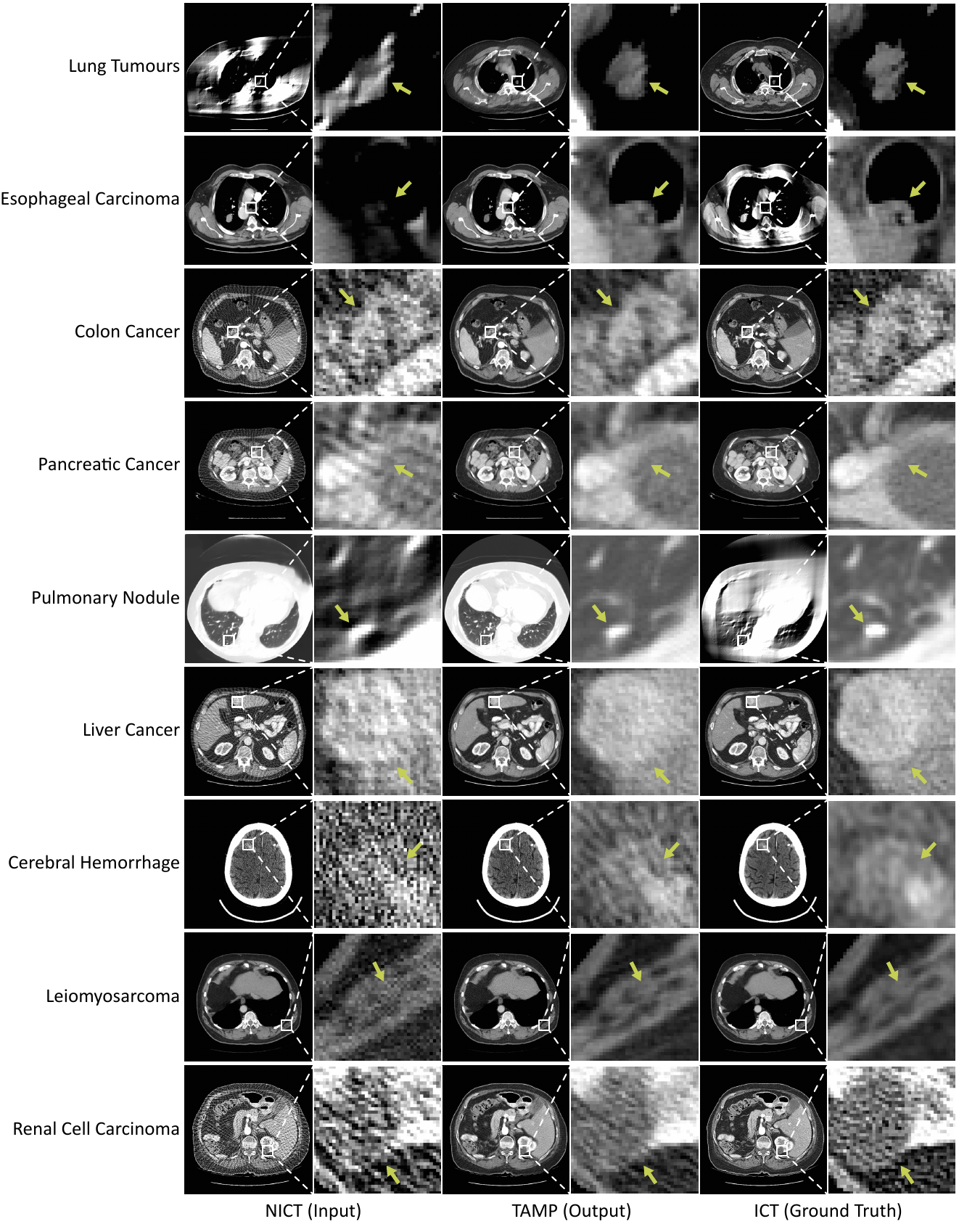}
\caption{Our TAMP demonstrates effective enhancement for NICTs with different lesions, highlighting its significant potential in assisting medical diagnosis across a wide range of clinical scenarios.}
\label{TAMP enhance NICT with lesions} 
\end{figure*}

Our TAMP achieves superior image quality improvement in diverse NICT enhancement tasks across different NICT settings (LDCT, LACT, SVCT) and body regions (head, chest, abdomen, lower-limbs), as shown in Fig. \ref{TAMP enhance head LDCT}-Fig. \ref{TAMP enhance lower-limbs SVCT}. Moreover, TAMP demonstrates equally effective enhancement for NICTs with lesions, highlighting its significant potential in assisting medical diagnosis across a wide range of clinical scenarios, as shown in Fig. \ref{TAMP enhance NICT with lesions}.

\subsection{Additional quantitative results}
The universal NICT enhancement performance of TAMP has been quantitatively evaluated across 27 tasks, including three datasets (AMOS22, AutoPET, COVID-19), three NICT settings (LDCT, LACT, SVCT), and three defect degrees (High, Mid, Low), using four evaluation metrics, i.e., PSNR, SSIM, LPIPS, and RMSE, as shown in Table \hyperref[quantitative_PSNR]{4-7}.

\begin{table*}
\centering
\caption{
Quantitative evaluation of PSNR (dB) for different methods across 27 tasks. The best results are highlighted in \textbf{bold}, and the second-best results are \underline{underlined}.
}\label{quantitative_PSNR}
\resizebox{\linewidth}{!}{
\begin{tabular}{cccccccccc}
\toprule
Task & RedCNN & WGAN-VGG & FBPConvNet & SPECIAL & TransCT & MITNet & ProCT & TAMP & TAMP-S \\
\midrule
AMOS-LDCT-High & 48.96±1.59 & 47.78±1.40 & 50.07±1.90 & 49.50±1.83 & 46.30±1.08 & 46.49±1.24 & 47.79±1.25 & {\ul 50.49±1.95} & \textbf{50.65±1.99} \\
AMOS-LDCT-Mid & 50.13±1.71 & 44.99±1.01 & 51.43±2.06 & 47.29±0.98 & 47.70±1.30 & 46.92±1.33 & 48.38±1.37 & {\ul 51.87±2.15} & \textbf{52.03±2.19} \\
AMOS-LDCT-Low & 51.38±1.91 & 45.18±1.10 & {\ul 52.40±2.19} & 52.11±2.26 & 48.44±1.44 & 47.53±1.43 & 48.61±1.44 & 51.94±1.88 & \textbf{53.05±2.31} \\
AMOS-LACT-High & 37.46±0.74 & 38.30±0.76 & 38.84±0.71 & 39.81±0.96 & 38.59±0.66 & 39.37±0.65 & 35.53±0.72 & {\ul 40.30±0.81} & \textbf{42.48±0.98} \\
AMOS-LACT-Mid & 37.46±1.00 & 37.91±0.84 & 40.89±1.02 & 41.49±1.46 & 39.40±0.77 & 40.29±0.83 & 36.54±0.52 & {\ul 41.67±0.88} & \textbf{44.59±1.23} \\
AMOS-LACT-Low & 40.21±1.40 & 38.85±1.03 & 41.81±1.45 & {\ul 44.03±1.63} & 40.96±1.17 & 41.54±0.96 & 37.83±0.64 & 42.78±1.15 & \textbf{47.49±1.56} \\
AMOS-SVCT-High & 41.50±1.04 & 39.23±0.84 & 40.74±0.78 & 41.65±0.88 & 40.51±1.00 & 41.62±0.92 & {\ul 41.84±0.74} & 41.69±1.14 & \textbf{44.20±1.17} \\
AMOS-SVCT-Mid & 43.49±1.25 & 41.31±1.02 & 43.30±1.19 & {\ul 44.72±1.39} & 41.54±1.07 & 42.82±0.97 & 43.63±0.91 & 44.09±1.54 & \textbf{47.21±1.58} \\
AMOS-SVCT-Low & 49.02±1.85 & 43.24±0.96 & 48.59±1.76 & 49.10±1.97 & 45.44±1.40 & 45.25±1.21 & 46.78±1.07 & {\ul 50.67±1.79} & \textbf{51.85±2.27} \\
APET-LDCT-High & 49.13±1.21 & 41.33±0.90 & {\ul 50.75±1.58} & 50.38±1.46 & 47.69±1.11 & 48.20±1.13 & 48.00±0.95 & 50.25±1.42 & \textbf{50.87±1.64} \\
APET-LDCT-Mid & 50.31±1.18 & 40.29±0.92 & {\ul 51.91±1.61} & 51.82±1.47 & 47.48±1.03 & 48.45±1.03 & 48.66±0.92 & 51.43±1.32 & \textbf{52.13±1.61} \\
APET-LDCT-Low & 50.75±0.88 & 40.30±0.80 & 52.61±1.43 & {\ul 52.74±1.41} & 50.02±1.03 & 49.28±1.00 & 48.70±0.91 & 52.29±1.20 & \textbf{53.10±1.54} \\
APET-LACT-High & 36.95±0.91 & 38.32±0.47 & 40.69±1.06 & 41.06±0.90 & 39.28±0.83 & 40.35±1.01 & 37.00±0.82 & {\ul 41.25±0.85} & \textbf{42.68±1.07} \\
APET-LACT-Mid & 38.82±0.82 & 37.91±0.56 & 41.71±1.01 & {\ul 42.72±0.99} & 40.13±0.94 & 41.75±0.95 & 37.36±0.63 & 42.40±0.86 & \textbf{44.56±1.12} \\
APET-LACT-Low & 39.80±1.00 & 38.39±0.71 & 44.22±1.10 & {\ul 45.71±1.12} & 42.07±1.16 & 42.30±1.15 & 38.41±0.67 & 43.44±1.29 & \textbf{47.50±1.15} \\
APET-SVCT-High & 41.31±0.76 & 39.66±0.72 & 40.06±1.36 & 42.74±1.02 & 41.35±0.81 & 42.37±0.90 & 40.39±0.66 & {\ul 43.15±0.94} & \textbf{44.31±1.19} \\
APET-SVCT-Mid & 43.85±1.07 & 40.53±0.85 & {\ul 47.16±1.26} & 46.43±1.24 & 43.11±1.07 & 44.42±1.07 & 43.12±0.80 & 46.04±1.13 & \textbf{47.18±1.37} \\
APET-SVCT-Low & 46.33±1.52 & 43.63±0.87 & \textbf{51.93±1.26} & 50.51±1.14 & 47.07±1.16 & 47.18±1.03 & 46.98±0.83 & 50.76±1.06 & {\ul 51.60±1.26} \\
COVID-LDCT-High & 44.46±0.65 & 40.18±0.56 & 42.44±0.82 & 45.37±0.82 & 42.33±0.91 & 41.82±0.76 & 42.97±1.09 & {\ul 46.55±0.78} & \textbf{46.90±0.82} \\
COVID-LDCT-Mid & 45.02±0.75 & 40.13±0.52 & 43.25±1.09 & 46.87±1.06 & 43.06±1.07 & 42.36±0.81 & 43.12±1.16 & {\ul 47.54±0.88} & \textbf{47.99±0.92} \\
COVID-LDCT-Low & 45.27±0.96 & 47.15±5.96 & 43.56±1.13 & 46.95±1.01 & 42.92±1.15 & 42.04±0.84 & 43.17±1.18 & {\ul 48.12±0.97} & \textbf{48.67±1.01} \\
COVID-LACT-High & 37.37±0.56 & 38.80±0.78 & 38.41±0.67 & 38.92±0.50 & 38.40±0.39 & 38.59±0.52 & 35.67±0.42 & {\ul 39.86±0.65} & \textbf{40.99±0.77} \\
COVID-LACT-Mid & 38.01±0.63 & 38.06±0.45 & 33.28±0.41 & 39.82±0.64 & 38.41±0.41 & 38.85±0.53 & 36.94±0.52 & {\ul 40.65±0.77} & \textbf{42.48±0.89} \\
COVID-LACT-Low & 38.37±0.96 & 37.77±0.77 & 39.14±0.86 & 41.24±0.79 & 39.02±0.68 & 38.94±0.48 & 37.45±0.58 & {\ul 42.15±0.81} & \textbf{44.60±1.07} \\
COVID-SVCT-High & 39.51±0.68 & 38.09±0.51 & 38.27±0.57 & 40.13±0.67 & 38.74±0.63 & 39.39±0.51 & 37.25±0.89 & {\ul 41.04±0.70} & \textbf{41.30±0.77} \\
COVID-SVCT-Mid & 39.93±0.97 & 37.93±0.41 & 39.82±0.92 & 41.17±0.95 & 39.25±0.80 & 40.38±0.72 & 39.31±1.17 & {\ul 42.48±0.99} & \textbf{42.84±1.02} \\
COVID-SVCT-Low & 42.44±1.02 & 39.35±0.60 & 43.25±1.35 & 43.92±1.26 & 41.23±1.13 & 41.16±0.79 & 42.31±1.18 & {\ul 45.66±1.10} & \textbf{46.04±1.10} \\
\bottomrule
\end{tabular}
}
\end{table*}

\begin{table*}
\centering
\caption{
Quantitative evaluation of SSIM (\%) for different methods across 27 tasks. The best results are highlighted in \textbf{bold}, and the second-best results are \underline{underlined}.
}\label{quantitative_SSIM}
\resizebox{\linewidth}{!}{
\begin{tabular}{cccccccccc}
\toprule
Task & RedCNN & WGAN-VGG & FBPConvNet & SPECIAL & TransCT & MITNet & ProCT & TAMP & TAMP-S \\
\midrule
AMOS-LDCT-High & 99.03±0.50 & 98.82±0.53 & 99.17±0.50 & 99.09±0.52 & 98.38±0.55 & 98.51±0.57 & 98.86±0.52 & {\ul 99.23±0.48} & \textbf{99.24±0.49} \\
AMOS-LDCT-Mid & 99.25±0.42 & 98.40±0.59 & 99.38±0.43 & 98.34±0.53 & 98.81±0.47 & 98.55±0.55 & 99.03±0.49 & {\ul 99.42±0.42} & \textbf{99.43±0.42} \\
AMOS-LDCT-Low & 99.43±0.39 & 97.83±0.60 & {\ul 99.50±0.38} & 99.47±0.42 & 98.96±0.45 & 98.90±0.51 & 99.10±0.48 & 99.46±0.37 & \textbf{99.54±0.37} \\
AMOS-LACT-High & 64.64±7.96 & 59.07±5.60 & 83.49±2.91 & 85.74±4.06 & 79.31±3.89 & 82.08±2.82 & 67.24±4.17 & {\ul 90.32±2.35} & \textbf{93.41±1.57} \\
AMOS-LACT-Mid & 78.79±5.82 & 71.86±5.48 & 89.80±2.94 & 91.17±4.18 & 85.04±3.68 & 87.67±2.87 & 78.67±4.02 & {\ul 94.50±1.47} & \textbf{96.57±1.00} \\
AMOS-LACT-Low & 89.19±4.16 & 85.42±5.04 & 90.83±3.52 & {\ul 96.06±1.45} & 91.31±2.65 & 92.14±1.86 & 85.57±2.99 & 95.70±1.54 & \textbf{98.35±0.60} \\
AMOS-SVCT-High & 93.12±1.78 & 88.47±2.81 & 90.90±1.70 & 93.29±1.51 & 90.93±2.21 & 92.49±1.73 & {\ul 93.86±1.14} & 93.67±1.66 & \textbf{96.60±0.88} \\
AMOS-SVCT-Mid & 96.43±1.13 & 94.30±1.68 & 96.10±1.13 & 97.24±0.98 & 94.52±1.53 & 95.44±1.12 & 97.20±0.73 & {\ul 97.73±0.71} & \textbf{98.49±0.57} \\
AMOS-SVCT-Low & 99.01±0.53 & 95.15±0.91 & 99.01±0.55 & 98.98±0.55 & 97.95±0.68 & 97.91±0.69 & 98.94±0.51 & {\ul 99.33±0.41} & \textbf{99.43±0.44} \\
APET-LDCT-High & 99.05±0.36 & 90.79±2.26 & {\ul 99.28±0.30} & 99.22±0.30 & 98.72±0.44 & 98.83±0.39 & 98.88±0.33 & 99.19±0.32 & \textbf{99.27±0.30} \\
APET-LDCT-Mid & 99.30±0.24 & 86.88±3.23 & {\ul 99.45±0.24} & 99.43±0.23 & 98.86±0.25 & 98.60±0.46 & 99.05±0.25 & 99.40±0.23 & \textbf{99.45±0.23} \\
APET-LDCT-Low & 99.34±0.22 & 86.99±3.47 & {\ul 99.55±0.19} & 99.54±0.17 & 99.27±0.20 & 99.12±0.23 & 99.06±0.25 & 99.52±0.18 & \textbf{99.56±0.18} \\
APET-LACT-High & 68.36±8.23 & 65.95±8.27 & 90.34±3.36 & 91.57±2.34 & 84.34±3.92 & 88.41±3.24 & 70.37±7.45 & {\ul 92.40±1.75} & \textbf{94.79±1.56} \\
APET-LACT-Mid & 86.49±3.58 & 76.60±6.00 & 93.32±2.05 & 94.76±1.59 & 88.97±3.14 & 92.68±1.99 & 77.00±6.58 & {\ul 95.48±1.32} & \textbf{97.04±0.89} \\
APET-LACT-Low & 90.68±2.78 & 85.51±4.26 & 97.23±0.98 & {\ul 97.62±0.76} & 93.61±1.95 & 94.18±1.59 & 87.25±3.74 & 96.00±2.16 & \textbf{98.53±0.44} \\
APET-SVCT-High & 93.59±1.47 & 89.36±2.37 & 94.99±1.20 & 95.16±1.27 & 93.19±1.56 & 94.58±1.34 & 94.33±1.16 & {\ul 96.11±0.98} & \textbf{96.79±0.90} \\
APET-SVCT-Mid & 96.79±0.74 & 90.87±2.65 & {\ul 98.41±0.42} & 98.17±0.49 & 96.28±0.87 & 97.06±0.69 & 97.19±0.57 & 98.11±0.49 & \textbf{98.42±0.45} \\
APET-SVCT-Low & 98.10±0.74 & 96.22±1.02 & \textbf{99.44±0.17} & 99.27±0.18 & 98.53±0.34 & 98.63±0.28 & 98.86±0.29 & 99.33±0.18 & {\ul 99.42±0.17} \\
COVID-LDCT-High & 97.69±0.50 & 92.73±0.82 & 95.76±0.94 & 98.15±0.39 & 96.09±0.68 & 95.31±0.69 & 97.41±0.43 & {\ul 98.46±0.30} & \textbf{98.59±0.27} \\
COVID-LDCT-Mid & 97.97±0.51 & 91.44±1.09 & 96.74±0.88 & 98.61±0.50 & 96.86±0.69 & 95.98±0.54 & 97.48±0.43 & {\ul 98.78±0.24} & \textbf{98.89±0.22} \\
COVID-LDCT-Low & 98.19±0.56 & 71.86±3.41 & 97.07±0.91 & 98.64±0.44 & 96.71±0.75 & 95.69±0.77 & 97.50±0.44 & {\ul 98.94±0.21} & \textbf{99.06±0.19} \\
COVID-LACT-High & 65.55±6.65 & 56.99±5.49 & 66.16±3.69 & 83.03±2.39 & 75.35±2.69 & 74.65±2.63 & 69.19±4.21 & {\ul 88.25±1.84} & \textbf{91.48±1.57} \\
COVID-LACT-Mid & 80.94±2.73 & 73.05±5.39 & 78.12±3.32 & 89.35±1.92 & 81.82±2.47 & 82.36±2.12 & 76.26±3.62 & {\ul 93.22±1.38} & \textbf{95.33±0.96} \\
COVID-LACT-Low & 85.90±3.49 & 72.88±5.37 & 85.45±3.20 & 93.08±1.43 & 86.66±2.25 & 83.85±1.88 & 82.32±3.10 & {\ul 95.49±0.86} & \textbf{97.40±0.60} \\
COVID-SVCT-High & 89.11±2.46 & 79.87±2.87 & 85.75±2.21 & 90.43±2.35 & 85.62±2.96 & 86.55±1.94 & 91.16±1.39 & {\ul 93.20±1.16} & \textbf{93.87±1.14} \\
COVID-SVCT-Mid & 92.70±1.89 & 78.70±3.25 & 92.32±2.10 & 94.28±2.14 & 90.34±2.68 & 90.87±1.85 & 94.88±0.98 & {\ul 96.17±0.72} & \textbf{96.51±0.69} \\
COVID-SVCT-Low & 95.76±1.07 & 87.82±1.53 & 97.01±1.04 & 97.35±0.96 & 94.93±1.53 & 94.29±1.07 & 97.30±0.48 & {\ul 98.25±0.33} & \textbf{98.37±0.32} \\
\bottomrule
\end{tabular}
}
\end{table*}

\begin{table*}
\centering
\caption{
Quantitative evaluation of LPIPS (\%) for different methods across 27 tasks. The best results are highlighted in \textbf{bold}, and the second-best results are \underline{underlined}.
}\label{quantitative_LPIPS}
\resizebox{\linewidth}{!}{
\begin{tabular}{cccccccccc}
\toprule
Task & RedCNN & WGAN-VGG & FBPConvNet & SPECIAL & TransCT & MITNet & ProCT & TAMP & TAMP-S \\
\midrule
AMOS-LDCT-High & 5.37±1.72 & 5.09±1.70 & 5.50±1.88 & 5.90±1.99 & 6.63±2.10 & 6.28±1.78 & 4.71±1.98 & \textbf{2.29±2.45} & {\ul 3.52±1.44} \\
AMOS-LDCT-Mid & 3.41±1.22 & 9.83±2.68 & 3.92±1.40 & 4.57±1.57 & 4.70±1.43 & 5.66±1.51 & 5.42±1.97 & \textbf{1.67±1.93} & {\ul 2.34±0.92} \\
AMOS-LDCT-Low & 2.48±0.93 & 5.16±1.31 & 2.76±1.04 & 3.87±1.46 & 3.64±1.06 & 5.96±1.79 & 6.29±2.15 & {\ul 1.74±0.65} & \textbf{1.66±0.63} \\
AMOS-LACT-High & 31.54±4.24 & 34.48±4.82 & 22.06±4.13 & 15.39±2.89 & 24.62±3.87 & 22.59±3.68 & 37.51±4.79 & \textbf{9.04±2.38} & {\ul 9.88±2.25} \\
AMOS-LACT-Mid & 17.42±3.78 & 21.49±3.63 & 11.61±2.53 & 8.23±2.04 & 16.56±2.97 & 15.53±2.94 & 19.64±2.99 & \textbf{4.78±2.01} & {\ul 5.39±1.20} \\
AMOS-LACT-Low & 8.16±1.99 & 12.70±2.44 & 5.73±1.53 & 4.67±1.25 & 10.45±2.24 & 10.29±1.93 & 13.88±3.08 & \textbf{2.83±1.89} & {\ul 3.02±0.72} \\
AMOS-SVCT-High & 15.94±3.58 & 20.79±3.85 & 17.78±3.73 & 15.26±3.15 & 18.81±3.49 & 16.87±3.34 & 15.21±3.11 & {\ul 15.24±4.13} & \textbf{8.00±1.77} \\
AMOS-SVCT-Mid & 12.60±2.99 & 14.78±2.98 & 13.53±2.96 & 10.20±2.69 & 13.94±2.88 & 15.14±3.15 & 9.50±2.67 & {\ul 5.90±1.14} & \textbf{5.02±1.32} \\
AMOS-SVCT-Low & 4.44±1.89 & 8.04±1.76 & 5.82±2.37 & 4.26±1.52 & 6.25±1.95 & 7.53±1.89 & 6.35±2.22 & \textbf{1.86±1.98} & {\ul 2.41±0.81} \\
APET-LDCT-High & 5.99±3.10 & 8.28±2.55 & 7.12±3.77 & 8.14±4.03 & 6.06±3.64 & 4.72±2.58 & 4.21±3.03 & {\ul 3.64±2.16} & \textbf{3.62±2.52} \\
APET-LDCT-Mid & 3.93±2.00 & 10.13±2.98 & 3.62±2.10 & 3.75±2.01 & 8.35±4.46 & 3.62±1.93 & 5.76±4.41 & {\ul 2.61±1.63} & \textbf{2.52±1.80} \\
APET-LDCT-Low & {\ul 2.79±1.49} & 9.54±2.78 & 4.28±2.63 & 4.31±2.35 & 3.16±1.76 & 4.53±2.44 & 6.94±5.46 & 3.05±2.28 & \textbf{1.68±1.21} \\
APET-LACT-High & 31.95±7.85 & 29.59±7.24 & 13.08±4.80 & 10.64±3.82 & 20.82±5.44 & 16.48±6.23 & 27.01±17.59 & {\ul 10.19±2.23} & \textbf{8.85±3.79} \\
APET-LACT-Mid & 12.94±4.12 & 18.20±4.73 & 8.69±3.18 & 6.88±2.29 & 13.56±4.23 & 11.93±4.47 & 19.84±8.96 & {\ul 6.67±2.08} & \textbf{5.28±2.14} \\
APET-LACT-Low & 7.63±2.12 & 12.36±3.16 & 4.02±1.35 & 3.97±1.46 & 8.26±2.74 & 11.10±4.99 & 14.78±5.34 & {\ul 3.69±1.44} & \textbf{3.19±1.35} \\
APET-SVCT-High & 12.99±4.84 & 14.31±4.36 & 11.69±4.47 & 12.36±5.01 & 14.66±5.23 & 13.41±5.20 & 11.38±7.19 & {\ul 7.59±2.33} & \textbf{7.31±2.61} \\
APET-SVCT-Mid & 10.40±4.11 & 11.64±3.76 & 6.96±2.88 & 7.61±3.22 & 10.89±3.84 & 9.78±3.80 & 8.92±6.14 & {\ul 4.85±1.64} & \textbf{4.54±1.77} \\
APET-SVCT-Low & 4.19±1.47 & 6.31±1.94 & 3.40±1.37 & 4.27±2.04 & 5.38±2.05 & 7.32±3.29 & 7.03±5.62 & \textbf{1.81±0.84} & {\ul 2.04±0.94} \\
COVID-LDCT-High & {\ul 2.76±1.14} & 8.33±1.25 & 3.95±1.61 & 4.17±1.41 & 5.92±2.00 & 13.49±3.11 & 10.12±3.98 & 3.19±2.34 & \textbf{2.55±1.24} \\
COVID-LDCT-Mid & {\ul 2.07±0.80} & 7.61±1.35 & 2.90±0.88 & 2.47±0.96 & 4.81±1.58 & 9.62±2.41 & 11.18±4.55 & 2.33±1.55 & \textbf{1.65±0.79} \\
COVID-LDCT-Low & {\ul 1.85±0.67} & 21.14±3.83 & 2.76±0.75 & 2.01±0.85 & 4.97±1.63 & 12.28±2.84 & 11.63±4.81 & 3.15±1.36 & \textbf{1.36±0.60} \\
COVID-LACT-High & 29.90±4.26 & 35.12±4.27 & 32.76±3.48 & 15.95±2.58 & 25.36±3.82 & 32.43±5.09 & 35.05±7.71 & {\ul 12.42±3.37} & \textbf{9.96±2.32} \\
COVID-LACT-Mid & 15.95±2.59 & 19.55±2.99 & 15.29±2.27 & 9.50±1.98 & 17.64±3.29 & 22.24±3.82 & 27.12±4.92 & {\ul 6.64±3.54} & \textbf{5.73±1.53} \\
COVID-LACT-Low & 9.07±1.87 & 18.46±2.89 & 8.17±1.35 & 5.71±1.22 & 12.36±2.74 & 18.20±3.61 & 19.68±6.55 & \textbf{2.82±2.02} & {\ul 3.37±0.92} \\
COVID-SVCT-High & 16.47±4.38 & 19.55±4.29 & 19.31±4.59 & 14.40±3.57 & 19.94±4.01 & 20.60±4.47 & 12.63±4.77 & \textbf{8.75±1.55} & {\ul 9.06±1.94} \\
COVID-SVCT-Mid & 11.17±3.13 & 15.78±3.23 & 12.76±3.04 & 11.86±3.15 & 16.02±3.52 & 18.72±4.03 & 10.54±4.23 & \textbf{5.87±1.22} & {\ul 6.31±1.46} \\
COVID-SVCT-Low & 5.17±1.94 & 11.93±2.45 & 6.33±2.43 & 6.40±2.20 & 8.04±2.37 & 12.60±3.29 & 10.24±4.30 & \textbf{2.41±1.85} & {\ul 3.26±0.94} \\
\bottomrule
\end{tabular}
}
\end{table*}

\begin{table*}
\centering
\caption{
Quantitative evaluation of RMSE (Hu) for different methods across 27 tasks. The best results are highlighted in \textbf{bold}, and the second-best results are \underline{underlined}.
}\label{quantitative_RMSE}
\resizebox{\linewidth}{!}{
\begin{tabular}{cccccccccc}
\toprule
Task & RedCNN & WGAN-VGG & FBPConvNet & SPECIAL & TransCT & MITNet & ProCT & TAMP & TAMP-S \\
\midrule
AMOS-LDCT-High & 14.86±2.92 & 16.96±2.87 & 13.18±3.20 & 14.05±3.25 & 19.98±2.56 & 19.61±2.93 & 16.88±2.63 & {\ul 12.58±3.15} & \textbf{12.37±3.19} \\
AMOS-LDCT-Mid & 13.03±2.81 & 23.22±2.76 & 11.33±3.08 & 17.81±2.20 & 17.08±2.69 & 18.69±3.01 & 15.82±2.70 & {\ul 10.80±3.05} & \textbf{10.61±3.09} \\
AMOS-LDCT-Low & 11.35±2.83 & 22.75±2.96 & {\ul 10.18±2.99} & 10.55±3.24 & 15.73±2.77 & 17.45±3.04 & 15.41±2.77 & 10.64±2.70 & \textbf{9.49±2.97} \\
AMOS-LACT-High & 55.05±4.59 & 49.98±4.25 & 46.95±3.76 & 42.11±4.68 & 48.33±3.60 & 44.18±3.24 & 68.80±5.71 & {\ul 39.72±3.72} & \textbf{31.00±3.55} \\
AMOS-LACT-Mid & 55.22±6.10 & 52.36±4.92 & 37.24±4.41 & 35.00±6.11 & 44.07±3.96 & 39.78±3.78 & 61.14±3.63 & {\ul 33.96±3.50} & \textbf{24.38±3.55} \\
AMOS-LACT-Low & 40.52±6.82 & 47.12±5.80 & 33.71±5.61 & {\ul 26.23±5.13} & 37.00±5.23 & 34.53±3.90 & 52.70±3.87 & 30.00±4.05 & \textbf{17.58±3.36} \\
AMOS-SVCT-High & 34.72±4.29 & 44.97±4.47 & 37.75±3.42 & 34.03±3.54 & 38.89±4.58 & 34.20±3.75 & {\ul 33.27±2.86} & 34.00±4.43 & \textbf{25.47±3.49} \\
AMOS-SVCT-Mid & 27.69±4.14 & 35.46±4.29 & 28.28±4.00 & {\ul 24.10±4.05} & 34.56±4.39 & 29.80±3.45 & 27.10±2.86 & 25.97±4.54 & \textbf{18.16±3.45} \\
AMOS-SVCT-Low & 14.84±3.40 & 28.37±3.20 & 15.57±3.41 & 14.76±3.61 & 22.19±3.60 & 22.60±3.24 & 18.91±2.48 & {\ul 12.27±2.86} & \textbf{10.87±3.27} \\
APET-LDCT-High & 14.46±1.97 & 35.35±3.56 & {\ul 12.07±2.10} & 12.57±2.03 & 17.03±2.09 & 16.07±1.99 & 16.41±1.75 & 12.74±2.00 & \textbf{11.92±2.15} \\
APET-LDCT-Mid & 12.60±1.65 & 39.85±4.10 & {\ul 10.57±1.92} & 10.65±1.75 & 17.44±1.99 & 15.58±1.76 & 15.20±1.55 & 11.11±1.67 & \textbf{10.30±1.85} \\
APET-LDCT-Low & 11.94±1.20 & 39.74±3.56 & 9.72±1.58 & {\ul 9.57±1.51} & 13.01±1.47 & 14.17±1.56 & 15.12±1.54 & 10.04±1.40 & \textbf{9.21±1.59} \\
APET-LACT-High & 58.55±6.46 & 49.76±2.74 & 38.11±4.67 & 36.45±3.71 & 44.69±4.19 & 39.59±4.40 & 58.15±5.54 & {\ul 35.65±3.48} & \textbf{30.31±3.61} \\
APET-LACT-Mid & 47.13±4.50 & 52.21±3.55 & 33.86±3.92 & {\ul 30.16±3.38} & 40.61±4.37 & 33.70±3.60 & 55.63±4.03 & 31.22±3.08 & \textbf{24.42±3.08} \\
APET-LACT-Low & 42.19±5.00 & 49.44±4.11 & 25.42±3.28 & {\ul 21.42±2.84} & 32.58±4.48 & 31.71±4.20 & 49.36±3.76 & 27.87±4.39 & \textbf{17.43±2.30} \\
APET-SVCT-High & 35.37±3.09 & 42.74±3.57 & 41.17±6.55 & 30.10±3.46 & 35.21±3.26 & 31.33±3.17 & 39.28±3.01 & {\ul 28.68±2.98} & \textbf{25.18±3.32} \\
APET-SVCT-Mid & 26.49±3.32 & 38.69±3.65 & \textbf{18.14±2.60} & 19.73±2.80 & 28.84±3.61 & 24.82±3.01 & 28.72±2.65 & 20.61±2.61 & {\ul 18.14±2.78} \\
APET-SVCT-Low & 20.07±3.73 & 27.12±2.68 & \textbf{10.48±1.47} & 12.32±1.57 & 18.31±2.45 & 18.05±2.10 & 18.42±1.70 & 11.96±1.42 & {\ul 10.89±1.49} \\
COVID-LDCT-High & 24.58±1.89 & 40.21±2.53 & 31.06±2.93 & 22.16±2.20 & 31.49±3.29 & 33.32±2.84 & 29.34±3.64 & {\ul 19.35±1.80} & \textbf{18.59±1.82} \\
COVID-LDCT-Mid & 23.07±2.12 & 40.41±2.37 & 28.41±3.57 & 18.71±2.55 & 29.00±3.58 & 31.34±2.85 & 28.84±3.78 & {\ul 17.28±1.85} & \textbf{16.43±1.84} \\
COVID-LDCT-Low & 22.48±2.66 & 21.07±8.95 & 27.42±3.61 & 18.54±2.43 & 29.52±3.89 & 32.53±3.06 & 28.69±3.83 & {\ul 16.18±1.92} & \textbf{15.20±1.91} \\
COVID-LACT-High & 55.57±3.59 & 47.20±4.17 & 49.36±3.75 & 46.47±2.68 & 49.27±2.19 & 48.25±2.96 & 67.51±3.25 & {\ul 41.73±3.08} & \textbf{36.69±3.25} \\
COVID-LACT-Mid & 51.64±3.64 & 51.26±2.66 & 88.84±4.25 & 41.95±3.05 & 49.27±2.31 & 46.84±2.87 & 58.39±3.41 & {\ul 38.16±3.37} & \textbf{30.96±3.15} \\
COVID-LACT-Low & 49.69±5.28 & 53.19±4.76 & 45.45±4.31 & 35.68±3.23 & 45.97±3.44 & 46.34±2.49 & 55.04±3.61 & {\ul 32.10±3.00} & \textbf{24.31±2.99} \\
COVID-SVCT-High & 43.49±3.40 & 51.14±2.97 & 50.09±3.23 & 40.46±3.14 & 47.49±3.37 & 44.01±2.59 & 56.51±5.87 & {\ul 36.45±2.91} & \textbf{35.39±3.12} \\
COVID-SVCT-Mid & 41.54±4.62 & 52.01±2.39 & 42.04±4.44 & 36.04±4.03 & 44.82±4.07 & 39.35±3.28 & 44.76±5.95 & {\ul 30.99±3.52} & \textbf{29.74±3.50} \\
COVID-SVCT-Low & 31.14±3.77 & 44.22±3.01 & 28.52±4.49 & 26.38±3.99 & 35.84±4.59 & 35.97±3.19 & 31.67±4.26 & {\ul 21.53±2.80} & \textbf{20.60±2.71} \\
\bottomrule
\end{tabular}
}
\end{table*}

\subsection{External-scene adaptation evaluation results}
We evaluated the performance of TAMP in adapting to sparse-view cone beam CT (CBCT) and low-dose Micro-CT enhancement tasks to assess its generalization capability as a foundation model.

\subsubsection{Sparse-view CBCT}
\begin{figure*}[ht] 
\centering
\includegraphics[width=\linewidth]{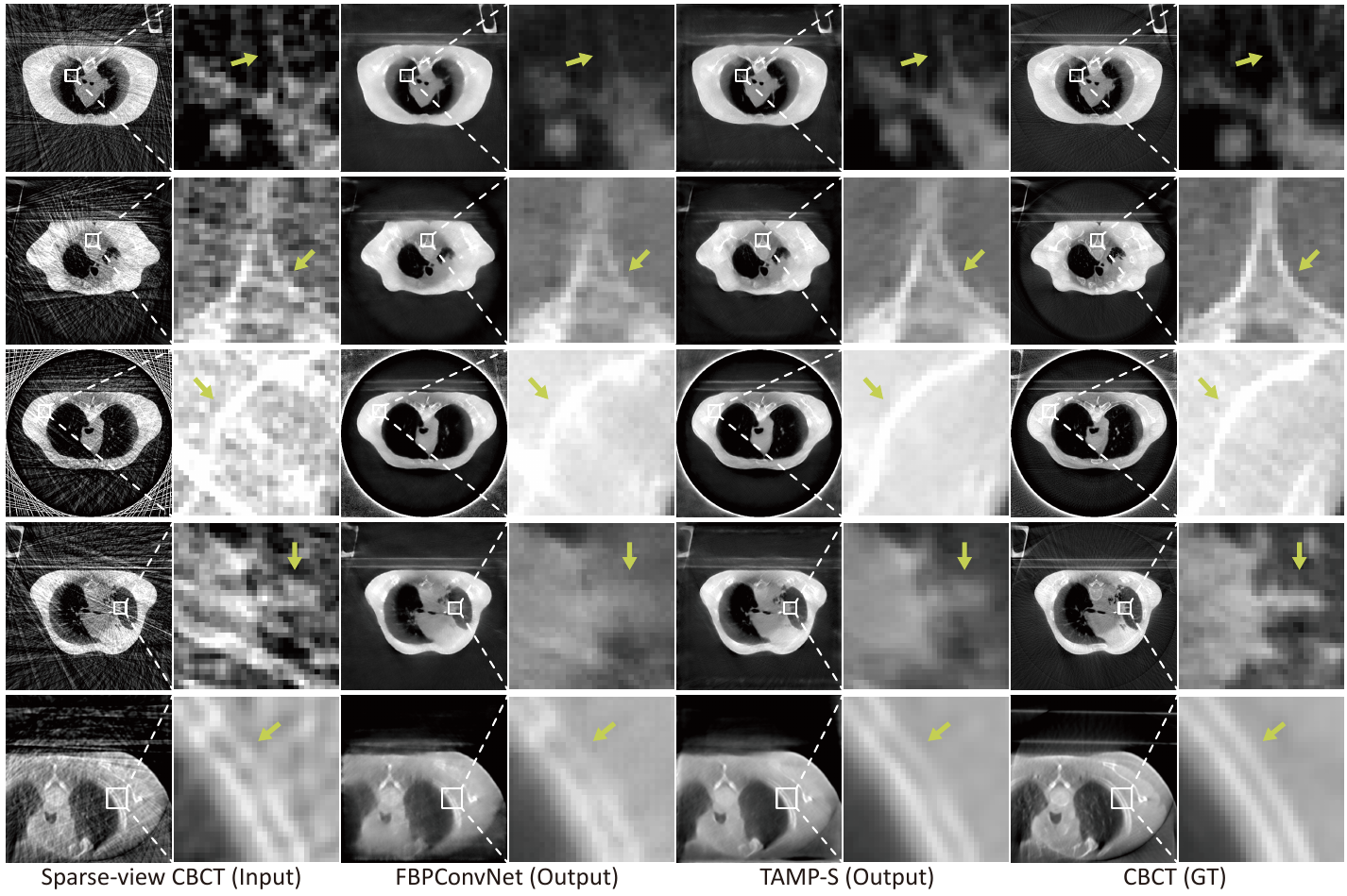}
\caption{
Our TAMP demonstrates effective adaptation capability in enhancing sparse-view CBCT, highlighting its significant potential in widespread clinical scenarios.
}
\label{sup_spare} 
\end{figure*}
We evaluate the adaptability of TAMP to the sparse-view CBCT enhancement task. The clinical Varian and Elekta datasets, sourced from the SPARE Challenge\cite{shieh2019spare}, are used for model training and testing. Specifically, each dataset contains multiple scans from five patients, with one scan used for training and the remaining scans used for testing. For each single scan, the CBCT FDK reconstruction is used as the reference for model inference, while one phase bin from the ten-phase sparse-view CBCT FDK reconstruction is used as the input for model inference. The TAMPS is adapted, and FBPConvNet is trained for comparison. The test results of FBPConvNet and TAMP-S are visually presented, as shown in Fig. \ref{sup_spare}. Compared with FBPConvNet, the sparse-view CBCT with TAMP-S demonstrates clearer and more accurate enhancement, showing smoother skeletal lines and clearer vascular structures.

\subsubsection{Limited-angle CBCT}
We evaluate TAMP on an external cone-beam CT dataset with limited-angle artifacts to validate its capability across axial CT modalities. The dataset comprises 4 cases (1576 slices) from radiotherapy workflows. For baseline methods (RED-CNN, FBPConvNet, ProCT), we trained specialized models from scratch on the cone-beam CT training set. TAMP was evaluated in both zero-shot and fine-tuned (TAMP-S) settings.

\begin{figure*}[h]
\centering
\includegraphics[width=\linewidth]{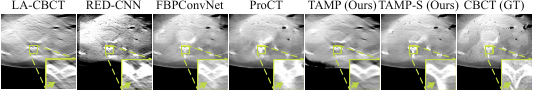}
\caption{Qualitative comparison on limited-angle cone-beam CT enhancement. TAMP effectively suppresses radial artifacts and TAMP-S further reconstructs corrupted anatomical details.}
\label{fig:sup_cbct}
\end{figure*}

As shown in Table~\ref{tab:sup_cbct} and Fig.~\ref{fig:sup_cbct}, TAMP demonstrates effective knowledge transfer from diagnostic CT to cone-beam CT. TAMP-S achieves PSNR of 30.59 dB and SSIM of 84.33\%, outperforming the best baseline ProCT (30.16 dB, 82.95\%) by +0.43 dB and +1.38\% respectively. Qualitatively, in the limited-angle CBCT input, vertebral bone structures are severely corrupted with radial wedge-shaped artifacts. The baseline methods fail to suppress these artifacts or reconstruct the damaged structures, while TAMP effectively suppresses artifacts and TAMP-S further reconstructs the corrupted vertebral bone structures approaching the ground truth.

\begin{table*}[h]
\centering
\caption{Quantitative evaluation on limited-angle cone-beam CT enhancement.}
\label{tab:sup_cbct}
\begin{tabular}{lcccc}
\toprule
Method & PSNR (dB) $\uparrow$ & SSIM (\%) $\uparrow$ & RMSE (Hu) $\downarrow$ & LPIPS (\%) $\downarrow$ \\
\midrule
LA-CBCT (Input) & 23.48$\pm$4.77 & 75.16$\pm$10.39 & 522.17$\pm$298.40 & 36.12$\pm$7.66 \\
RED-CNN & 21.67$\pm$6.35 & 62.46$\pm$8.94 & 599.97$\pm$259.62 & 40.78$\pm$13.07 \\
FBPConvNet & 26.64$\pm$4.77 & 79.75$\pm$9.21 & 369.38$\pm$226.09 & 26.09$\pm$8.49 \\
ProCT & 30.16$\pm$5.54 & 82.95$\pm$8.51 & 252.77$\pm$170.80 & 32.04$\pm$10.87 \\
TAMP & 25.12$\pm$5.51 & 71.39$\pm$10.43 & 415.73$\pm$220.76 & 37.26$\pm$9.18 \\
TAMP-S & \textbf{30.59$\pm$5.87} & \textbf{84.33$\pm$7.86} & \textbf{244.97$\pm$173.44} & \textbf{25.10$\pm$8.69} \\
\bottomrule
\end{tabular}
\end{table*}

\subsubsection{Low-dose Micro-CT}
\begin{figure*}[ht] 
\centering
\includegraphics[width=\linewidth]{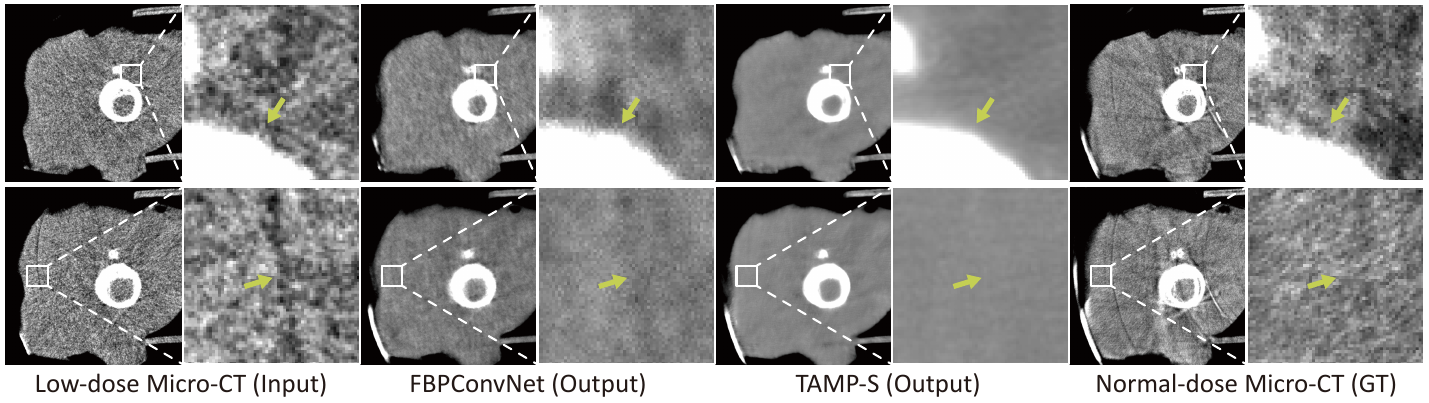}
\caption{
Our TAMP demonstrates effective adaptation capability in enhancing low-dose Micro-CT, highlighting its significant potential in widespread clinical scenarios.
}
\label{sup_micro-ct} 
\end{figure*}
We evaluate TAMP's adaptability for low-dose Micro-CT enhancement tasks. A chicken drumstick was scanned using both low-dose (80 kVp) and normal-dose (120 kVp) settings, with subsequent reconstruction into micro-CT image volumes via the advanced MARS spectral CT equipment at Rensselaer Polytechnic Institute. After registration and cropping, both low-dose and normal-dose Micro-CT volumes contain 5 bins of 280×1216×1216 pixels. The first bin's 5 slices are used for training, while the remaining slices are reserved for testing. The TAMPS is adapted, and FBPConvNet is trained for comparison. The test results of FBPConvNet and TAMP-S are visually presented, as shown in Fig. \ref{sup_micro-ct}. Compared with FBPConvNet, the low-dose Micro-CT with TAMP-S demonstrates clearer and more accurate enhancement, showing sharper edges, and the ring artifacts are more thoroughly suppressed.

\subsection{Universal enhancement validation}
To validate TAMP's universal enhancement capability in multi-scenario deployment settings that represent real-world clinical practice, we conducted mixed test cohort evaluation and oracle ensemble comparison.

\subsubsection{Mixed test cohort evaluation}
We constructed a mixed test cohort by pooling all 27 diverse NICT enhancement scenarios (3 datasets $\times$ 3 NICT types $\times$ 3 degradation degrees) from the main manuscript test sets. For baseline methods (RED-CNN, FBPConvNet, ProCT), we trained 27 specialized models, each targeting one specific scenario. To simulate real-world deployment where scene information is unavailable, we evaluated each test scenario by applying all 27 specialized models and averaging their performance metrics. In contrast, TAMP employs a single universal model and directly tested across all 27 mixed scenarios without adaptation, representing the zero-shot generalization capability of our foundation model.

\begin{figure*}[h]
\centering
\includegraphics[width=\linewidth]{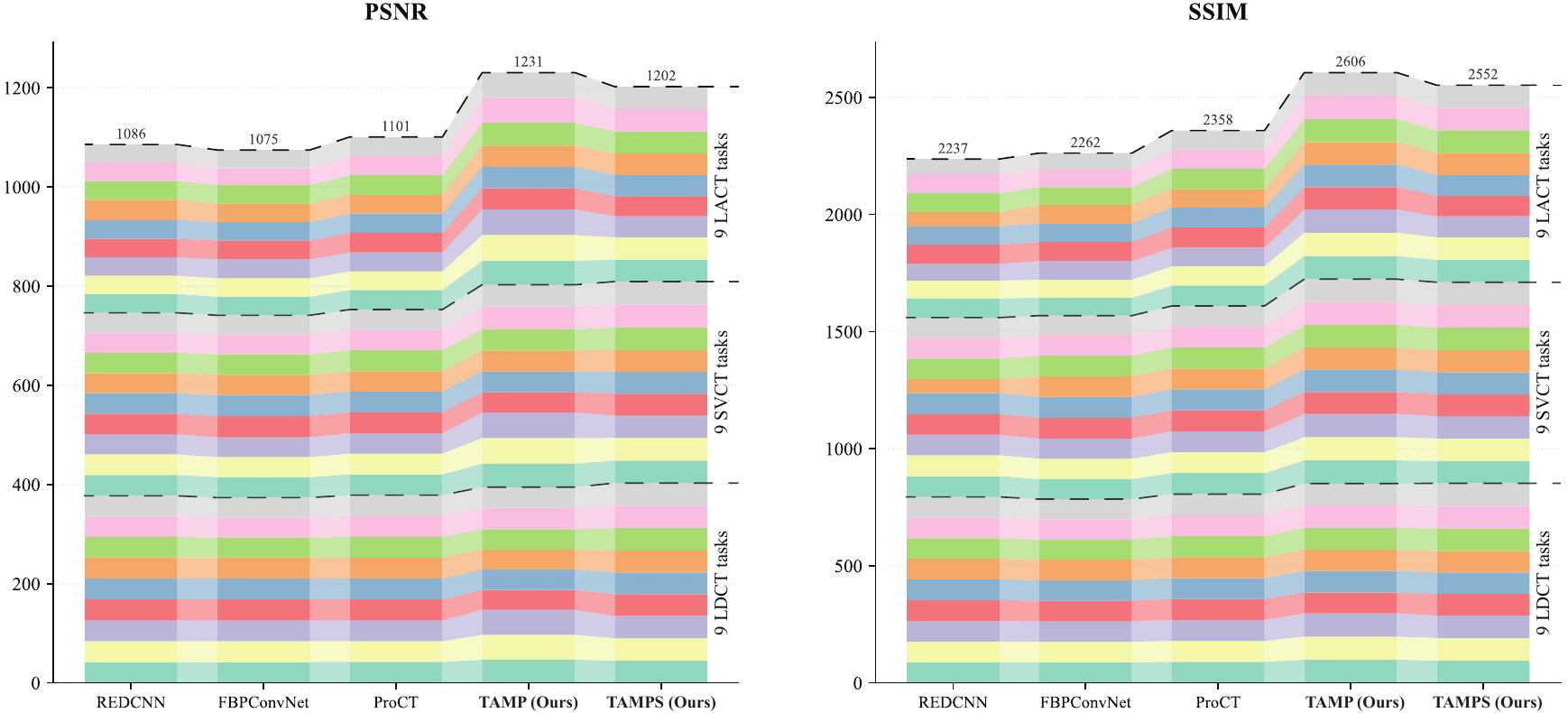}
\caption{Stacked bar chart showing PSNR and SSIM performance across 27 NICT enhancement scenarios. Each bar represents one method, with 27 segments stacked to show individual scenario contributions. TAMP (single universal model) achieves the highest cumulative performance without requiring scene-specific adaptation.}
\label{fig:sup_mixed}
\end{figure*}

As shown in Table~\ref{tab:sup_mixed} and Fig.~\ref{fig:sup_mixed}, TAMP achieves cumulative PSNR of 1230.56 dB and SSIM of 2605.66\% across all 27 scenarios, outperforming the best baseline ProCT by 11.79\% and 10.48\% respectively. The consistent performance superiority across heterogeneous scenarios demonstrates that TAMP's physics-driven pretraining learns universal representations that effectively capture diverse artifact patterns, enabling robust generalization without requiring specialized models for each scenario combination.

\begin{table*}[h]
\centering
\caption{Mixed test cohort evaluation: cumulative performance across 27 NICT enhancement scenarios.}
\label{tab:sup_mixed}
\begin{tabular}{lcc}
\toprule
Method & Cumulative PSNR (dB) $\uparrow$ & Cumulative SSIM (\%) $\uparrow$ \\
\midrule
RED-CNN & 1087.53 & 2312.45 \\
FBPConvNet & 1100.82 & 2359.21 \\
ProCT & 1100.73 & 2359.02 \\
TAMP & \textbf{1230.56} & \textbf{2605.66} \\
\bottomrule
\end{tabular}
\end{table*}

\subsubsection{Oracle ensemble comparison}
To compare TAMP against the theoretical upper bound of specialized model approaches, we constructed an oracle ensemble evaluation. The oracle ensemble assumes perfect prior knowledge: for each test image, an oracle knows its exact (dataset, NICT type, degradation degree) combination and selects the corresponding perfectly-matched specialized model from the 27 trained models. This represents the theoretical performance upper bound of specialized approaches, which is unattainable in real-world deployment where scene information is typically unavailable.

As shown in Table~\ref{tab:sup_oracle}, TAMP achieves PSNR of 46.16 dB and SSIM of 97.35\%, outperforming the best oracle baseline (FBPConvNet) by +1.67 dB and +2.81\% respectively. The superior performance demonstrates that TAMP's physics-driven pretraining on 10.8M diverse SimNICT samples learns more generalizable enhancement patterns than scene-specific features from limited data, while achieving 25-61\% lower standard deviations across scenarios for enhanced robustness.

\begin{table*}[h]
\centering
\caption{Oracle ensemble comparison: TAMP vs perfectly-matched specialized models.}
\label{tab:sup_oracle}
\begin{tabular}{lcccc}
\toprule
Method & PSNR (dB) $\uparrow$ & SSIM (\%) $\uparrow$ & RMSE (HU) $\downarrow$ & LPIPS (\%) $\downarrow$ \\
\midrule
Oracle RED-CNN & 43.90$\pm$4.80 & 91.75$\pm$11.20 & 30.33$\pm$16.51 & 9.73$\pm$9.58 \\
Oracle FBPConvNet & 44.49$\pm$5.02 & 94.54$\pm$5.95 & 28.64$\pm$16.91 & 7.86$\pm$5.56 \\
Oracle ProCT & 42.71$\pm$4.40 & 91.45$\pm$10.90 & 33.96$\pm$16.99 & 13.42$\pm$10.59 \\
TAMP & \textbf{46.16$\pm$3.30} & \textbf{97.35$\pm$2.33} & \textbf{21.67$\pm$8.11} & \textbf{4.97$\pm$3.63} \\
\bottomrule
\end{tabular}
\end{table*}

\subsection{Ablation study}

\subsubsection{Component ablation}
To verify the contributions of the MITNet architecture, model pretraining, DDEL training strategy, and the four loss functions in various NICT enhancement tasks, we conducted an ablation study, with the results shown in Table \hyperref[Ablation study]{8}. The ablation study systematically evaluates three key components:

\textbf{1) MITNet architecture contribution:} By replacing RED-CNN's single-scale channel residual convolution network with our multi-scale integrated transformer architecture (MITNet), the model shows improvements across all three tasks, with particularly notable enhancement in LACT task (2.02 dB PSNR improvement from 37.51 to 39.53 dB), moderate improvement in SVCT task (0.10 dB improvement from 43.96 to 44.06 dB), and minimal improvement in LDCT task (0.04 dB improvement from 45.63 to 45.67 dB). This demonstrates that the multi-scale transformer architecture is especially effective for complex geometric artifacts (LACT) where large-scale features are critical, while providing consistent but modest improvements for noise-dominated tasks (LDCT and SVCT).

\textbf{2) DDEL training strategy contribution:} By gradually incorporating the four loss components of DDEL, the model achieves substantial PSNR improvements. The largest single improvement comes from adding SSIM loss, which provides 2.79 dB improvement for LDCT (from 45.67 to 48.46 dB), 2.46 dB for LACT (from 39.53 to 41.99 dB), and 3.95 dB for SVCT (from 44.06 to 48.01 dB), validating that structural similarity constraints are crucial for all NICT types. The VGG loss contributes modest but consistent improvements of 0.16 dB for LDCT, 0.25 dB for LACT, and 0.28 dB for SVCT. The projection domain loss provides the final boost with 0.85 dB for LDCT, 2.03 dB for LACT, and 0.50 dB for SVCT, with particularly strong contribution to LACT reconstruction due to its geometric nature.

\textbf{3) Large-scale pretraining contribution:} By initializing the model weights with pre-trained TAMP and using LoRA fine-tuning, we achieved final PSNR values of 51.62, 45.68, and 50.26 dB for the three tasks, representing improvements of 2.15 dB, 1.41 dB, and 1.47 dB respectively over training from scratch. This consistent 1.4-2.2 dB improvement across all tasks confirms the substantial benefits of large-scale physics-driven pretraining for universal NICT enhancement capabilities, with the largest benefit observed for LDCT tasks where the diverse noise patterns in pretraining data provide robust initialization.

\begin{table*}[htbp]
\caption{Ablation study of TAMP. The first two rows compare the networks of RED-CNN and MITNet. The subsequent rows illustrate the impact of the loss components in DDEL. The final rows demonstrate the effect of initializing network weights with pre-trained TAMP weights. PSNR is used as the evaluation metric.}
\label{Ablation study}
\resizebox{\linewidth}{!}{
\begin{tabular}{lcccccccc}
\toprule
\multirow{2}{*}{Pre-train} & \multirow{2}{*}{MITNet} & \multicolumn{4}{c}{DDEL} & \multicolumn{3}{c}{Task} \\
\cmidrule(lr){3-6} \cmidrule(lr){7-9}
                                    & & Proj loss & VGG loss & SSIM loss & MSE loss     & APET-LDCT-High & APET-LACT-High & APET-SVCT-High \\
\midrule
            &              &               &               &               & \checkmark & 45.63±1.20    & 37.51±0.80    & 43.96±1.13 \\
            & \checkmark   &               &               &               & \checkmark & 45.67±0.97    & 39.53±1.11    & 44.06±0.95 \\
            & \checkmark   &               &               & \checkmark    & \checkmark & 48.46±1.11    & 41.99±1.04    & 48.01±1.04 \\
            & \checkmark   &               & \checkmark    & \checkmark    & \checkmark & 48.62±0.95    & 42.24±1.18    & 48.29±0.96 \\
            & \checkmark   & \checkmark    & \checkmark    & \checkmark    & \checkmark & 49.47±0.94    & 44.27±1.01    & 48.79±1.02 \\
\checkmark  & \checkmark   & \checkmark    & \checkmark    & \checkmark    & \checkmark & 51.62±1.07    & 45.68±0.98    & 50.26±1.04 \\
\bottomrule
\end{tabular}
}
\end{table*}

\subsubsection{Training data size ablation}
To validate the effect of training data size, we pre-trained TAMP variants on 100k, 1M, and 10.8M image pairs respectively, and evaluated their performance across all 27 benchmark tasks. As shown in Table~\ref{tab:data_scale_ablation}, the results demonstrate consistent and substantial improvement with increased data scale across all three NICT categories. Scaling from 100k to 1M image pairs yields significant improvements of 3.35 dB, 3.17 dB, and 2.83 dB for LDCT, LACT, and SVCT respectively. Further scaling to the full 10.8M dataset brings additional substantial gains of 4.11 dB, 4.08 dB, and 3.43 dB respectively, demonstrating that large-scale pretraining continues to improve performance without saturation. This scaling behavior validates the importance of large-scale physics-driven pretraining for establishing effective foundation models.

\begin{table*}[h]
\centering
\caption{Effect of pretraining data size on NICT enhancement performance. PSNR (dB) is reported as mean$\pm$std.}
\label{tab:data_scale_ablation}
\begin{tabular}{lccc}
\toprule
\textbf{Training Data Size} & \textbf{APET-LDCT-High} & \textbf{APET-LACT-High} & \textbf{APET-SVCT-High} \\
\midrule
100k & 44.16$\pm$0.76 & 38.43$\pm$0.82 & 44.00$\pm$0.75 \\
1M & 47.51$\pm$0.75 & 41.60$\pm$0.91 & 46.83$\pm$0.89 \\
10.8M & 51.62$\pm$1.07 & 45.68$\pm$0.98 & 50.26$\pm$1.04 \\
\bottomrule
\end{tabular}
\end{table*}

\subsubsection{Model size ablation}
We investigated the impact of model capacity by training three variants with systematically reduced channel dimensions while maintaining the same network depth and multi-scale architecture. Table~\ref{tab:sup_model_arch} shows the detailed architectural configurations of each variant. All variants were trained on the same 100k image pairs with identical hyperparameters. As shown in Table~\ref{tab:sup_model_perf}, the Base model (7.48M) achieves PSNR of 44.16, 38.43, 44.00 dB for LDCT, LACT, SVCT, substantially outperforming Tiny (0.52M) by 4.83, 1.16, 4.08 dB respectively. The substantial improvements in LDCT and SVCT ($>$4 dB) demonstrate that TAMP's transformer architecture effectively leverages increased capacity to learn complex geometric artifact patterns.

\begin{table*}[h]
\centering
\caption{Detailed architectural configurations of three model variants for model size ablation.}
\label{tab:sup_model_arch}
\renewcommand{\arraystretch}{1.2}
\resizebox{\textwidth}{!}{
\begin{tabular}{l|c|c|c}
\Xhline{1.2pt}
\textbf{Layer Name} & \textbf{MITNet-Base} & \textbf{MITNet-Small} & \textbf{MITNet-Tiny} \\
\Xhline{1.2pt}
Embedding-1 & Conv 3$\times$3, stride 1, ch=2 & Conv 3$\times$3, stride 1, ch=1 & Conv 3$\times$3, stride 1, ch=1 \\
\hline
DFE-1 & $\left[\makecell{\text{RSTB, dim=2} \\ \text{depth=[2]}  \\ \text{heads=[1]}}\right]\times 1$ & $\left[\makecell{\text{RSTB, dim=1} \\ \text{depth=[2]}  \\ \text{heads=[1]}}\right]\times 1$ & $\left[\makecell{\text{RSTB, dim=1} \\ \text{depth=[2]}  \\ \text{heads=[1]}}\right]\times 1$ \\
\hline
Embedding-2 & Conv 4$\times$4, stride 2, ch=8 & Conv 4$\times$4, stride 2, ch=4 & Conv 4$\times$4, stride 2, ch=2 \\
\hline
DFE-2 & $\left[\makecell{\text{RSTB, dim=8} \\ \text{depth=[4,4]}  \\ \text{heads=[2,2]}}\right]\times 2$ & $\left[\makecell{\text{RSTB, dim=4} \\ \text{depth=[4,4]}  \\ \text{heads=[2,2]}}\right]\times 2$ & $\left[\makecell{\text{RSTB, dim=2} \\ \text{depth=[4,4]}  \\ \text{heads=[1,1]}}\right]\times 2$ \\
\hline
Embedding-4 & Conv 6$\times$6, stride 4, ch=32 & Conv 6$\times$6, stride 4, ch=16 & Conv 6$\times$6, stride 4, ch=8 \\
\hline
DFE-4 & $\left[\makecell{\text{RSTB, dim=32} \\ \text{depth=[6,6,6]}  \\ \text{heads=[4,4,4]}}\right]\times 3$ & $\left[\makecell{\text{RSTB, dim=16} \\ \text{depth=[6,6,6]}  \\ \text{heads=[4,4,4]}}\right]\times 3$ & $\left[\makecell{\text{RSTB, dim=8} \\ \text{depth=[6,6,6]}  \\ \text{heads=[2,2,2]}}\right]\times 3$ \\
\hline
Embedding-8 & Conv 10$\times$10, stride 8, ch=128 & Conv 10$\times$10, stride 8, ch=64 & Conv 10$\times$10, stride 8, ch=32 \\
\hline
DFE-8 & $\left[\makecell{\text{RSTB, dim=128} \\ \text{depth=[8,8,8,8]}  \\ \text{heads=[8,8,8,8]}}\right]\times 4$ & $\left[\makecell{\text{RSTB, dim=64} \\ \text{depth=[8,8,8,8]}  \\ \text{heads=[8,8,8,8]}}\right]\times 4$ & $\left[\makecell{\text{RSTB, dim=32} \\ \text{depth=[8,8,8,8]}  \\ \text{heads=[4,4,4,4]}}\right]\times 4$ \\
\hline
Fusion-4 & Upsample + Conv, ch=80 & Upsample + Conv, ch=40 & Upsample + Conv, ch=20 \\
\hline
Fusion-2 & Upsample + Conv, ch=44 & Upsample + Conv, ch=22 & Upsample + Conv, ch=11 \\
\hline
Fusion-1 & Upsample + Conv, ch=96 & Upsample + Conv, ch=48 & Upsample + Conv, ch=24 \\
\hline
Enhance & Conv 3$\times$3, ch=1 & Conv 3$\times$3, ch=1 & Conv 3$\times$3, ch=1 \\
\hline
Params (M) & 7.48 & 1.95 & 0.52 \\
\hline
FLOPs (G) & 96.1 & 24.5 & 6.4 \\
\Xhline{1.2pt}
\end{tabular}
}
\end{table*}

\begin{table*}[h]
\centering
\caption{Model size ablation: performance comparison on APET test cases (PSNR in dB).}
\label{tab:sup_model_perf}
\begin{tabular}{lccc}
\toprule
Model Size & LDCT-High & LACT-High & SVCT-High \\
\midrule
MITNet-Tiny & 39.33$\pm$0.92 & 37.27$\pm$0.70 & 39.92$\pm$0.99 \\
MITNet-Small & 40.26$\pm$0.55 & 37.56$\pm$0.74 & 40.14$\pm$0.55 \\
MITNet-Base & 44.16$\pm$0.76 & 38.43$\pm$0.82 & 44.00$\pm$0.75 \\
\bottomrule
\end{tabular}
\end{table*}

\subsection{Downstream clinical validation}
We conducted three downstream clinical validation experiments to validate that TAMP-enhanced images improve performance across critical diagnostic applications including pulmonary nodule segmentation, pulmonary nodule malignancy diagnosis, and radiomics feature analysis.

\subsubsection{Pulmonary nodule segmentation}
Pulmonary nodule segmentation is a critical component in lung cancer diagnosis, as accurate delineation of nodule boundaries directly impacts malignancy assessment and treatment planning. We conducted pulmonary nodule segmentation experiments on the LIDC-IDRI dataset \cite{armato2011lidc} to evaluate how TAMP enhancement improves automated lesion delineation accuracy.

We selected CT slices with expert consensus nodule masks, splitting them into 8,553, 856, 2,139 slices for training, validation, test respectively. ICT images were synthetically degraded into three NICT types (LDCT, SVCT, LACT) and subsequently enhanced by TAMP. A U-Net \cite{ronneberger2015unet} segmentation model was trained on ICT images with their corresponding masks, then tested on both NICT-degraded and TAMP-enhanced images to quantify segmentation accuracy improvements.

TAMP enhancement effectively restores segmentation accuracy degraded by NICT artifacts, validating improved diagnostic utility for lesion delineation. As shown in Table~\ref{tab:sup_seg}, TAMP improves Dice from 0.676, 0.527, 0.451 to 0.690, 0.662, 0.621 for LDCT, SVCT, LACT, corresponding to absolute gains of +0.014, +0.135, +0.170. The substantial improvements in SVCT and LACT demonstrate that TAMP's learned representations capture geometric artifact patterns and effectively restore boundary information corrupted by angular undersampling.

\begin{table*}[h]
\caption{Our enhancement improves pulmonary nodule segmentation on LIDC-IDRI (w/o vs. w/ our method).}
\label{tab:sup_seg}
\centering
\begin{tabular}{lcccccc}
\toprule
\multirow{2}{*}{Input} & \multicolumn{3}{c}{Dice} & \multicolumn{3}{c}{IoU} \\
\cmidrule(lr){2-4} \cmidrule(lr){5-7}
& w/o Enhance & w/ Ours & $\Delta$ & w/o Enhance & w/ Ours & $\Delta$ \\
\midrule
LDCT & 0.676 & 0.690 & +0.014 & 0.527 & 0.544 & +0.017 \\
SVCT & 0.527 & 0.662 & +0.135 & 0.380 & 0.513 & +0.133 \\
LACT & 0.451 & 0.621 & +0.170 & 0.314 & 0.468 & +0.154 \\
\bottomrule
\end{tabular}
\end{table*}

\subsubsection{Pulmonary nodule malignancy diagnosis}
Accurate diagnosis of nodule malignancy levels is critical for lung cancer screening, as it directly impacts clinical decision-making for follow-up and treatment. We conducted pulmonary nodule malignancy diagnosis experiments on the LIDC-IDRI dataset to evaluate how TAMP enhancement improves malignancy assessment.

We formulated a five-class malignancy classification task at the 2D slice level, where nodules are classified into expert consensus ratings: 1 (highly unlikely for cancer), 2 (moderately unlikely), 3 (indeterminate likelihood), 4 (moderately suspicious), and 5 (highly suspicious for cancer). A ResNet-18 classifier was trained on 8,458 ICT training slices with their malignancy labels, then tested on 2,139 test slices across seven scenarios: ICT baseline, three NICT types (LDCT, SVCT, LACT), and three TAMP-enhanced types to quantify classification accuracy improvements.

TAMP enhancement effectively recovers diagnostic accuracy degraded by NICT artifacts, validating improved clinical utility for malignancy assessment. As shown in Table~\ref{tab:sup_diag}, TAMP improves accuracy from 0.817, 0.740, 0.758 to 0.831, 0.843, 0.839 for LDCT, SVCT, LACT, corresponding to absolute gains of +0.014, +0.103, +0.081. The substantial improvements in SVCT and LACT demonstrate that TAMP's learned representations effectively restore diagnostically critical features corrupted by geometric artifacts, maintaining strong discriminative ability (AUC $>$ 0.96) for reliable lung cancer screening.

\begin{table*}[h]
\caption{Our enhancement improves pulmonary nodule malignancy diagnosis on LIDC-IDRI (w/o vs. w/ our method).}
\label{tab:sup_diag}
\centering
\begin{tabular}{lcccccc}
\toprule
\multirow{2}{*}{Input} & \multicolumn{3}{c}{Accuracy} & \multicolumn{3}{c}{AUC} \\
\cmidrule(lr){2-4} \cmidrule(lr){5-7}
& w/o Enhance & w/ Ours & $\Delta$ & w/o Enhance & w/ Ours & $\Delta$ \\
\midrule
LDCT & 0.817 & 0.831 & +0.014 & 0.962 & 0.967 & +0.005 \\
SVCT & 0.740 & 0.843 & +0.103 & 0.932 & 0.966 & +0.034 \\
LACT & 0.758 & 0.839 & +0.081 & 0.939 & 0.964 & +0.025 \\
\bottomrule
\end{tabular}
\end{table*}

\subsubsection{Radiomics feature analysis}
Radiomics analysis is highly sensitive to image quality, and feature preservation is critical for reliable downstream clinical applications. We evaluated how TAMP affects quantitative radiomics features to assess its clinical value in preserving quantitative imaging biomarkers.

We extracted 93 standard 2D radiomics features from ICT baseline, three NICT types (LDCT, SVCT, LACT), and TAMP-enhanced images using PyRadiomics on the same LIDC-IDRI lung nodules. Feature stability was quantified by computing Pearson correlations between NICT (or TAMP) features and ICT reference features, where stable features achieving $r>0.85$ are considered reliable for downstream clinical analysis.

TAMP enhancement partially restores radiomics feature preservation degraded by NICT artifacts, validating improved quantitative imaging biomarker reliability. As shown in Table~\ref{tab:sup_radio}, TAMP improves mean correlation from 0.646, 0.722, 0.721 to 0.685, 0.741, 0.750 for LDCT, SVCT, LACT, corresponding to absolute gains of +0.039, +0.019, +0.029. The modest improvements compared to segmentation and diagnosis tasks demonstrate that while TAMP's learned representations effectively restore visual quality and diagnostic features, quantitative radiomics features remain more sensitive to reconstruction artifacts, indicating the need for careful interpretation in radiomics-based clinical analysis.

\begin{table*}[h]
\caption{Our enhancement improves radiomics feature preservation (w/o vs. w/ our method).}
\label{tab:sup_radio}
\centering
\begin{tabular}{lcccccc}
\toprule
\multirow{2}{*}{Input} & \multicolumn{3}{c}{Mean Correlation ($r$)} & \multicolumn{3}{c}{Stable Rate ($r>0.85$)} \\
\cmidrule(lr){2-4} \cmidrule(lr){5-7}
& w/o Enhance & w/ Ours & $\Delta$ & w/o Enhance & w/ Ours & $\Delta$ (\%) \\
\midrule
LDCT & 0.646 & 0.685 & +0.039 & 40.9\% & 41.9\% & +1.0\% \\
SVCT & 0.722 & 0.741 & +0.019 & 51.7\% & 52.7\% & +1.0\% \\
LACT & 0.721 & 0.750 & +0.029 & 48.4\% & 52.7\% & +4.3\% \\
\bottomrule
\end{tabular}
\end{table*}

\subsection{TAMP model analysis}

\begin{figure*}[thbp] 
\centering
\includegraphics[width=\linewidth]{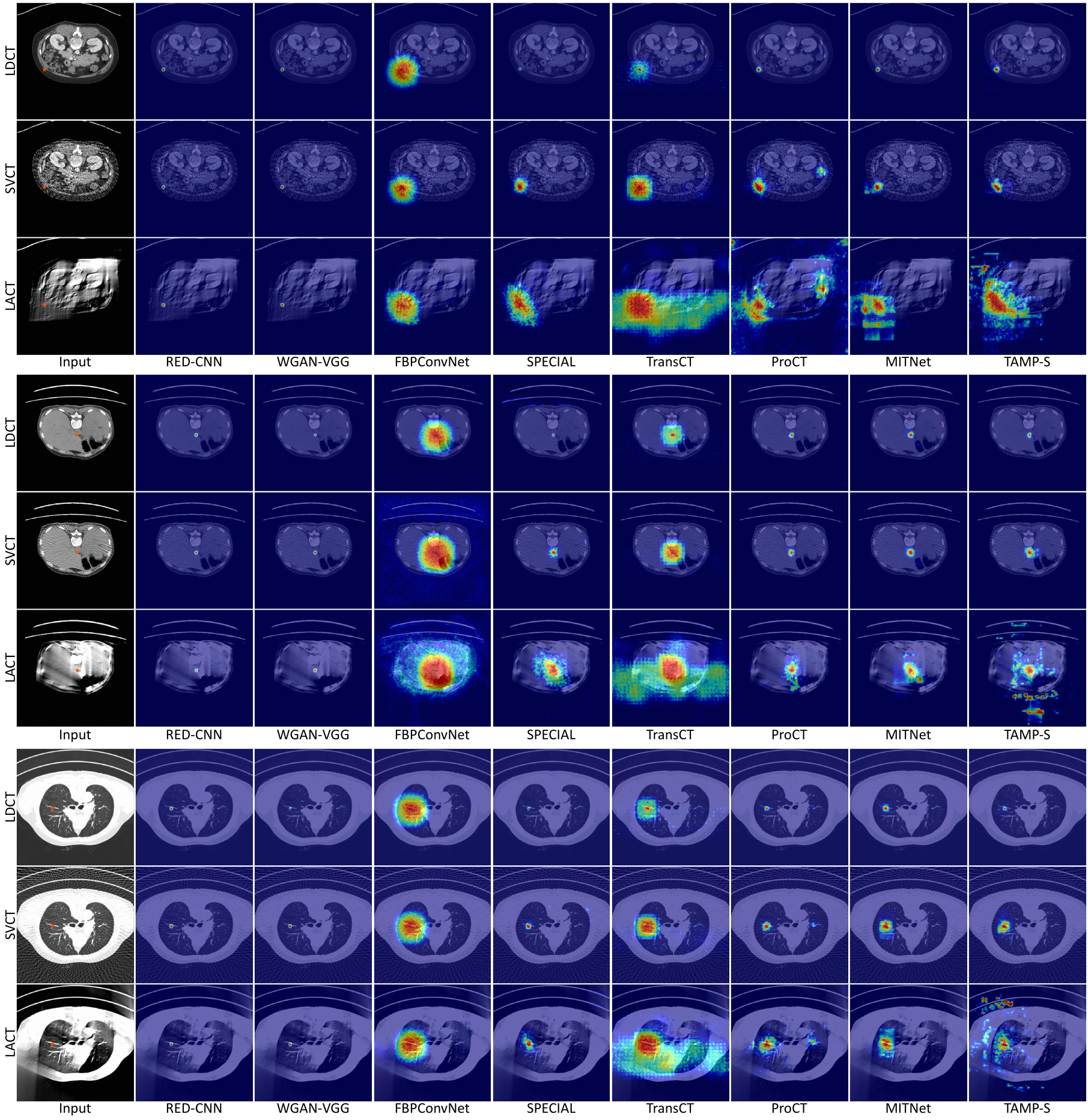}
\caption{The effective receptive fields of different methods for inferring various types of NICT images. TAMP-S effectively extracts artifact features of different scales and shapes.
}
\label{erf} 
\end{figure*}

TAMP-S effectively extracts artifact features of various scales and shapes with its multi-scale integrated transformer network through pre-training and adaptation. As shown in Fig. \ref{erf}, we use Effective Receptive Fields (ERF) to visually indicate the input image regions that receive focused attention during specific processing units, enabling a detailed analysis of model behavior and feature extraction characteristics. 
Our TAMP-S demonstrates precise ERF coverage of relevant regions, such as the concentrated ERF of the detailed region in LDCT and the wedge-shaped artifact in LACT, allowing it to accurately characterize a wide range of NICT defects and capture the most effective features for image enhancement. In contrast, single-scale convolutional networks (e.g., RED-CNN and WGAN-VGG) exhibit uniformly small-scale ERFs that fail to represent large-scale artifact structures. Although multi-scale U-Nets (e.g., FBPConvNet and SPECIAL) adapt to defects at different scales, their ERFs are approximately elliptical, indicating limited adaptability to precise artifact shapes. Furthermore, although multi-scale transformer networks (e.g., TransCT and ProCT) can accommodate artifacts of varying scales and shapes, their ERF coverage does not align well with the areas of interest, as evidenced by the inclusion of irrelevant regions in LACT.

\bibliography{sn-bibliography}
\end{document}